\documentclass[aps,prd,twocolumn,10pt,superscriptaddress,amssymb,amsmath,nofootinbib]{revtex4-2}
\usepackage[utf8]{inputenc}
\usepackage{lmodern}
\usepackage[T1]{fontenc}
\usepackage{graphicx} 
\usepackage[pdftex,breaklinks,colorlinks,
linkcolor=Blue,
citecolor=teal,
anchorcolor=red,
urlcolor=cyan,
pdfencoding=auto]{hyperref}
\usepackage[dvipsnames]{xcolor}
\usepackage{mathtools}
\usepackage{mathrsfs}
\usepackage{booktabs}
\usepackage{longtable}
\usepackage{soul}
\usepackage{orcidlink}
\usepackage[mathscr]{euscript}
\newcommand{\beq}{\begin{equation}\begin{aligned}}
\newcommand{\eeq}{\end{aligned}\end{equation}}

\usepackage{hyperref}
\hypersetup{colorlinks=true}

\def\equationautorefname~#1\null{%
	Eq.~(#1)\null
}
\def\figureautorefname~#1\null{%
	Fig.~#1\null
}
\def\tableautorefname~#1\null{%
	Table.~#1\null
}
\def\sectionautorefname~#1\null{%
	Section #1\null
}
\def\appendixautorefname~#1\null{%
	Appendix #1\null
}

\newcommand{\NCUa}{Department of physics, Nanchang University, Nanchang, 330031, China}
\newcommand{\NCUb}{Center for Relativistic Astrophysics and High Energy Physics, Nanchang University, Nanchang, 330031, China}

\begin{document}

\title{
Massive scalar fields in eccentric regime: Detectability and constraints from LISA observations of extreme mass-ratio inspirals}

\author{Tieguang Zi\,\orcidlink{0000-0003-0046-2056}}
\email{zitieguang@ncu.edu.cn}
\affiliation{\NCUa}
\affiliation{\NCUb}

\author{Shailesh Kumar\,\orcidlink{0000-0001-7072-9452}}
\email{ shaileshkumar.1770@gmail.com}
\affiliation{Department of Physics, Indian Institute of Technology, Kharagpur 721 302, India}

\author{Genliang Li \orcidlink{?}}
\email{ligenliang20@mails.ucas.ac.cn}
\affiliation{College of Computer and Big data, Putian University, Putian 351100, Fujian, China}

\author{Jun-Kun Zhao \orcidlink{0000-0002-5835-3920}}
\email{junkunzhao@itp.ac.cn}
\affiliation{Institute of Theoretical Physics, Chinese Academy of Sciences, Beijing 100190, China}

\author{Bao-Min Gu \orcidlink{0000-0002-5221-2625}}
\email{gubm@ncu.edu.cn}
\affiliation{\NCUa}
\affiliation{\NCUb}
\author{Fu-Wen Shu \orcidlink{0000-0002-0752-6457}}
\email{shufuwen@ncu.edu.cn}
\affiliation{\NCUa}
\affiliation{\NCUb}

\begin{abstract}
Extreme mass-ratio inspirals (EMRIs) are among the most promising sources for future space-based gravitational wave (GW) observatories and provide sensitive probes of additional fundamental fields in the strong-gravity regime. In the present work, we investigate eccentric equatorial EMRIs around Kerr black holes in the presence of a massive scalar field. We assume that the inspiralling object carries a scalar charge, therefore, emits scalar radiation in addition to GWs. By solving the massive scalar perturbation equation in the frequency domain, we compute the relativistic scalar energy fluxes at the event horizon and at infinity and incorporate them into an adiabatic inspiral model. We investigate the impact of the scalar charge and scalar field mass on the orbital evolution and gravitational waveforms by analysing the accumulated phase differences and waveform mismatches relative to both general relativity (GR) and the massless scalar limit. Our results show that the massive scalar radiation can produce significant GW dephasing, with the effect becoming increasingly pronounced for more eccentric orbits. In addition, the scalar flux is suppressed for larger scalar field masses as fewer scalar harmonics satisfy the propagation condition at infinity. Finally, using both Fisher-information-matrix forecasts and Bayesian parameter estimation, we assess the capability of the Laser Interferometer Space Antenna to constrain the scalar charge and scalar field mass. The two inference approaches gives consistent constraints, demonstrating that eccentric EMRIs provide a promising avenue for probing massive scalar fields and scalar-tensor extensions of gravity in the strong-field regime.
\end{abstract}

\maketitle

\tableofcontents


\section{Introduction}
The growing precision and number of gravitational wave (GW) observations have made compact binaries powerful laboratories for testing gravity \cite{LIGOScientific:2016aoc, LIGOScientific:2017bnn,LIGOScientific:2016sjg, LIGOScientific:2018dkp}, motivating studies of how deviations from general relativity (GR) may modify binary evolution, the resulting GW signals and further the distinguishability of such effects from standard astrophysical signatures \cite{Gair:2012nm, Amaro-Seoane:2014ela, Berti:2015itd, Cardenas-Avendano:2024mqp, Cardoso:2019rvt}. Among the GW sources accessible to future space-based detectors, extreme mass-ratio inspirals (EMRIs) are one of the most important and powerful probes of new gravitational physics, including precision waveform modelling \cite{LISAConsortiumWaveformWorkingGroup:2023arg}, characteristics beyond-vacuum General Relativity (GR), and additional fundamental fields \cite{Babak:2017tow, Amaro-Seoane:2014ela, Cardenas-Avendano:2024mqp}. An EMRI is a binary system that consists of a stellar-mass object (called secondary) orbiting a supermassive black hole (SMBH, called primary) with the mass-ratio lying in the range $\epsilon \equiv m_p/M=10^{-4}-10^{-7}$, where $q$ is the mass-ratio with $m_p$ being the mass of the secondary and $M$ being the mass of the primary SMBH. In such systems, the stellar-mass body can complete $10^{4}-10^{6}$ orbital cycles before plunging \cite{Amaro-Seoane:2014ela, Babak:2017tow, Cardenas-Avendano:2024mqp}. As a result, any small deviations from GR can accumulate over the inspiral dynamics and generate measurable phase differences or dehasings in the GW signal. Given this, the phase sensitivity of low-frequency detectors (in mHz range) such as LISA \cite{2017arXiv170200786A, LISA:2022kgy, LISA:2024hlh, Barsanti:2024kul}, TianQin \cite{TianQin:2015yph, TianQin:2020hid}, Taiji \cite{Ruan:2018tsw} and DECIGO \cite{Seto:2001qf, Kawamura:2020pcg} make EMRIs an exceptional testbed for testing/probing gravity and constraining fundamental fields. In particular, EMRI observations have been extensively carried out as probes of black hole environments and constraints on modified theories of gravity~\cite{Han:2016zee, Han:2018hby, Zi:2021pdp, Zi:2021pdp, Speri:2022upm, Cardoso:2022whc, Shen:2023pje, Spiers:2023cva, Zi:2023omh, Zi:2023omh, Rahman:2023sof, Duque:2023seg, Figueiredo:2023gas, Kumar:2024utz, Kumar:2024our, Speeney:2024mas, Zi:2024itp, Duque:2024mfw, Kumar:2024dql, Shen:2025svs, Gliorio:2025cbh, Kumar:2025jsi, Kumar:2025njz, Li:2025ffh, Duque:2025yfm, Zi:2025jxy, Rahman:2025mip, HegadeKR:2025dur, HegadeKR:2025rpr, Zhong:2026jnp, Zi:2026zpw}.

A large class of theories beyond GR suggests that gravity is not acting alone; it may be accompanied by additional channels or scalar degrees of freedom, that may include a wide range of extensively studied possibilities, particularly those of interest to us involving scalar fields, describing gravitational interactions in theories beyond GR \cite{Sotiriou:2013qea, Julie:2019sab}. {\color{black}Early studies analyzed coupled gravitational and scalar perturbations in dynamical Chern–Simons gravity and subsequently applied this framework to quasicircular EMRIs~\cite{Molina:2010fb,Pani:2011xj}. Later works investigated massive scalar floating and sinking resonances~\cite{Cardoso:2011xi}, quasicircular EMRIs in scalar–tensor theories~\cite{Yunes:2011aa}, and matter-induced effective scalar masses around black holes~\cite{Cardoso:2013fwa}. Fujita and Cardoso subsequently extended the floating-resonance analysis to eccentric equatorial orbits~\cite{Fujita:2016yav}.}
Focusing on EMRIs, if the stellar-mass object carries a nonzero scalar charge, we know that its dynamics not only emit the standard tensor GWs, but also radiate scalar energy, consequently modifying the orbital evolution of EMRI dynamics. These changes will accumulate throughout the inspiral and can leave observable imprints in GW signals. In this line of research, recent works have thoroughly studied this effect for a massless scalar field \cite{Maselli:2020zgv, Maselli:2021men, Barsanti:2022ana, Zhang:2022rfr, Guo:2022euk, Spiers:2023cva, DellaRocca:2024pnm, Speri:2024qak, Zi:2025lio, Gliorio:2026yvh}, giving the scalar field a mass changes the game entirely \cite{Barsanti:2022vvl, Xie:2024xex}. A nonzero mass completely reshapes how the scalar waves propagate and how energy escapes the system \cite{Barsanti:2022vvl}. As a result, the orbital breakdown and the GW phase look radically different from both standard GR and massless cases. Investigating such deviations in direct GW observations gives us an important pathway to test the strong-gravity field regions that further enables us to discover massive scalar fields or place tight constraints on their existence with future GW observations.

On the other hand, most studies of EMRIs in both GR and beyond GR have focused on quasi-circular inspirals, largely because circular orbits provide a simpler framework for waveform modeling and parameter estimation \cite{Maselli:2020zgv, Maselli:2021men, Compere:2021zfj, Barsanti:2022vvl, Xie:2024xex, DellaRocca:2024pnm, Warburton:2024xnr}. Nevertheless, as astrophysical systems are expected to carry some nonzero eccentricity, in view of this, eccentric orbits have attracted growing attention in recent studies of binary dynamics, especially for EMRIs \cite{Decanini:2015yba, Barsanti:2022ana, Zhang:2022rfr, Speri:2024qak, Zi:2025lio, Gliorio:2026yvh}. In contrast to circular motion, eccentric orbits are characterized by distinct radial and azimuthal frequencies, leading to a richer harmonic structure in the emitted radiation. The resulting waveform, along with effects from additional channels/sources, contains a larger set of harmonics, that encode relevant information about the underlying spacetime and orbital dynamics. Moreover, the compact object repeatedly traverses a broad range of orbital radii during each cycle, probing different regions of the strong-field geometry. The wider coverage of spacetime can increase sensitivity to even small deviations/departures from GR and to additional degrees of freedom, indicating effects of non-GR theories or galactic environments \cite{Cardoso:2019rou, Cardoso:2021wlq, Speeney:2022ryg, Figueiredo:2023gas, Rahman:2023sof, Zhang:2023ceh, Zi:2024jla, Zhang:2024ugv, Qiao:2024gfb, Zi:2024lmt, Mitra:2025tag, Muguruza:2026hqn}. Such considerations make eccentric EMRIs promising systems for investigating the effects of massive scalar fields. At the same time, the theoretical description becomes considerably more involved. In eccentric motion, the combined radial and azimuthal motion generates a rich spectrum of gravitational and scalar modes \cite{Barsanti:2022ana}, whereas the presence of a massive scalar field modifies the perturbation dynamics/equations through its mass-dependent effective potential and then associated radiations. Consequently, despite the recent interest in eccentric EMRIs, a systematic relativistic study of their dynamics and GW signatures in the presence of massive scalar radiation remains relatively unexplored, which is the main context of this present study. {\color{black}In addition, as mentioned earlier, massive scalar fields can also generate narrow resonances when an orbital harmonic approaches the frequency of a scalar quasi-bound state. In the superradiant regime, the resonant horizon flux extracts rotational energy from the black hole spin and may produce floating orbits, whereas non-superradiant resonances may accelerate the inspiral, producing sinking orbits \cite{Cardoso:2011xi, Fujita:2016yav}. These resonances remain relevant for eccentric equatorial orbits but require careful mode-by-mode treatment. This also motivates one to investigate the resonant massive-scalar fluxes for eccentric EMRIs and assess their impact on the orbital evolution.}

In this work, we study the impact of a massive scalar field on eccentric EMRIs orbiting Kerr black holes and evaluate the prospects for constraining the properties of the scalar field with future low-frequency detector like LISA. We consider a scalar-tensor scenario, where the Kerr geometry well approximates the spacetime background and the inspiralling object carries a scalar charge which sources a massive scalar perturbation \cite{Barsanti:2022vvl, Saravani:2019xwx}. Working within the black hole perturbation theory \cite{Teukolsky:1972my, Teukolsky:1973ha, Pound:2021qin} and the adiabatic approximation \cite{Hughes:1999bq, Fujita:2004rb, Khalvati:2025znb}, we compute both tensor gravitational fluxes and massive scalar fluxes for eccentric equatorial inspirals. We solve the scalar perturbation equation in the Kerr background using a frequency-domain decomposition and compute the scalar energy and angular momentum fluxes for the parameter space taken in this work. The generated fluxes are represented using Chebyshev interpolants \cite{Speri:2024qak}, which are then used in the orbital evolution computation. Thus, the resulting fluxes enable rapid evolution of eccentric EMRIs starting from their initial orbital configurations to the last stable orbit (LSO). 
{\color{black}We also investigate resonant massive-scalar fluxes from eccentric EMRIs and assess whether the resulting floating- and sinking-orbit signatures can be distinguished in the orbital evolution.}
Further, we quantify the impact of the massive scalar field through accumulated GW dephasings and waveform mismatches relative to both GR and the corresponding massless scalar theory, including a detailed comparison. We then perform a Fisher Information Matrix (FIM) analysis and Bayesian parameter estimation, with the latter implemented through Markov chain Monte Carlo (MCMC) sampling, to evaluate the ability of LISA to measure or constrain the scalar charge and scalar field mass \cite{Vallisneri:2007ev}. The FIM provides an estimate of the expected statistical uncertainties, whereas the Bayesian analysis explores the full posterior distributions and enables a more robust assessment of parameter constraints. Briefly, as an outcome, we find that massive scalar radiation alters the evolution of eccentric EMRIs and generates measurable waveform dephasings. The impact depends on the scalar charge and scalar field mass and is typically enhanced by eccentricity. This allows future LISA observations to place meaningful constraints on the massive scalar fields considered in this work.

The remainder of this paper is organized as follows. In Sec. (\ref{method}) we present the theoretical framework, derive the massive scalar perturbation equations, and describe the computation of scalar and gravitational fluxes for eccentric EMRIs. We also outline the inspiral evolution scheme, waveform construction, and parameter-estimation methodology. In Sec. (\ref{method}) we compare scalar and gravitational fluxes, investigate the resulting dephasings and waveform mismatches, and derive projected constraints on the scalar charge and scalar field mass with LISA observations, including the parameter estimation through FIM. Finally, Sec. (\ref{result}) summarizes our conclusions and discusses future directions.

\section{Method}\label{method}
\subsection{Action and perturbative equations}
The generic action for the family of scalar–tensor theories of gravity can be written as \cite{Saravani:2019xwx, Barsanti:2022vvl}
\begin{align}\label{action}
S[\mathbf{g}_{\mu\nu},\psi, \mathbf{\Psi}] = S_{0}[\mathbf{g}_{\mu\nu},\psi]+\alpha S_{c}[\mathbf{g}_{\mu\nu},\psi]+S_{\textup{mat}}[\mathbf{g}_{\mu\nu},\psi,\mathbf{\Psi}]\;,
\end{align}
where $S_0[\mathbf{g}_{\mu\nu},\psi]$ is the action related to the spacetime background
\begin{align}
S_0[\mathbf{g}_{\mu\nu},\psi] = \int d^{4}x\frac{\sqrt{-\mathbf{g}}}{16\pi}\Big(R-\frac{1}{2}\partial_{\mu}\psi \partial^{\mu}\psi 
-\frac{1}{2} \mu^2 \psi^2 \Big),
\end{align}
and $R$ is the Ricci scalar described by metric $g_{\mu\nu}$ and $\mu$ is the mass of the scalar field. 
The action of the matter field $\mathbf{\Psi}$ is determined by $S_{\rm mat}$.
The term $\alpha S_c$ denotes the non-minimal coupling between the scalar field and the spacetime background,  which describes the interactions suppressed by a characteristic energy scale in the physical unit \cite{Saravani:2019xwx}, and the constant $\alpha$ has dimensions of $\rm(mass)^n$ with $n>1$. 
For the case of $\alpha=0$, the theory in Eq.~\eqref{action} satisfies the no-hair theorem, where  SMBH can be characterized by the Kerr metric.
On the other hand, the spacetime geometry of SMBH in a large class of theories is determined by the action in Eq.~\eqref{action}, smoothly connecting to the Kerr solution when the parameter $\alpha$ approaches zero. The deviation from a Kerr black hole in such gravity theories is subjected to a dimensionless parameter defined by the  parameter $\alpha$ and mass-ratio $\epsilon=m_p/M$, that is,
\begin{equation}
\begin{aligned}\label{eq:xi:approxi}
\xi\equiv \frac{\alpha}{M^n} = \epsilon^n \frac{\alpha}{m_p^n} = \epsilon^n\xi_p\;.
\end{aligned}
\end{equation}
In an EMRI configuration, a secondary object of mass ($m_p$) inspirals around a massive black hole (MBH) of mass ($M$), where orbital dynamics can be treated within the black hole perturbation theory. One can obtain a parameter $\xi_p=\alpha/m_p^n$, which is constrained to $\xi_p\leq1$, from GW and electromagnetic observations of astrophysical stellar-mass objects~\cite{Silva:2020acr, Saffer:2021gak, Wang:2021jfc, Lyu:2022gdr, Sanger:2024axs}. Thus, we obtain the approximate inequality $\xi \leq \epsilon^n \ll 1$, implying that the background spacetime of the MBH can be well approximated by the Kerr geometry, since deviations from Kerr are strongly suppressed by the small parameter $\xi$. In summary, when $\alpha=0$, the theory reduces to GR, and the MBH in an EMRI system is described by the Kerr solution with a constant scalar field, $\psi=\mathrm{constant}$.
For $\alpha\neq0$ but $\xi \ll 1$, the background spacetime remains approximately Kerr with a small deviation of order $\mathcal{O}(\xi) = \mathcal{O}(\epsilon^n \xi_p)$, and EMRI dynamics can be dealt with the perturbation method at the adiabatic order of $\epsilon$.

For a typical EMRI system, the characteristic size of the secondary object is much smaller than the characteristic curvature scale of the background spacetime. Therefore, within the skeleton approach developed in Refs.~\cite{Eardley:1975fgi, Maselli:2020zgv}, the inspiralling object can be modelled as a point particle. The action $S_{\textup{mat}}$ of the matter field is replaced with the particle's action $S_p$, which is determined by an integral of the scalar function $m(\psi)$ over the worldline of the point-particle $y^\mu(\lambda)$ in a reference frame
\begin{equation}
S_p = -\int m(\psi)\sqrt{g_{\mu\nu}\frac{dy^\mu}{d\lambda}\frac{dy^\nu}{d\lambda}}\;,
\end{equation}
where the scalar function $m(\psi)$ is related to the value of the scalar field at the location of the particle,  $u^\mu=dy^\mu/d\lambda$ is four-velocity and $\lambda$ is proper time for the moving point-particle.

One can expand the metric of MBH and the scalar field in terms of mass-ratio $\epsilon$
\begin{equation}
g_{\mu\nu} = g_{\mu\nu}^{(0)} + \epsilon h_{\mu\nu}^{(1)} +\cdots\;, \quad \psi= \psi^{(0)} +\epsilon \psi^{(1)}+\cdots\;, 
\end{equation}
Note that we only consider the leading dissipative contribution of EMRI dynamics in our work, which is encoded with the linear order of gravitational and scalar perturbations $h_{\mu\nu}^{(1)}$ and $\psi^{(1)}$. Following \cite{Barsanti:2022ana, Barsanti:2022vvl},  varying the action and expanding in the EMRI mass-ratio, the corrections proportional to $\xi=\alpha/M^n\ll1$ are subleading at the adiabatic order considered here. The leading gravitational and scalar perturbation equations therefore reduce to a set of equations
\begin{eqnarray}
G^{\mu\nu}[h^{(1)}_{\mu\nu}]  = 8\pi m_p \int \frac{\delta^{(4)} (x-x(\tau))}{\sqrt{-g}} \frac{dx^\mu}{d\tau}\frac{dx^\nu}{d\tau} d\tau \label{eq:grav:pert}\;,\\
(\nabla_\mu \nabla^\mu -\mu^2)\psi^{(1)} = -4\pi q_s m_p \int \frac{\delta^{(4)} (x-x(\tau))}{\sqrt{-g}} d\tau  \label{eq:scal:pert}\;.
\end{eqnarray}
For notational simplicity, we omit this superscript of $(1)$  in the following sections. Furthermore, the gravitational field equation in Eq.~\eqref{eq:grav:pert} has the same form as that in GR. The massive scalar field equation in Eq.~\eqref{eq:scal:pert} has an extra source term, where the scalar charge and mass induced by the secondary object determine the magnitude of the scalar perturbation. Therefore, the scalar charge ($q_s$) and the scalar-mass ($\mu$) are the key parameters controlling the deviations of the EMRI dynamics from GR. In fact, the scalar field equation can be mapped onto the parameter space of a broad class of modified gravity theories, consequently, tighter constraints on the scalar charge ($q_s$) from future LISA observations can be translated into improved bounds on the parameters characterizing these theories \cite{Maselli:2021men, Julie:2022huo, Speri:2024qak}.

\subsection{Gravitational and massive scalar fluxes in EMRIs}
In the adiabatic approximation, EMRI orbital evolution is governed by gravitational and scalar fluxes obtained from the perturbative Eqs.~\eqref{eq:grav:pert} and \eqref{eq:scal:pert} within the Teukolsky formalism~\cite{Teukolsky:1972my, Teukolsky:1973ha}. Accurate Kerr fluxes are computed by summing multipolar modes for inclined circular, equatorial eccentric, and generic orbits~\cite{Maselli:2020zgv, Barsanti:2022vvl, Barsanti:2022gjv, Guo:2022euk, Spiers:2023cva, Speri:2024qak, Li:2025ffh, Zi:2025lio}, using homogeneous Teukolsky solutions and the point-particle source, with implementations available in \texttt{Mathematica}, \texttt{C++}, and the Black Hole Perturbation Toolkit (\texttt{BHPT})~\cite{BHPT}. In this work, we employ a \texttt{Mathematica}-based approach to compute the gravitational energy and angular-momentum fluxes, $(\dot{E},\dot{L})^{H,\infty}_{g}$, at the Kerr horizon and infinity with arbitrary numerical precision~\cite{BHPT, Hughes:2021exa, Barsanti:2022ana}. Although the secondary carries a scalar charge \(q_s\), the back-reaction of the stress-energy tensor associated with the scalar field on the metric perturbation is subleading at the adiabatic order considered \cite{Barsanti:2022ana, Barsanti:2022vvl}; therefore, the gravitational perturbation retains the standard point-particle source in \autoref{eq:grav:pert}, while the scalar charge affects the orbital evolution through the massive-scalar flux discussed below.

We now turn to the massive-scalar perturbation sourced by an eccentric equatorial orbit in the Kerr background. Using Fourier and spheroidal-harmonic decompositions, the scalar field \(\psi\) is separated into radial and angular functions, $R_{\ell mn}(r,\omega)$ and $S_{\ell mn}(r,\omega)$,
\begin{equation}
\psi(t,r,\theta,\phi)=\int d\omega \frac{R_{\ell mn}(r,\omega)}{\sqrt{r^2+a^2}} S_{\ell mn}(\theta,\omega) e^{im\phi}e^{-i\omega t}\;,
\end{equation}
where the radial function $R_{\ell mn}$ can be simplified as the Schr$\ddot{\rm o}$edinger-type equation
\begin{equation}\label{eq:scal:radial}
\frac{d^2 R^s_{\ell mn} (r,\omega)}{dr^2_\star}+ V(r)R^s_{\ell mn} (r,\omega) 
= T^s_{\ell mn}\;.
\end{equation}
The tortoise coordinate $r_\star$ defined by $dr_\star/dr=(r^2+a^2)/\Delta$, $\Delta=r^2+a^2-2Mr$ and the effective radial potential is given by~\cite{Teukolsky:1973ha, Barsanti:2022vvl}
\begin{equation}
V(r)=\Big[\omega-\frac{am}{\rho^2}\Big]^2-\frac{\Delta}{\rho^8}\Big[\rho^4\lambda +2Mr^3+a(\Delta-2Mr)+\frac{\mu^2}{\rho^6}\Big]\;,
\end{equation}
where $\rho=r^2+a^2$ and the eigenvalue $\hat{\lambda}$ of angular function $S_{\ell mn}$ is given by $\lambda=\hat{\lambda}+2ma(\omega-\sqrt{\omega^2-\mu^2})$.
For a secondary body moving on the eccentric and equatorial orbits, the source term is given in terms of the delta-distribution function
\begin{equation}
\begin{aligned}
T^s_{\ell mn} = -4\pi & q_s m_p \frac{\Delta}{u^t\sqrt{a^2+r^2}}S_{\ell mn}(\theta,\phi)|_{\theta=\pi/2}
\nonumber \\ &\times\delta(r-r_p)\delta(\omega-\omega_{mn})\;,
\end{aligned}
\end{equation}
where  $u^t$ is the time component of the four velocity $u^\mu$, $\omega_{mn}$ is the orbital frequency of eccentric orbits on the equatorial plane, which is computed in the subsection.  

With the Green-function method, we can obtain the non-homogeneous solution in Eq.~\eqref{eq:scal:radial}, where one can get two homogeneous solutions $R_{\ell mn}^{\rm in,out}$ by imposing in-going and out-going conditions near the horizon and at infinity. The boundary conditions for solving homogeneous solutions are discussed  in appendix~\ref{subsec:scalar_boundary_conditions}.
Then, the general solutions $R_{\ell mn}^{\rm in,out}$ are evaluated with the integral of $R_{\ell mn}^{\rm in, out}$ over the source term $T^s_{\ell mn}$ numerically \cite{Barsanti:2022ana}.
The massive scalar energy fluxes at horizon and infinity can be determined with the asymptotic behaviors of two general solutions
\begin{equation}\label{eq:edots}
\dot{E}_s^{H,\infty} = \frac{1}{16\pi}\sum_{\ell=1}^{\ell=\infty}\sum_{m=-\ell}^{m=\ell}
\sum_{n=-\infty}^{n=\infty}
\omega_{mn} k_{\mp} |Z_{\ell mn}^{H,\infty}|^2   \;,
\end{equation}
where $k_+=\sqrt{\omega_{mn}^2-\mu^2}$,  $k_-=\omega_{mn}-m\Omega_H$, $\Omega_H=a/(2Mr_H)$, $r_H=M+\sqrt{M^2-a^2}$  and the quantities $Z_{\ell mn}^{H,\infty}$ are given by
\begin{equation}
\begin{aligned}
Z_{\ell mn}^{H,\infty} = R_{\ell mn}^{\rm in,out}({r_\star\longrightarrow\mp\infty })\int_{-\infty}^{+\infty}
\frac{R_{\ell mn}^{\rm in,up}T^s_{\ell mn} dr_\star}{W}
\end{aligned}
\end{equation}
where $W=R_{\ell mn}^{\rm in}dR_{\ell mn}^{\rm out}/dr_\star - R_{\ell mn}^{\rm out}dR_{\ell mn}^{\rm in}/dr_\star$ is the Wronskian. 
The homogeneous solutions are solved by imposing the outgoing and ingoing boundary conditions, the specific method can be found in Appendix (\ref{subsec:scalar_boundary_conditions}).
The angular momentum fluxes for scalar field can be obtained with the energy fluxes
\begin{equation}
\dot{L}_s^{H,\infty} = \frac{m}{\omega_{mn}}\dot{E}_s^{H,\infty}\;.
\end{equation}
Within the adiabatic approximation, the total GW flux is determined by the summation of  gravitational and massive scalar fluxes near the horizon and at infinity
\begin{equation}\label{flux:total}
\dot{\mathcal{C }}_{\rm tot} = \dot{\mathcal{C }}^{H}_s +  \dot{\mathcal{C }}^{\infty}_s + \dot{\mathcal{C }}^{H}_g +  \dot{\mathcal{C }}^{\infty}_g\;,
\end{equation}
where $\mathcal{C}\in[E,L]$ is the orbital energy and angular momentum, the quantities with dots mean their changing rate.

\subsection{Source term  perturbed by scalar-charged secondary object}
In this paper, we focus on the geodesics of eccentric equatorial orbits of the secondary body moving around a Kerr black hole. In the Boyer-Lindquist coordinates, the geodesic equations are written as \cite{Hughes:1999bq}
\begin{equation}
\begin{aligned}
\Sigma^2 \left(\frac{dr}{d\tau}\right)^2 &=V_r(r)= T^2-\Delta[r^2+(L-aE)^2]\;,\\
\Sigma^2 \frac{dt}{d\tau} &=V_t(r) = a(L-aE) +\frac{T(a^2+r^2)}{\Delta} \;,\\
\Sigma^2 \frac{d\phi}{d\tau} &=V_\phi(r)= L-aE +\frac{Ta}{\Delta} \;,\\
\theta(\tau) &= 0\;,
\end{aligned}
\end{equation}
where $\Sigma=r^2$ and $T=E(r^2+a^2)-aL$, $(E, L)$ is the orbital energy and angular momentum per unit mass $m_p$, respectively. Non-circular bounded orbits with eccentricity $e$ have two turning points, given by the apastron $r_a=p/(1-e)$ and periastron $r_{\rm per}=p/(1+e)$ with the semi-latus rectum $p$, the radial effective potential satisfies $V_r(r_a)=V_r(r_{\rm per})=0$. The separation $r_p$ of orbital motion is expressed in terms of an angle $\chi$
\begin{equation}
r_p(\chi)=\frac{p}{1+\cos\chi}\;,
\end{equation}
where $\chi$ is a monotonic parameter and changes from 0 to $\pi$, corresponding to the orbital motion between the periastron and the apastron. The secondary moves repeatedly in radial direction from two turning points, which has a radial period $T_r=t(\chi=2\pi)=2t(\pi)$, and the accumulated azimuthal phase over one radial cycle is $\Delta\phi=\phi(2\pi)$ in time $T_r$.
Therefore, one can get two orbital frequencies 
\begin{equation}
\Omega_r = \frac{2\pi}{T_r}\;,\quad \Omega_\phi = \frac{\Delta\phi}{T_r},
\end{equation}
and $\omega_{mn}$ is related to the EMRI phase with the linear combination of $\Omega_\phi$ and $\Omega_r$ 
\begin{equation}
\omega_{mn} = m\Omega_\phi + n \Omega_r\;,
\end{equation}
where $(m,n)$ takes integer values. Note that two quantities $(T_r,\Delta\phi)$ of the radial and azimuthal coordinates can be determined with integrals over angle $\chi$, which are computed with the following two equations
\begin{equation}
\begin{aligned}
\phi(\chi)& =\int_0^\chi d\chi' \frac{V_\phi(\chi' ,p,e)}{J(\chi' ,p,e) V_r(\chi',p,e)}\;,\\
t(\chi) &=\int_0^\chi d\chi'   \frac{V_t(\chi' ,p,e)}{J(\chi',p,e) V_r(\chi',p,e)}\;,
\end{aligned}
\end{equation}
where these potential functions are as follows \cite{Glampedakis:2002ya}
\begin{equation}
\begin{aligned}
V_r(\chi,p,e)  &= a^2+x^2+2axE - \frac{2Mx^2}{p} (3+e\cos\chi)\;,\\
V_t(\chi,p,e) &= a^2E-\frac{2axM}{p}(1+e\cos\chi)+\frac{Ep^2}{(1+e\cos\chi)^2}\;,\\
V_\phi(\chi,p,e)  &= x+aE-\frac{2Mx}{p}(1+e\cos\chi)\;,\\
J(\chi,p,e) &= 1-\frac{2M}{p}(1+e\cos\chi)+\frac{a^2}{p^2}(1+e\cos\chi)^2\;,
\end{aligned}
\end{equation}
where $x=L-aE$.

Now, we look at the general solution corresponding to scalar-field perturbation. For this, in order to express the integral of two homogeneous radial solutions over the source term, we define them in a simple form
\begin{equation}\label{integral:ZHf}
\psi_{\ell m\omega}^{H,\infty} = \int_{-\infty}^{+\infty} \frac{R_{\ell mn}^{\rm in,up}T^s_{\ell mn} dr_\star}{W}\;,
\end{equation}
the source term $T^s_{\ell mn}$ is obtained by projecting the source into spheroidal harmonic modes~\cite{Barsanti:2022ana}

\begin{align}
\Hat{T}^s_{\ell mn} = -4q_s m_p \int_{-\infty}^{+\infty}  \frac{\delta[r-r_p(t)] }{u^t} S_{\mathcal{\Pi}} e^{i[\omega_{mn}t-m\phi(t)]}dt\;,
\end{align}
where $S_{\mathcal{\Pi}}$ is the value $S(\pi/2,\omega_{mn})$ of the angular function in the equatorial plane. 
Inserting the above equation into the integral Eq.~\eqref{integral:ZHf}, one obtains the following equation
\begin{equation}\label{eq:psi:IHf}
\psi_{\ell m\omega}^{H,\infty} = \int_{-\infty}^{+\infty} \mathcal{I}^{H,\infty}[r_p(t)] e^{i[\omega_{mn}-m\phi(t)]} dt\;,
\end{equation}
with
\begin{equation}
\mathcal{I}^{H,\infty}[r_p(t)] = \Bigg[\frac{-4q_sm_p}{\sqrt{r^2+a^2}}\frac{R_{\ell mn}^{\rm in,up} }{W} \frac{ S_{\mathcal{\Pi}} }{u^t}\Bigg] \Bigg{|}_{r=r_p(t)} \;.
\end{equation}
The integral function in Eq.~\eqref{eq:psi:IHf} can be rewritten as
\begin{equation}
\mathcal{A}^{H,\infty}(t) = \mathcal{I}^{H,\infty}[r_p(t)] e^{-im[\phi(t) - \Omega_\phi t]}\;
\end{equation}
with the azimuthal fundamental frequency $\Omega_\phi$. Because eccentric orbits have period $T_r$ in the radial direction, these functions can be expanded as a Fourier series 
\begin{equation}
\mathcal{A}^{H,\infty}(t) =  \sum_{n=-\infty}^{+\infty} \tilde{\mathcal{A}}^{H,\infty}_n e^{-in\Omega_r t}\;,
\end{equation}
where $\Omega_r$ is the fundamental radial frequency. Using the Fourier expansion of $\mathcal{A}^{H,\infty}(t)$, the time integral of $\psi_{\ell m\omega}^{H,\infty}$ is changed as follows
\begin{equation}
\psi_{\ell m\omega}^{H,\infty} =  \sum_{n=-\infty}^{+\infty} \tilde{\psi}_{\ell m n}^{H,\infty} 
\delta(\omega-\omega_{mn})\;,
\end{equation}
and the integrals can be transformed as 
\begin{equation}
\tilde{\psi}_{\ell m n}^{H,\infty} = \Omega_r \int_0^{T_r} \mathcal{A}^{H,\infty}(t) e^{in\Omega_r} dt\;.
\end{equation}
For an eccentric EMRI radial orbit, the secondary object is moving between the apastron and the periastron, one can adopt the radial parameterized scheme to avoid the numerical divergences; then two integrals of homogeneous solutions over source term on eccentric orbits in Eq.~\eqref{integral:ZHf} can be redefined as
\begin{equation}
\tilde{\psi}_{\ell m n}^{H,\infty} =  \Omega_r \int_0^{2\pi}d\chi 
\frac{V_t(\chi)}{J(\chi)V_r(\chi)^{1/2} } \mathcal{A}^{H,\infty} e^{i\omega_{mn}t(\chi)-m\phi(\chi)}.
\end{equation}
The two quantities above represent the amplitudes of the scalar radiation and are used to compute the scalar fluxes by integrating the source term generated by the secondary object moving on eccentric geodesic orbits. The details of these integrations can also be found in earlier studies ~\cite{Barsanti:2022vvl, Barsanti:2022ana}.

\subsection{Adiabatic inspiral on eccentric orbits}
In this subsection, we employ the adiabatic evolution scheme to compute the inspiral trajectory from an initial geodesic orbit to the LSO in Kerr spacetime. According to the flux-balance law \cite{Hughes:2021exa}, the loss of orbital energy and angular momentum in an EMRI is approximately balanced by the sum of the gravitational and massive scalar fluxes, which is written with Eq.~\eqref{flux:total} as
\begin{equation}
\dot{\mathcal{C}}_{\rm GW} = - \dot{\mathcal{C}}_{\rm tot}\;,
\end{equation}
where $\mathcal{C}\in[E,L]$. Under the influence of the massive scalar and gravitational radiation, the adiabatic evolution of orbital semi-latus rectum and eccentricity is determined by
\begin{equation}
\begin{aligned}\label{edotpdot}
\dot{p}&=(\dot{E}\partial L/\partial e - \dot{L}\partial E/\partial e )/\mathcal{H}\;, \quad \\
\dot{e}&=(\dot{E}\partial L/\partial p - \dot{L}\partial E/\partial p )/\mathcal{H}\;,
\end{aligned}
\end{equation}
where $\mathcal{H}=(\partial E/\partial p) (\partial L/\partial e) -(\partial E/\partial e) (\partial L/\partial p)$.

Using the above two evolutions in Eq.~\eqref{edotpdot}, we can get the trajectory of the secondary object. 
So it is necessary to compute the full relativistic EMRI fluxes along adiabatic eccentric inspirals; however, generating the gravitational and massive scalar fluxes by summing over a large number of spherical-harmonic modes is computationally expensive and requires substantial computational resources. To efficiently obtain these fluxes on the non-rectangular grid spanned by the orbital ($p, e$), we follow the approach of Ref.~\cite{Speri:2024qak}. In this method, the energy and angular momentum fluxes associated with both gravitational and massive scalar radiation are represented through interpolation on a three-dimensional grid constructed from 13 Chebyshev-Gauss-Lobatto (CGL) nodes. The orbital eccentricity and the MBH spin are sampled uniformly over the ranges $e\in[0,0.5]$ and $a/M\in[-0.99,0.99]$, respectively. However, the sampling nodes of orbital semi-latus rectum is more dense close to the separatrix $p_{\rm sep}$ in Kerr spacetime, an alternative variable is introduced to define $p$ direction in the sampling grid for convenience
\begin{equation}
u=(1+e)\Bigg[\frac{\Omega_\phi(a,p,e)}{\Omega_\phi(a,p_{\rm sep},e)}\Bigg]^{2/3}\;.
\end{equation}
We evaluate 17 CGL nodes in the $u$ direction, the value of $u$ is taken uniformly  $[0.08,0.97]$.
On every grid point,  we need to compute the scalar and gravitational total fluxes summing over different modes $(\ell,m,n)$
\begin{equation}
\begin{aligned}
\dot{\mathcal{C}}_{\Lambda}^{\beta} = \sum_{\ell mn} \dot{\mathcal{C}}^{\beta}_{\Lambda,\ell mn} = \sum_{\ell=\ell_{\rm min}}^{\ell=\ell_{\rm max}} \sum_{m=-\ell}^{m=\ell}
\left(\dot{\mathcal{C}}^{\beta}_{\Lambda,\ell m0}+2\sum_{n=1}^{n_{\rm max}}\dot{\mathcal{C}}^{\beta}_{\Lambda,\ell mn}\right)\;,
\end{aligned}
\end{equation}
where $\Lambda\in\{s,g\}$ denotes the massive scalar and gravitational flux, $\beta\in\{H,\infty\}$ is the EMRI flux near the horizon and at infinity.
Here, the index $\ell_{\rm min}=1,2$ represents the scalar and gravitational radiation and the maximum for both cases is $\ell_{\rm max}=10$.
The criterion for choosing the maximum mode index is as follows. The value of $n_{\rm max}$ is determined when the inclusion of three consecutive $n$-modes changes the fluxes by a fractional amount smaller than the mass-ratio ($\epsilon$). We require that the absolute difference of total flux between $\ell_{\rm max}$ and
$\ell_{\rm max}-1$ be less than $10^{-4}$. 
For the associated details, readers are encouraged to refer to Ref.~\cite{Hughes:2021exa, Barsanti:2022ana}. 
After computing two types of EMRI flux on the grids of $2871$ points,
assuming that these fluxes are all smooth functions of $(a,u,e)$ in our sampling grids, we plan to design eight interpolation functions, $(\dot{E}_{\rm s,int}^{\infty,H},\dot{L}_{\rm s,int}^{\infty,H}, \dot{E}_{\rm g,int}^{\infty,H},\dot{L}_{\rm g,int}^{\infty,H})$, based on Chebyshev polynomials to efficiently compute the gravitational and massive scalar fluxes at infinity and near the horizon that include the energy and angular momentum fluxes. In the process of generating EMRI flux on the sampling grids, as stated before, we set index to $\ell_{\rm max}=10$ for computing the scalar and gravitational fluxes, the computation of flux summation is simplified because the indices $(\ell, m)$ and $(\ell, -m)$ are both equal under invariance of the symmetry transformation, $ \dot{\mathcal{C}}^{\beta}_{\Lambda,\ell mn}=\dot{\mathcal{C}}^{\beta}_{\Lambda,\ell -m-n}$ summing over $n$. In fact, the EMRI flux summation for highly eccentric trajectories requires a larger value of $(\ell_{\rm max},n_{\rm max})$ to achieve convergence.

To verify the validity of our interpolation scheme, we compare gravitational fluxes evaluated from the interpolation functions $\dot{E}_{\rm g,int}^{\infty,H}$ with direct perturbative calculations obtained using the \texttt{Teukolsky} package over our sampling grids. We then evaluate the interpolation error by comparing the interpolated and directly computed tensor fluxes at different points within the EMRI parameter space. Similarly, we can assess the precision of massive scalar flux interpolations for any locations in the parameter space of 2873 grid points.
Table~\ref{tab:flux_interp} presents four interpolation error for computing different EMRI fluxes at a fixed grid of $a/M=0.8546,p/M=9.8467, e=0.4328$. These errors are also consistent with interpolation error analysis in previous paper~\cite{Speri:2024qak}, and whose propagation effect on EMRI phase evolution are assessed in Sec.~\ref{result:inspiral}. A full comparison of interpolation error of massive-scalar flux summing over maximum overtones $n_{\rm max}$ can be found in appendix~\ref{sec:interpolation_error}.
With these fluxes obtained with the Chebyshev-interpolation method, one can generate the gravitational and massive scalar flux data fast in $\texttt{FEW}$ package, then evolve orbital parameters with Eq.~\eqref{edotpdot} to obtain the relativistic trajectory. {\color{black}Additionally, we separately investigate the effects of narrow massive-scalar resonances on eccentric-orbit evolution, including possible floating- and sinking-orbit behaviour, in Appendix \ref{sec:resonant_scalar_fluxes}.}

\begin{table}[htbp]
\centering
\caption{Chebyshev interpolation of the energy and angular momentum fluxes and the estimates of their absolute interpolation errors. Two EMRI fluxes are computed with the interpolation functions and relativistic method at any off-grid point, for two sets of parameters $\mathcal{P}_a=\{a/M=0.8546,p/M=9.8467, e=0.4328\}$ and $\mathcal{P}_b=\{a/M=0.95,p/M=10.1930405906075, e=0.4081632653061225\}$ keeping same with Ref.~\cite{Speri:2024qak}.
}
\begin{tabular}{c|cc}
\hline\hline
Parameter setting & Interpolated function & Abs. error \\
\midrule
$\mathcal{P}_a$ &   $\dot{E}_{\rm g,int}^{\infty}$    & $6.82 \times 10^{-4}$ \\
 &   $\dot{E}_{\rm g,int}^{H}$                & $7.53 \times 10^{-4}$ \\
 &   $\dot{E}_{\rm s,int}^{\infty}$           & $3.64 \times 10^{-4}$ \\
&    $\dot{E}_{\rm s,int}^{H}$                & $5.78 \times 10^{-4}$ \\
\hline
$\mathcal{P}_b$ &
 $\dot{E}_{\rm g,int}^{\infty}$    & $7.65 \times 10^{-4}$ \\
 &   $\dot{E}_{\rm g,int}^{H}$                & $7.48 \times 10^{-4}$ \\
 &   $\dot{E}_{\rm s,int}^{\infty}$           & $3.21 \times 10^{-4}$ \\
&    $\dot{E}_{\rm s,int}^{H}$                & $2.53 \times 10^{-4}$ \\
\hline\hline
    \bottomrule
  \end{tabular}
  \label{tab:flux_interp}
\end{table}

In order to examine the impact of a massive scalar field on the accumulated GW phase of EMRIs observed by LISA, we generate inspiral trajectories driven by the combined gravitational and scalar fluxes in both the massive and massless scalar (or GR) scenarios. For a fixed observation time, the quadrupolar dephasing between the massive scalar and massless-scalar (or GR) cases is defined as
\begin{equation}
\begin{aligned}
\delta \Psi^{\rm GR, mScal}_{\phi,r} & = \Psi^{\rm mScal}_{\phi,r} -\Psi^{\rm GR}_{\phi,r} \\
&= 2\int_0^{t_{\rm obs}} \Big[\Omega^{\rm q_s\neq0, \bar{\mu}_s\neq0}(t) -\Omega^{\rm q_s=0, \bar{\mu}_s=0}(t)\Big]dt \;,\\
\delta \Psi^{\rm Scal, mScal}_{\phi,r}& = \Psi^{\rm mScal}_{\phi,r} -\Psi^{\rm Scal}_{\phi,r} \\ & = 
2\int_0^{t_{\rm obs}} \Big[\Omega^{\rm q_s\neq0, \bar{\mu}_s\neq0}(t) -\Omega^{\rm q_s\neq0, \bar{\mu}_s=0}(t)\Big]dt \;,
\end{aligned}
\end{equation}
As the massive scalar mass $\mu$ is scaled by an MBH mass $M$, we plan to use a dimensionless parameter $\bar{\mu}_s=\mu M$ in the following sections for convenience. Further, the quantities $\delta\Psi_{r,\phi}$ with superscripts ``GR, scal, mScal'' are obtained by EMRI fluxes in the standard GR, massless scalar and massive scalar theory.
For two components of the GW dephasing, the azimuthal phase deviation is generally much larger than the radial one, and the GW dephasing in the LISA detection should satisfy $\delta \Psi \gg \delta \Psi_{r}$ and $\delta \Psi \simeq \delta \Psi_{\phi}$.
The threshold GW dephasing discernible by LISA, assuming a signal-to-noise ratio (SNR) of $\rho_{\rm SNR}=30$, is $\delta\Psi \sim 0.1$~\cite{Bonga:2019ycj}. We focus mainly on the characteristic value of azimuthal dephasing $\delta\Psi_\phi=0.1$, as an indicator of the effect of a massive scalar field on EMRI dynamics.

\subsection{Waveform and parameter estimation}\label{wave:constraint}
In this paper, we compute the tensor GW flux and the massive scalar field flux, as well as the resulting EMRI inspiral dynamics, within a fully relativistic framework. 
For the construction of the observable strain, we adopt the quadrupole waveform approximation. 
This hybrid treatment captures the dominant accumulated phase difference induced by the scalar charge and the scalar field mass through their impact on EMRI fluxes. A fully relativistic waveform model would indeed improve the quantitative accuracy of the mismatch calculation and the associated Fisher-matrix forecasts together with Bayesian analysis. However, such refinements are not expected to change the main qualitative conclusions on the detectability of the effect of the massive scalar field studied here~\cite{Katz:2021yft,Mitra:2025tag}.

Here, we describe the quadrupole waveform model to generate EMRI signals efficiently. As mentioned earlier, in our analysis, the orbital evolution is obtained from the flux-balance equations 
in Eq.~\eqref{edotpdot}, while the waveform strain is constructed using the leading-order quadrupole formula. This hybrid implementation is computationally efficient and captures the 
dominant phase accumulation driven by the modified inspiral dynamics, which provides observables for constraining the massive scalar charge and scalar field mass.
We emphasize that the quadrupole waveform used here is an efficient approximate waveform model rather than a fully relativistic EMRI template. Since the detector is located in the 
radiation zone, the observed signal can be decomposed into two transverse polarizations. 
By inserting the orbital evolution of the secondary object into the quadrupole formula, 
the two GW polarizations are given by
\begin{align}\label{amplitude}
h_+ \equiv \sum_n A_n^+ = \sum_n &-\Big[1+(\hat{L}\cdot\hat{n})^2\Big]\Big[a_n \cos2\gamma -b_n\sin2\gamma\Big] \nonumber\\
& +c_n\Big[1-(\hat{L}\cdot\hat{n})^2\Big], \\
h_\times \equiv \sum_n A_n^\times = \sum_n &2(\hat{L}\cdot\hat{n})\Big[b_n\cos2\gamma+a_n\sin2\gamma\Big]\;,\nonumber
\end{align}
where ($\hat{n}, \hat{L}$) is the unit vector along the source direction and the orbital angular momentum, respectively, and
their dot product can be given by four location angles  $(\theta_L, \phi_L,\theta_K, \phi_K)$
\begin{eqnarray}
\hat{L}\cdot\hat{n} = \cos\theta_S\cos\theta_L + \sin\theta_S \sin\theta_L \cos(\phi_S-\phi_L).
\end{eqnarray}
A more detailed description of the angles $(\theta_L, \phi_L,\theta_K, \phi_K)$ can be found in \cite{Barack:2003fp}.
The coefficients ($a_{n}, b_{n}, c_{n}$) can be written as eccentricity-dependent functions \cite{Peters:1963ux}
\begin{equation}
\begin{aligned}
a_n =~ &-n \mathcal{A} \Big[J_{n-2}(ne)-2eJ_{n-1}(ne)+\frac{2}{n}J_n(ne) \\
& +2J_{n+1}(ne) -J_{n+2}(ne)\Big]\cos(n\Phi), \\
b_n =~ &-n \mathcal{A}(1-e^2)^{1/2}\Big[J_{n-2}(ne)-2J_{n}(ne)+J_{n+2}(ne)\Big]\\
& \times\sin(n\Phi), \\
c_n =~& 2\mathcal{A}J_n(ne)\cos(n\Phi) \hspace{1mm} ; \hspace{1mm} \mathcal{A} = ~ (2\pi M \nu )^{2/3}m_p/d,
\end{aligned}
\end{equation}
where $J_n$ are the Bessel functions of the first kind. In our configuration, the angel $\cos(\gamma) = \cos(\pi/4)\cos(\Psi_r)$ represents the direction of the orbital pericenter ~\cite{Barack:2003fp}, $d$ is the distance from the EMRI system to the detector. The frequency $\nu$ is related to the azimuthal orbital frequency $2\pi\nu=\Omega_\phi$, and the time-dependent phase of GW is approximately given by $\Phi=\Psi_\phi$. 
Within the low-frequency consideration, the amplitudes of GW, responded by LISA-like detectors, can be written as  
\begin{equation}\label{antenna}
h_{\textup{I,II}} = \sum_n \frac{\sqrt{3}}{2} \Big[F^+_{\textup{I,II}} (t)A^+_n(t) + F^\times_{\textup{I,II}} (t)A^\times_n(t) \Big]\;,
\end{equation}
where the quantities $F^+_{\textup{I,II}}$ are the antenna pattern functions, their detailed expressions can be found in Ref.~\cite{Cutler:1997ta}.

A quantitative assessment of the difference of two EMRI waveforms is the mismatch by computing their overlap between the waveforms $h_1$ and $h_2$ with the frequency-domain signal
\begin{equation}\label{overlap}
\begin{aligned}
\mathcal{M} \equiv 1-\mathcal{O}(h_{1}, h_{\rm 2})
= 1- \frac{(h_1 \vert h_2)}{\sqrt{(h_1 \vert h_1)(h_2 \vert h_2)}} \;,
\end{aligned}  
\end{equation}
with a inner product 
\begin{equation}
(h_1|h_2) = 2 \int^{f_{\rm high}}_{f_{\rm low}} \frac{\tilde h_1^*(f)\tilde h_2(f)+ \tilde h_1(f) \tilde h_2^*(f)}{S_n(f)}df\;,
\end{equation}
where $S_n(f)$ is the noise-weighted power spectral density of the LISA-like space-borne GW detector~\cite{Robson:2018ifk}, \(\tilde h_i(f)\) is the Fourier transform of \(h_i(t)\), and $f_{\rm low}=10^{-4}\rm Hz$, $f_{\rm high}=f_{p={p_{\rm LSO}}}$ is the frequency at the LSO of the Kerr background.
The inner product can give the SNR of the GW signal in a GW detector
\begin{equation}
\rho_{\rm SNR} = \sqrt{(h_1|h_1)}\;.
\end{equation}
For prospective EMRI observations, a signal is typically regarded detectable when the SNR satisfies ($\rho \gtrsim 20$)~\cite{Babak:2017tow, Fan:2020zhy}. The distinguishability between two waveforms from different EMRI systems can then be estimated using the standard rule-of-thumb criterion $\mathcal{M} \gtrsim 1/(2\rho^2)$~\cite{Flanagan:1997kp, Lindblom:2008cm}. Therefore, in the following discussion, we assume a threshold mismatch of $\mathcal{M}_c\sim0.001$.

Finally, we perform the FIM analysis to estimate the EMRI parameters and to forecast the constraints on the ultralight scalar charge and scalar field mass. 
In GW parameter inference, the FIM provides the leading-order estimate of statistical uncertainties under the assumption of stationary Gaussian noise, high SNR, 
and weak or approximately uniform priors~\cite{Vallisneri:2007ev}. 
In this regime, the posterior distribution can be approximated by a multivariate Gaussian 
centered on the injected parameters, with a covariance matrix given by the inverse of the FIM. 
We consider the full parameter space
\begin{equation}\label{eq:parameters:vector}
\boldsymbol{\theta} = 
\{M, m_p, a, p, e, q_s, \bar{\mu}_s,\Phi_{\phi,0},\Phi_{r,0}, \theta_S, \phi_S, \theta_L, \phi_L, d \} \;,
\end{equation}
where $q_s$ is the scalar charge related to the family of scalar-tensor gravitational theories. The  parameters $\Phi_{\phi,0}$ and $\Phi_{r,0}$ are the initial GW phase.

Using the hybrid quadrupole waveform model specified by the parameter vector 
in Eq.~\eqref{eq:parameters:vector}, we compute the FIM as
\begin{equation}
\Gamma_{ij} \equiv 
\left( \frac{\partial h}{\partial \theta_i} \,\bigg|\, 
       \frac{\partial h}{\partial \theta_j} \right),
\qquad i,j = 1, \dots, 14\, .
\end{equation}
The covariance matrix is then 
approximated by the inverse of the Fisher matrix,
\begin{equation}\label{eq:cov:matrix}
\Sigma_{ij} \equiv 
\langle \delta\theta_i \, \delta\theta_j \rangle
= \left( \Gamma^{-1} \right)_{ij}.
\end{equation}
The corresponding \(1\sigma\) statistical uncertainty of each parameter is obtained 
from the diagonal elements of the covariance matrix
\begin{equation}\label{sigma:fim}
\sigma_{\theta_i} = \sqrt{\Sigma_{ii}} \, .
\end{equation}

The Fisher information matrix provides a preliminary estimate of LISA's ability
to constrain the massive scalar field, in particular the scalar charge \(q_s\)
and the dimensionless scalar-field mass \(\bar{\mu}_s\). The diagonal elements
of the covariance matrix \(\Sigma_{ij}\) determine the expected statistical
uncertainties, while the off-diagonal elements encode correlations among the
intrinsic and extrinsic parameters. This allows us to identify possible
parameter degeneracies in the EMRI waveform model.

To go beyond the local Gaussian approximation inherent in the Fisher analysis, we further perform Bayesian parameter estimation using MCMC sampling. The scalar and gravitational energy fluxes for eccentric EMRIs are generated with Chebyshev interpolation, which enables efficient waveform generation within the \texttt{FEW} framework \cite{Barack:2003fp, Chua:2017ujo, Chua:2020stf, Katz:2021yft}. We then use the parallel-tempered ensemble sampler \texttt{Eryn}~\cite{Karnesis:2023ras}. In
the runs considered here, we employ three temperatures and 16 walkers per
temperature. The convergence of each chain is assessed by requiring the total
chain length to exceed \(50\hat{\tau}\), where \(\hat{\tau}\) denotes the mean integrated autocorrelation time~\cite{Goodman:2010dyf}.

For a data stream $s$ in the second-generation LISA TDI response,
the posterior distribution is
\begin{equation}
    p(\boldsymbol{\theta}|s)
    \propto
    p(s|\boldsymbol{\theta})p(\boldsymbol{\theta}),
\end{equation}
where \(p(\boldsymbol{\theta})\) is the prior distribution. Assuming stationary
Gaussian noise, the log-likelihood is given by
\begin{equation}
    \ln p(s|\boldsymbol{\theta})
    =
    -\frac{1}{2}
    \left<
    s-h(\boldsymbol{\theta})
    \middle|
    s-h(\boldsymbol{\theta})
    \right>
    +{\rm const.},
    \label{eq:mcmc_likelihood}
\end{equation}
where \(h(\boldsymbol{\theta})\) is the EMRI waveform evaluated at the parameter
point \(\boldsymbol{\theta}\), and \(\langle\cdot|\cdot\rangle\) denotes the
noise-weighted inner product. This Bayesian analysis provides posterior constraints on $q_s$ and $\bar{\mu}_s$, and enables a cross vaildation with the Fisher-matrix forecasts. We outline the prior distributions of the injected values of the source parameters in Table~\ref{tab:priors}. 
Note that the influence of scalar mass on EMRI flux depends on a factor of $k_+=\sqrt{\omega_{mn}^2-\mu^2}$, we therefore construct an additional interpolation  $(\dot{E}_{\rm s, int}^{\beta,a},\dot{L}_{\rm s, int}^{\beta,a})$ as a function of parameters $(p,e,\bar{\mu}_s)$, this interpolation error of massive scalar fluxes is reported in Appendix~\ref{app:error:mcmc}.

\begin{table}[htbp]
\caption{Prior distributions adopted in the MCMC analyses presented in this work. Barred quantities denote the injected parameter values used to generate the fiducial signal shown in Fig.~\ref{fig:MCMC:ecc0d1}. Throughout the analysis, we fix \(\delta=0.01\). The notation \(U(x_1,x_2)\) denotes a uniform distribution over the interval \([x_1,x_2)]\).}
\label{tab:priors}
\centering
\begin{tabular}{ccc}
\toprule
Parameter & \qquad & Prior distribution \\	
\midrule
$\ln M$ & \qquad & $U[(1 - \delta) \cdot \ln \bar{M}, \, (1 + \delta) \cdot \ln \bar{M}]$ \\
$\ln m_p$ & \qquad & $U[(1 - \delta) \cdot \ln \bar{m}_p, \, (1 + \delta) \cdot \ln  \bar{m}_p]$ \\
$a/M$ & \qquad & $U[(1 - \delta) \cdot \bar{a}, \, (1 + \delta) \cdot \bar{a}]$ \\
$p/M$ & \qquad & $U[(1 - \delta) \cdot \bar{p}, \, (1 + \delta) \cdot \bar{p}]$ \\
$e$ & \qquad & $U[(1 - \delta) \cdot \bar{e}, \, (1 + \delta) \cdot \bar{e}]$ \\
$d$ & \qquad & $U[0.01, \, 20]$ \\
$\cos \theta_S$ & \qquad & $U[-0.99999, \, 0.99999]$ \\
$\phi_S$ & \qquad & $U[0, \, 2 \pi]$ \\
$\cos \theta_K$ & \qquad & $U[-0.99999, \, 0.99999]$ \\
$\phi_K$ & \qquad & $U[0, \, 2 \pi]$ \\
$\Phi_{\phi_0}$ & \qquad & $U[0, \, 2 \pi]$ \\
$\bar{\mu}_s$ & \qquad & $U[10^{-3}, \, 0.1]$ \\
$q_s$ & \qquad & $U[0.0, \, 0.3]$ \\
\bottomrule
\end{tabular}
\end{table}

\begin{figure*}[htb!]
\centering
\includegraphics[width=6.85in, height=5.15in]{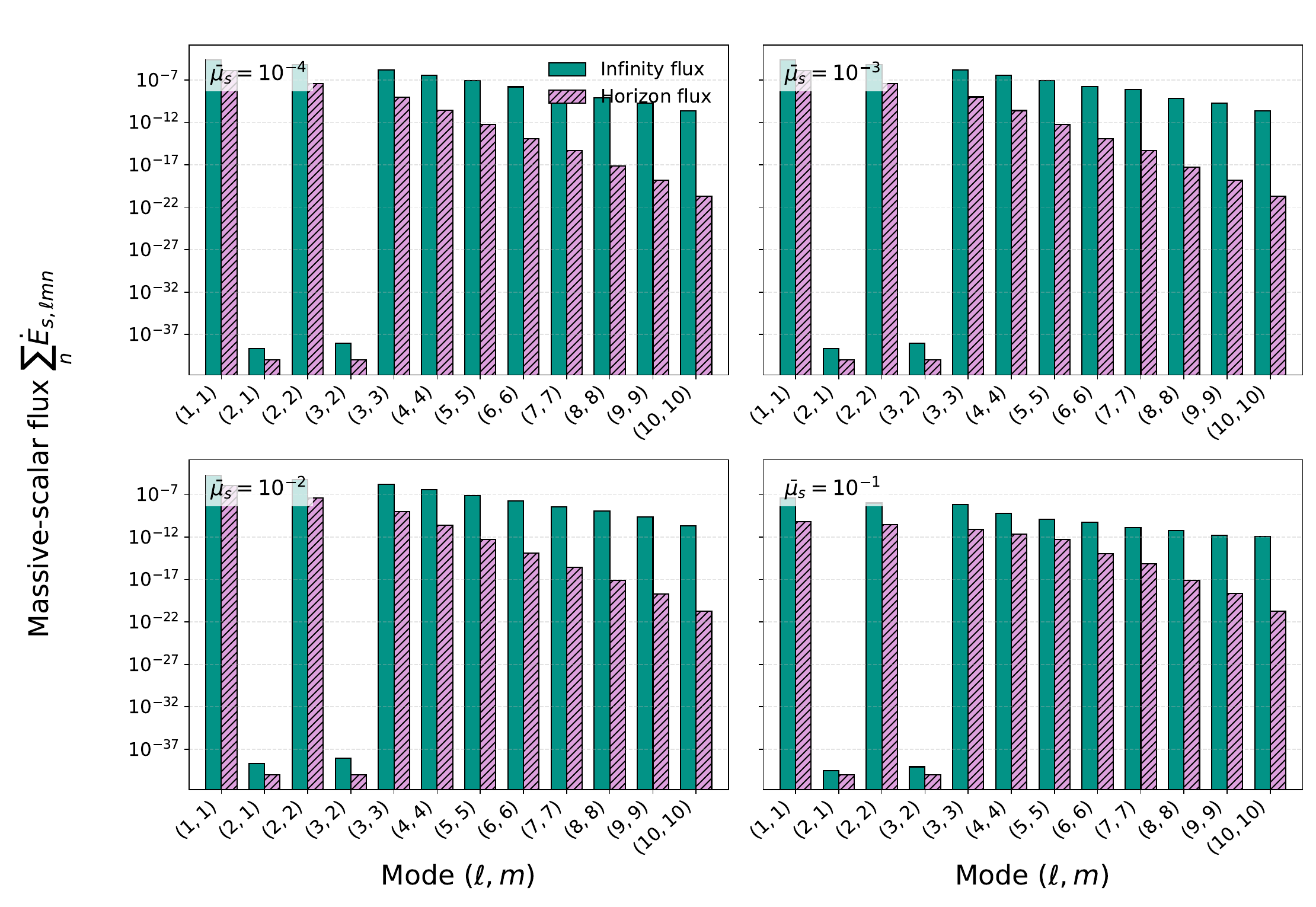}
\caption{Bar plot of the massive scalar energy fluxes (in units of mass-ratio $\epsilon^2$) at infinity and at the horizon, for massive-scalar modes with some specified angular indices $(\ell, m)$, summed over the $n$ harmonics. Other parameters are take as follows: $a/M=0.9, p/M=10.0, q_s=0.1,\bar{\mu}_s=\{10^{-4},10^{-3},10^{-2},10^{-1}\}$.
}\label{fig:bar:energyflux}
\end{figure*}

\begin{table*}[htb]
\footnotesize
\centering
\begin{tabular}{c|@{\hskip 5pt}c@{\hskip 5pt}|c@{\hskip 8pt}|c@{\hskip 8pt}|c}
\hline\hline
\multicolumn{5}{c}{$a/M = 0.1$} \\ 
\hline
$p/M$ & $\tilde{\mu}_s $ & $e=0.1$ & $e=0.3$ & $e=0.5$ \\ 
\hline
7 
 & $0.001$  & 8.63324$\times 10^{-6}$ (1.44569$\times 10^{-7}$) 
 & 8.47068$\times 10^{-6}$ (2.79939$\times 10^{-7}$)
 & 7.71304$\times 10^{-6}$ (5.39683$\times 10^{-7}$) \\
& $0.01$ & 8.29262$\times 10^{-6}$ ($1.45457\times 10^{-7}$) 
& 8.20492$\times 10^{-6}$ (2.82456$\times 10^{-7}$)
& 7.56706$\times 10^{-6}$ (5.49684$\times 10^{-7}$) \\
  & $0.1$ & 2.03265$\times 10^{-11}$ (9.80737$\times 10^{-8}$) 
  & 1.32374$\times 10^{-8}$ (2.24415$\times 10^{-7}$) 
  & 1.02282$\times 10^{-7}$ (4.86167$\times 10^{-7}$) \\
\hline
15 & $0.001$ & 5.83654$\times 10^{-7}$ (-4.27585$\times 10^{-10}$) 
& 5.70521$\times 10^{-7}$ (-7.39763$\times 10^{-11}$) 
& 5.07148$\times 10^{-7}$ (5.09544$\times 10^{-10}$) \\
& $0.01$ & 3.23984$\times 10^{-7}$ (-4.26526$\times 10^{-10}$) 
& 3.22146$\times 10^{-7}$ (-7.26485$\times 10^{-11}$) 
& 3.06424$\times 10^{-7}$ (5.20331$\times 10^{-10}$) \\
& $0.1$ & 9.29114$\times 10^{-20}$ (-1.09964$\times 10^{-10}$) 
& 5.17913$\times 10^{-14}$ (5.53533$\times 10^{-11}$) 
& 3.77578$\times 10^{-12}$ (4.50707$\times 10^{-10}$) \\
\hline\hline
\multicolumn{5}{c}{$a/M = 0.3$} \\ 
\hline
7 & $0.001$ & 8.32241$\times 10^{-6}$ (-8.71677$\times 10^{-8}$)
& 8.26236$\times 10^{-6}$ (-3.22139$\times 10^{-8}$) 
& 7.25608$\times 10^{-6}$ (7.68163$\times 10^{-8}$) \\
& $0.01$ & 8.05652$\times 10^{-6}$ (-8.91698$\times 10^{-8}$)
& 7.87606$\times 10^{-6}$ (-3.18496$\times 10^{-8}$) 
& 7.01226$\times 10^{-6}$ (8.06283$\times 10^{-8}$)
\\
  & $0.1$ &5.39907$\times 10^{-10}$  (-5.66653$\times 10^{-8}$)
  & 1.50259$\times 10^{-8}$ (-1.51707$\times 10^{-8}$)
  & 1.27195$\times 10^{-7}$ (7.87882$\times 10^{-8}$) \\
\hline
15 & $0.001$ & 5.75831$\times 10^{-7}$ (-3.46141$\times 10^{-9}$) 
& 5.60036$\times 10^{-7}$ (-3.49315$\times 10^{-9}$) 
& 4.96139$\times 10^{-7}$ (-3.08827$\times 10^{-9}$) \\
& $0.01$ & 4.96139$\times 10^{-7}$ (-3.08827$\times 10^{-9}$) 
& 3.14883$\times 10^{-7}$ (-3.41372$\times 10^{-9}$) 
& 2.94899$\times 10^{-7}$ (-3.05832$\times 10^{-9}$) \\
  & $0.1$ & 9.62478$\times 10^{-19}$  (-1.41849$\times 10^{-10}$) 
  & 7.91298$\times 10^{-14}$ (-1.17007$\times 10^{-9}$) 
  & 4.75083$\times 10^{-12}$ (-1.27156$\times 10^{-9}$) \\
\hline\hline
\multicolumn{5}{c}{$a/M = 0.8$} \\ 
\hline
7 & $0.001$   
& 7.59479$\times 10^{-6}$ (-5.76934$\times 10^{-7}$)
& 7.48081$\times 10^{-6}$ (-5.77683$\times 10^{-7}$)
&6.54745$\times 10^{-6}$ (-5.04591$\times 10^{-7}$) \\
& $0.01$ & 7.27031$\times 10^{-6}$ (-5.82724$\times 10^{-7}$) 
& 7.04967$\times 10^{-6}$ (-5.753462$\times 10^{-7}$) 
& 6.12126$\times 10^{-6}$ (-4.97351$\times 10^{-7}$) \\
  & $0.1$ & 1.01976$\times 10^{-9}$ (-3.79634$\times 10^{-7}$)
  & 4.92871$\times 10^{-8}$ (-4.08837$\times 10^{-7}$) 
  & 1.86172$\times 10^{-7}$ (-3.82228$\times 10^{-7}$) \\
\hline
15 & $0.001$ & 5.56932$\times 10^{-7}$ (-1.23619$\times 10^{-8}$) 
& 5.32865$\times 10^{-7}$ (-1.28867$\times 10^{-8}$) 
& 4.62999$\times 10^{-7}$ (-1.23875$\times 10^{-8}$) \\
& $0.01$ & 2.985692$\times 10^{-7}$ (-1.21032$\times 10^{-8}$)
& 2.93617$\times 10^{-7}$ (-1.28651$\times 10^{-8}$)
& 2.6021$\times 10^{-7}$ (-1.20286$\times 10^{-8}$) \\
& $0.1$ & 1.18866$\times 10^{-18}$ (-3.41361$\times 10^{-9}$) 
& 1.78114$\times 10^{-13}$ (-4.45443$\times 10^{-9}$) 
& 6.91821$\times 10^{-12}$ (-5.47473$\times 10^{-9}$) \\
\hline\hline
\end{tabular}
\caption{Massive scalar fluxes (in units of mass-ratio $\epsilon^2$)  at infinity and the horizon of MBH are computed by summing over the harmonics of $n_{\rm max}=10$  for the fundamental $(\ell, m) = (1,1)$ mode, the values in the parenthesis are the scalar fluxes near the horizon. For each spin of MBH, we consider different values of the 
orbital semi-latus rectum $p/M$, scalar mass $\bar{\mu}_s$, 
and eccentricity $e$.}
\label{tab:massivescalar:fluxes}
\end{table*}

\begin{figure*}[htb!]
\centering
\includegraphics[width=3.5in, height=2.5in]{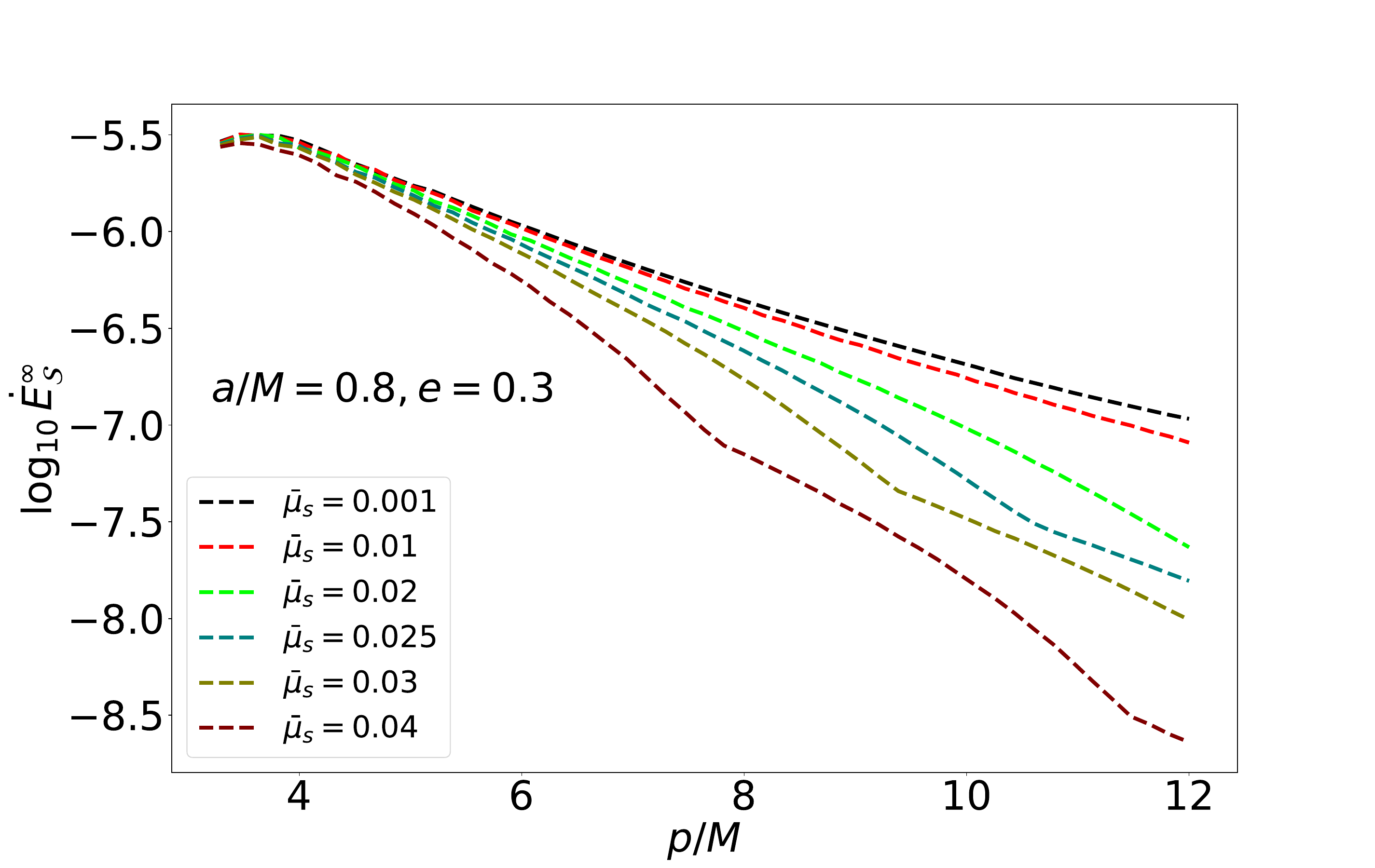}
\includegraphics[width=3.5in, height=2.5in]{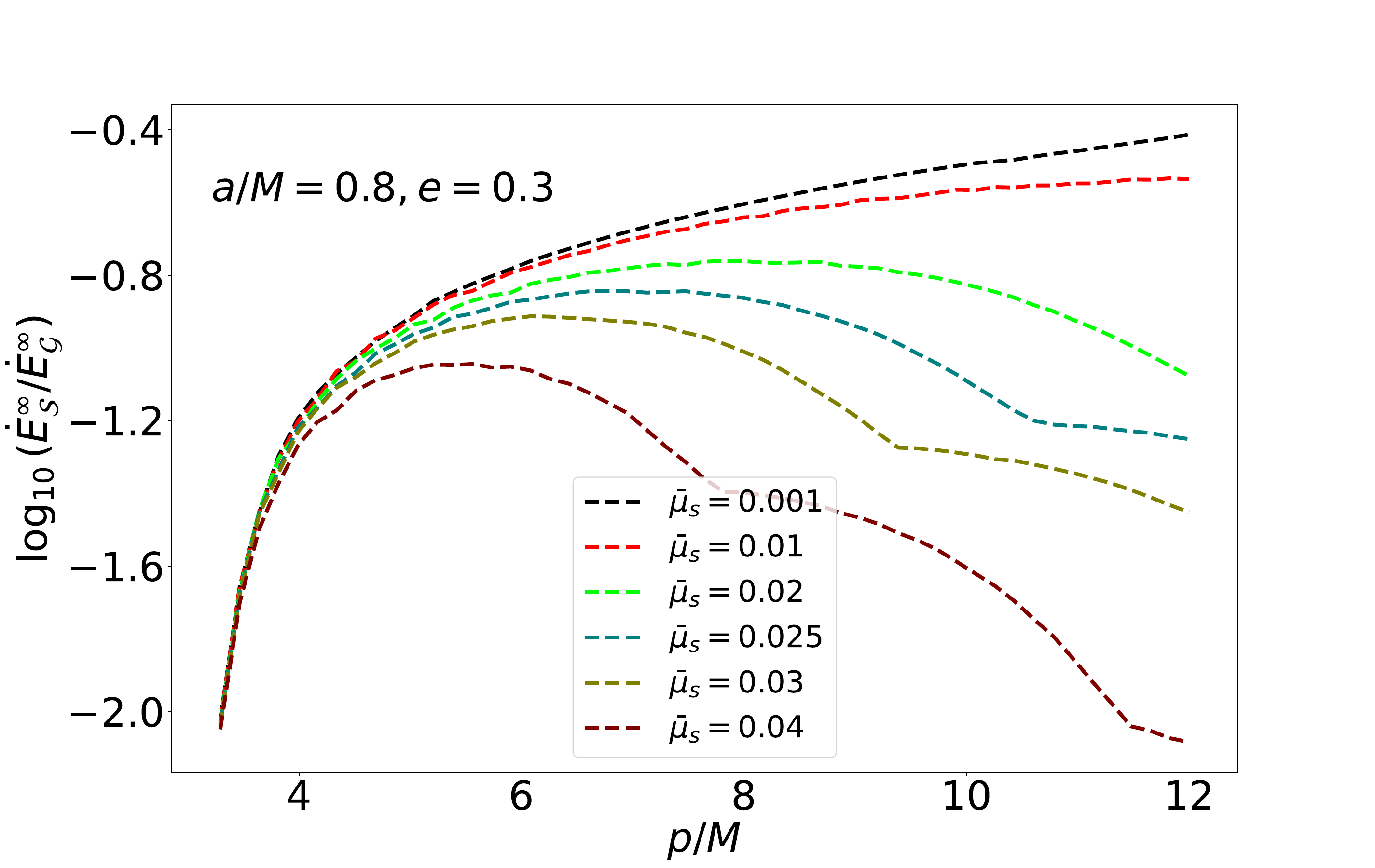}
\includegraphics[width=3.5in, height=2.5in]{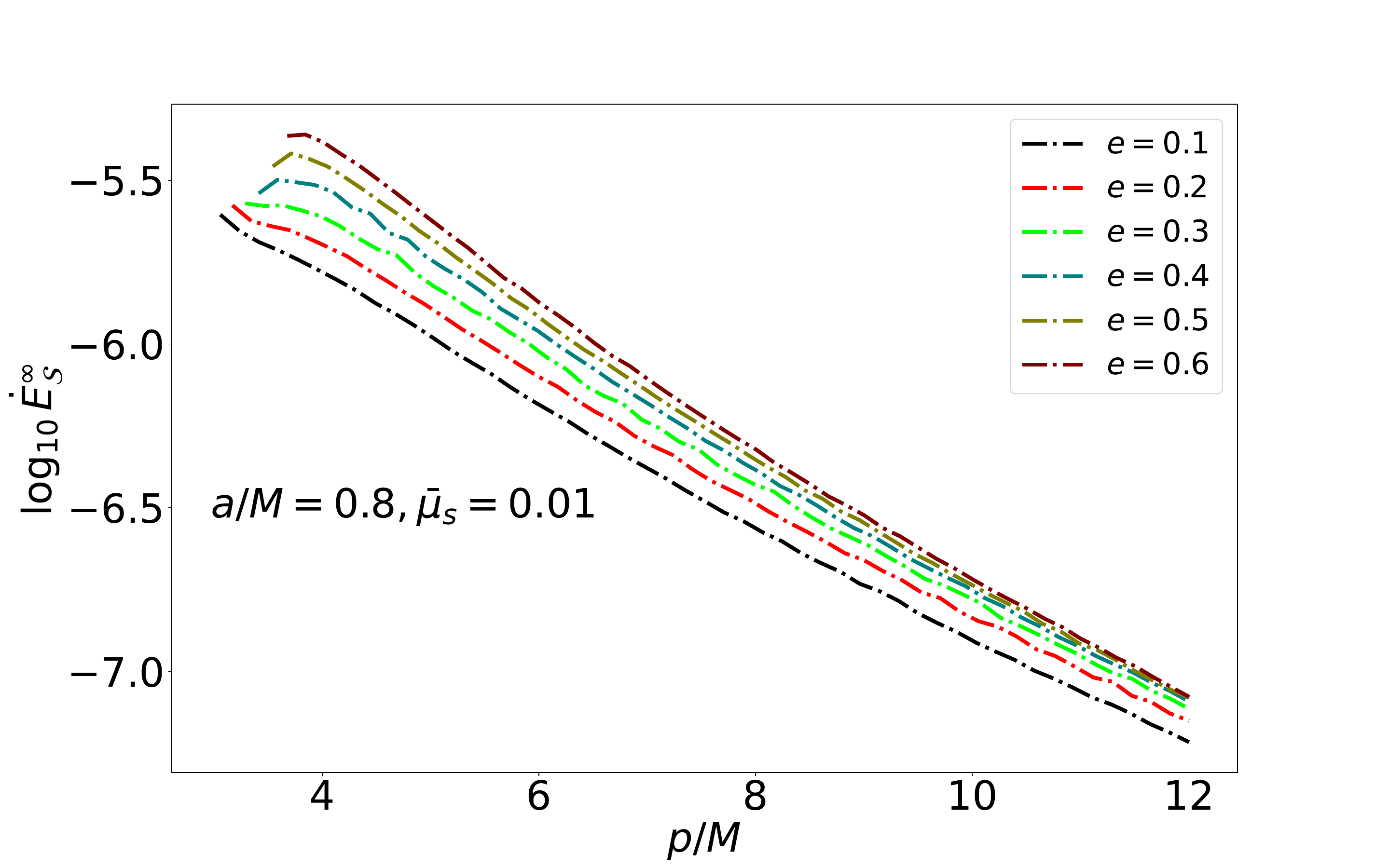}
\includegraphics[width=3.5in, height=2.5in]{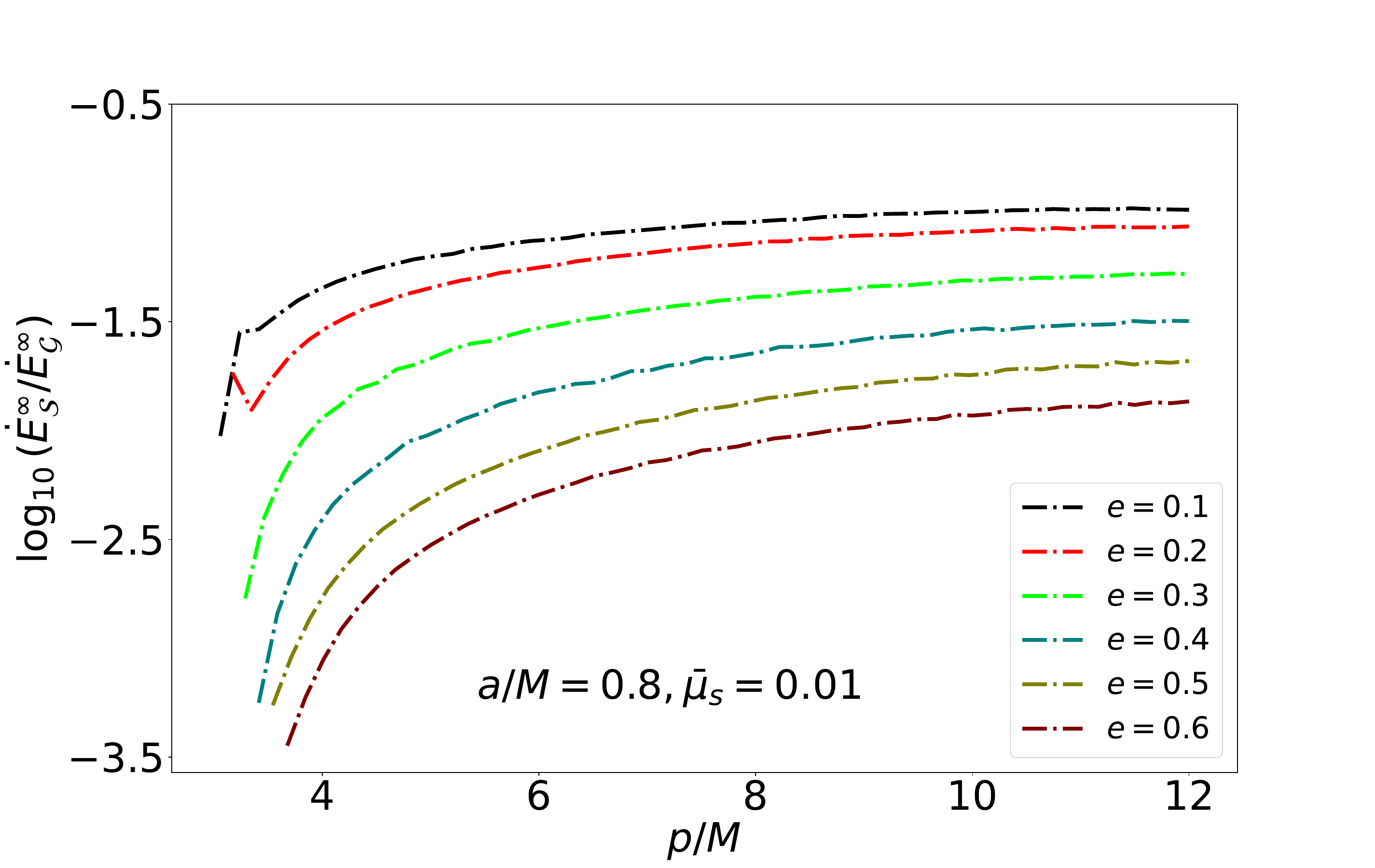}
\caption{Massive scalar fluxes and ratios between them, and  gravitational flux at the infinity as a function of orbital semi-latus rectum $p/M$ are  plotted for a spinning MBH $a/M=0.8$ and fixed scalar-charge $q_s=0.1$. The top two plots show the fluxes influenced by massive scalar field with a mass range $\bar{\mu}_s\in\{0.001,0.01,0.02,0.025,0.03,0.04\}$ for a fixed initial eccentricity $e=0.3$. The bottom two figures present the fluxes influenced by  the initial eccentricities $e\in\{0.001,0.01,0.02,0.025,0.03,0.04\}$ for a fixed scalar mass $\bar{\mu}_s=0.01$.
}\label{fig:fluxes}
\end{figure*}

\begin{figure*}[htb!]
\centering
\includegraphics[width=3.5in, height=2.5in]{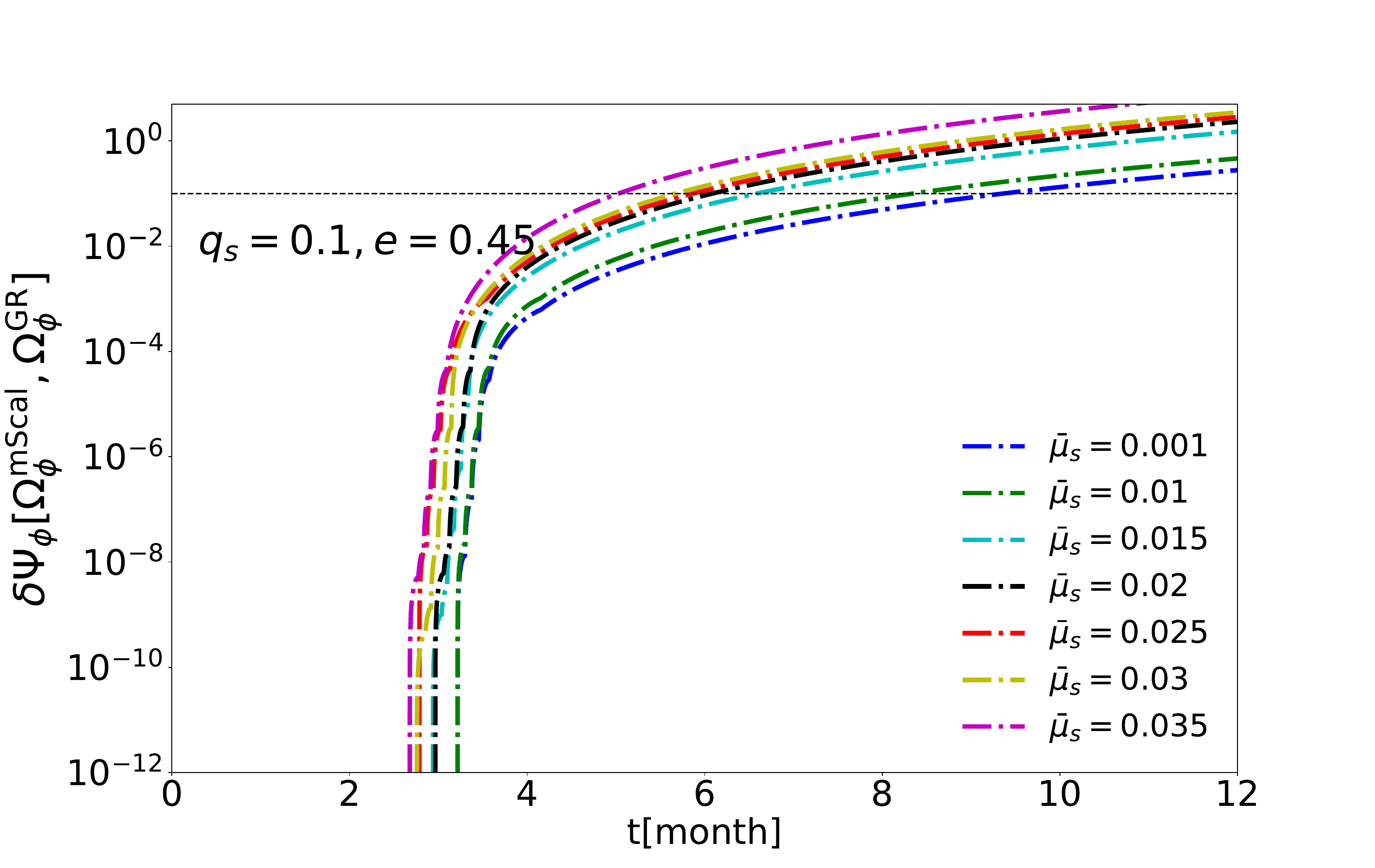}
\includegraphics[width=3.5in, height=2.5in]{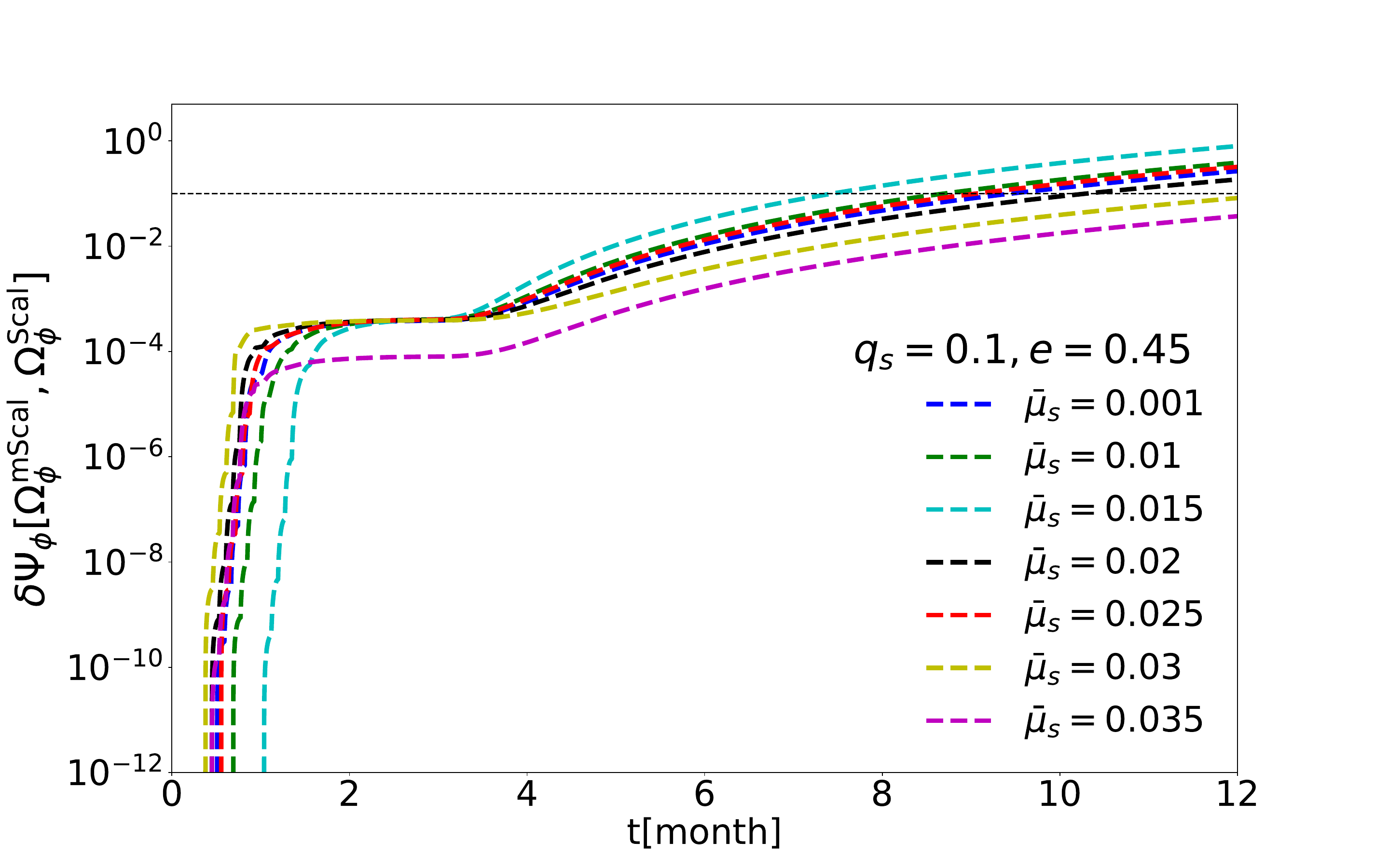}
\includegraphics[width=3.5in, height=2.5in]{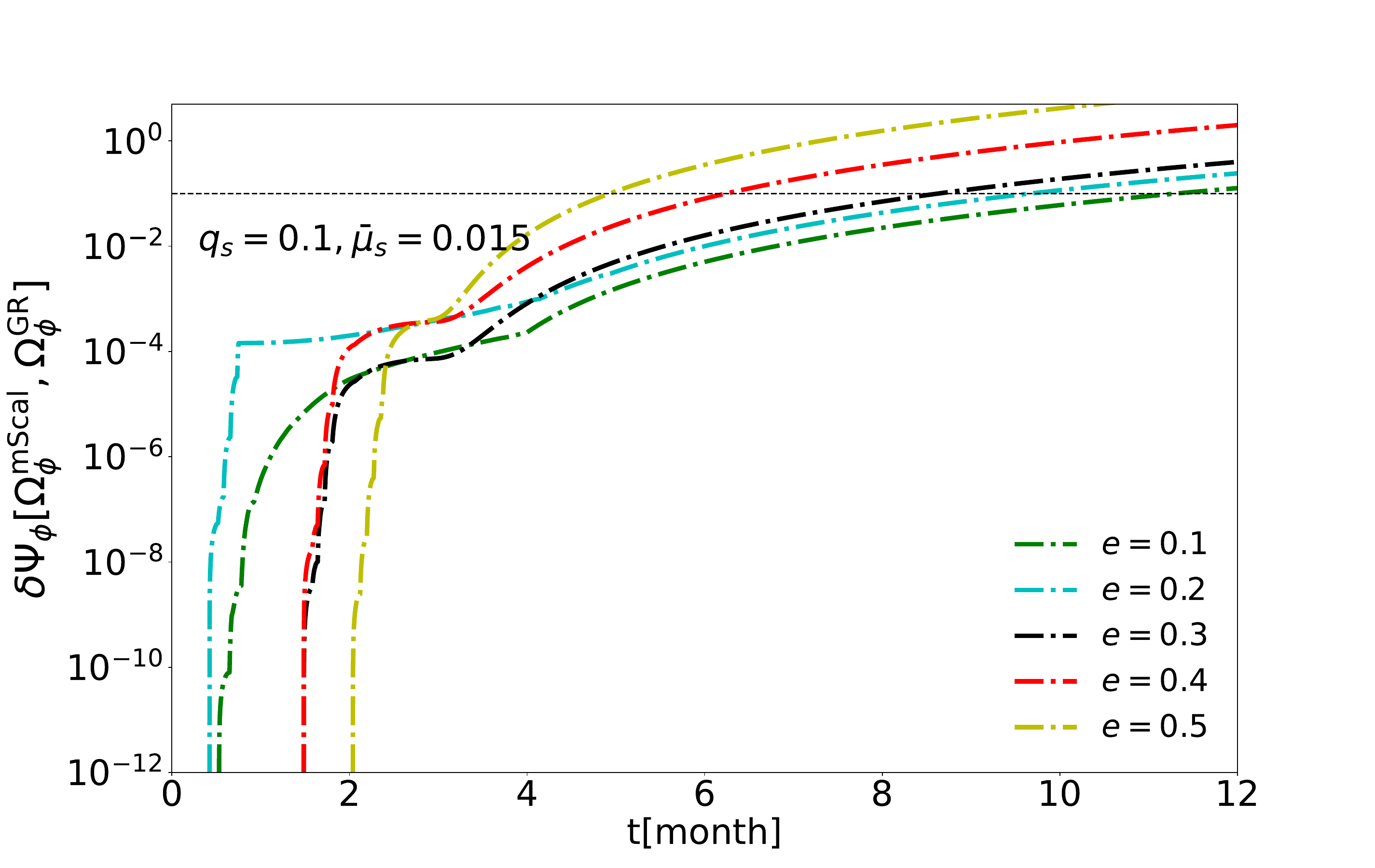}
\includegraphics[width=3.5in, height=2.5in]{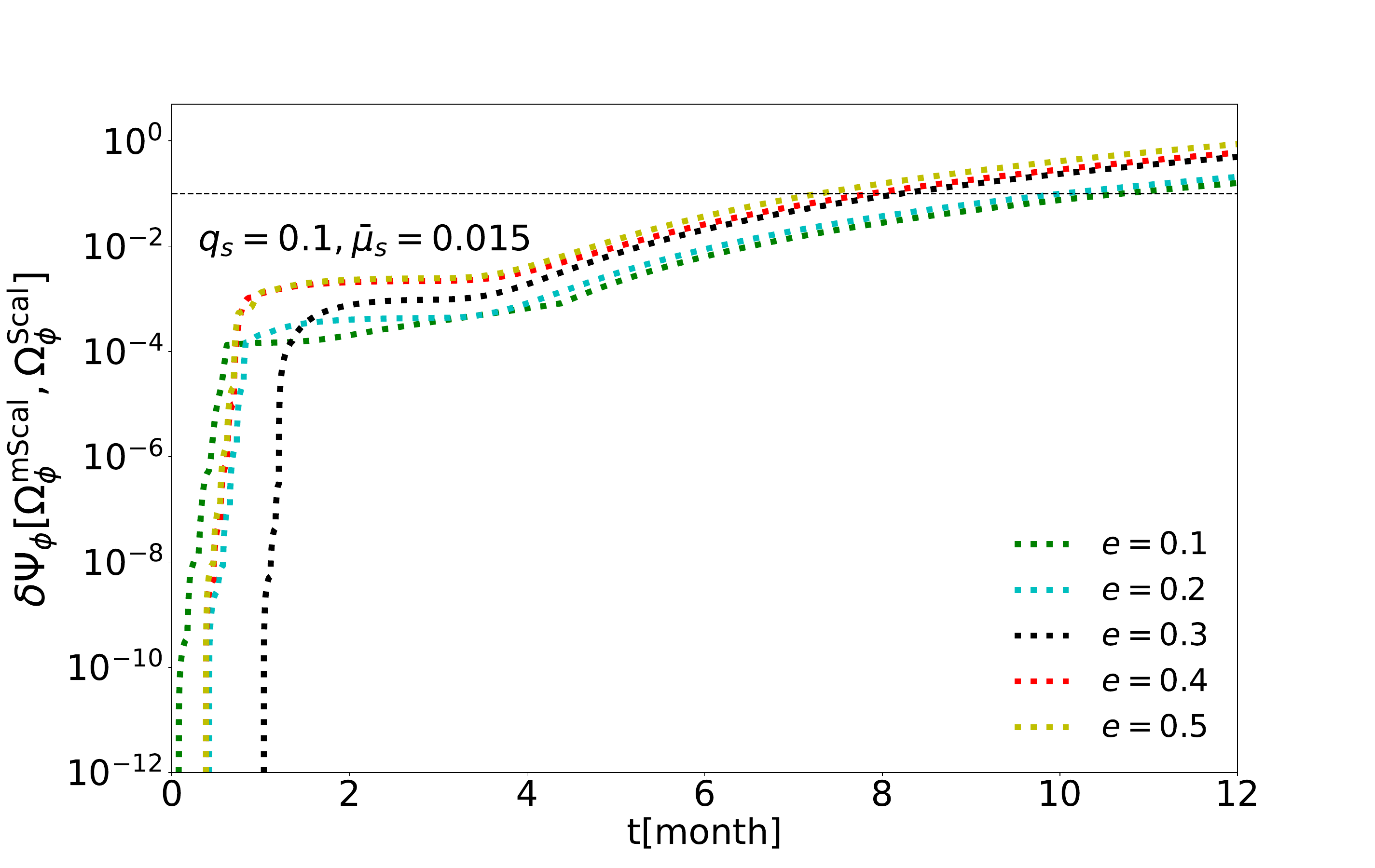}
\caption{Accumulated azimuthal dephasings for an EMRI inspiralling during the last one year before plunge into the Kerr MBH as a function of time are plotted with a fixed scalar charge $q_s=0.1$. The top panels consider the initial eccentricity $e=0.45$ and some scalar masses $\bar{\mu}_s=\{0.001,0.01,0.015,0.02,0.025,0.03,0.035\}$; whereas, the bottom panels show five orbital eccentricities $e=\{0.1,0.2,0.3,0.4,0.5\}$ for a fixed scalar mass $\bar{\mu}_s=0.015$, in which the evolution is derived by the fluxes from the standard GR and massive scalar gravitational theories case for the left panels, and the right panels consider the massive and massless scalar gravitational theories case.}\label{fig:dephasing}
\end{figure*}

\begin{figure*}[htb!]
\centering
\includegraphics[width=6.5in, height=4.95in]{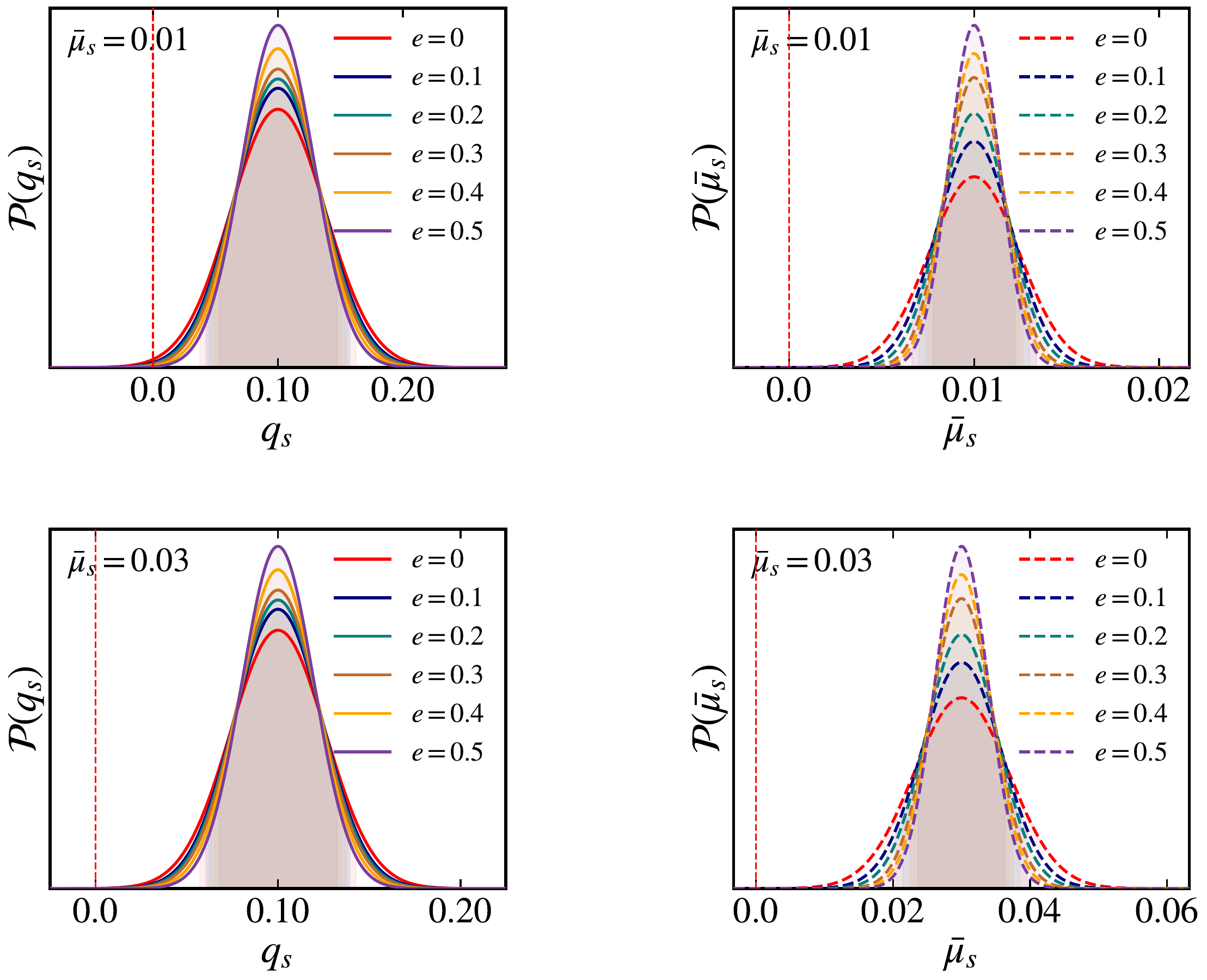}
\caption{The marginal distributions of scalar charge and mass $(q_s,\bar{\mu}_s)$ from two dimensional elements in FIM are plotted for different orbital eccentricities $e=\{0,0.1,0.2,0.3,0.4,0.5\}$, considering scalar charge $q_s=0.1$, MBH-spin $a/M=0.8$,
scalar mass $\bar{\mu}_s=0.01$ in the top two panel. and $\bar{\mu}_s=0.03$ in the bottom two panels. 
The vertical red dashed lines are the GR theory for the case of $q_s=0$ and $\bar{\mu}_s=0$. Eccentric insprial depends on the scalar fluxes by summing harmonic mode up to $\ell_{\rm max}=3$, $n_{\rm max}=5$, and the summation of gravitational fluxes are  computed up to harmonic mode of $\ell_{\rm max}=5, n_{\rm max}=10$.
For the generatation of EMRI waveforms, Other parameters are keeping same with Fig.~\ref{fig:mismatch:observtime}.
}\label{fig:postporb:charge:mass}
\end{figure*}

\begin{figure*}[htb!]
\centering
\includegraphics[width=6.5in, height=4.95in]{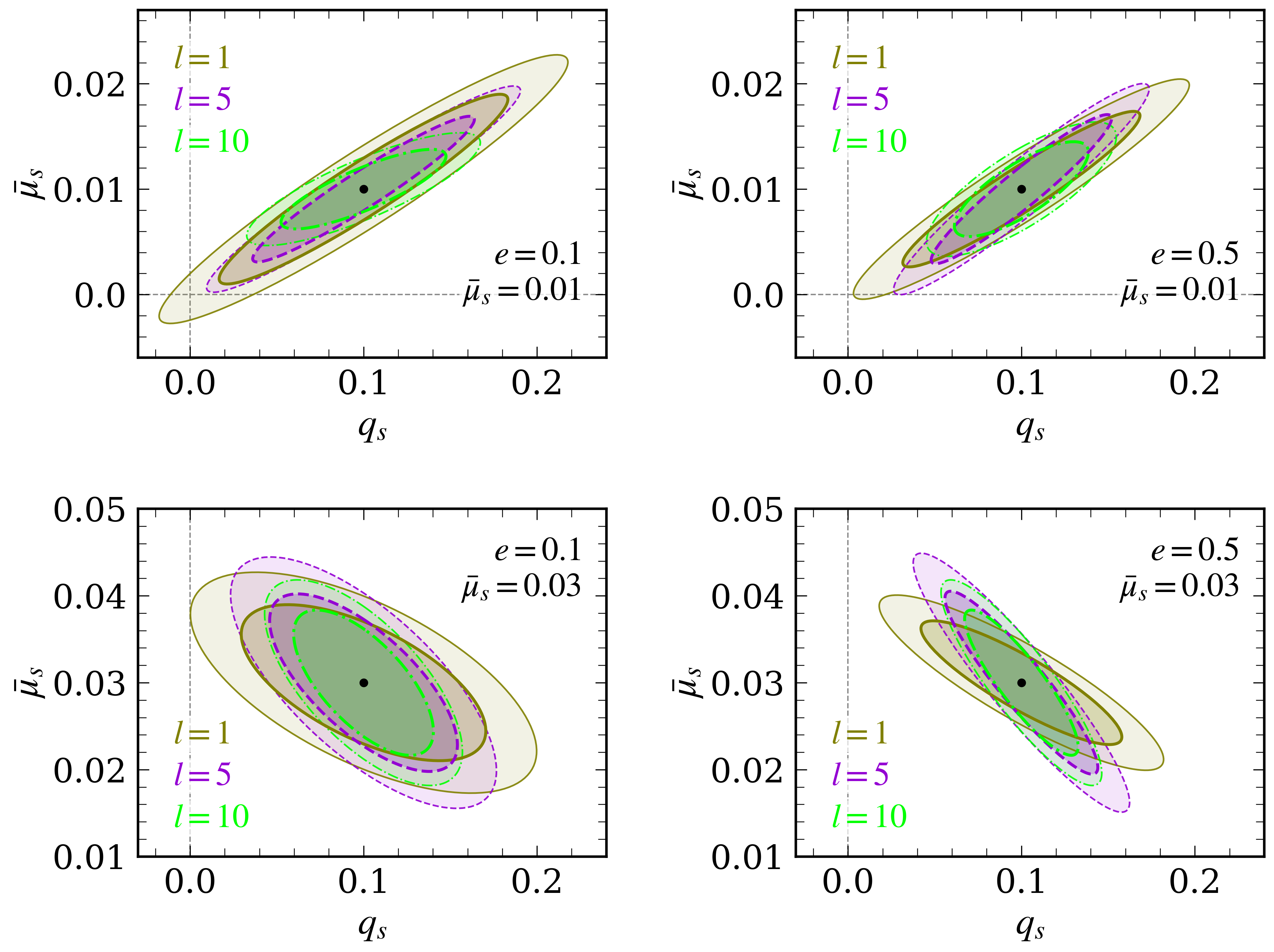}
\caption{Fiducial intervals at the level of $68\%$ and  $90\%$ for the posterior distribution of scalar charge and mass $(q_s,\bar{\mu}_s)$ are plotted for different orbital eccentricities $e=\{0.1,0.5\}$, given scalar charge $q_s=0.1$, MBH-spin $a/M=0.8$,
scalar mass $\bar{\mu}_s=0.01$ in the top two panels and $\bar{\mu}_s=0.03$ in the bottom two panels. The vertical gray dashed lines are the GR theory for the case of $q_s=0$ and $\bar{\mu}_s=0$. We consider the contribution of scalar fluxes at different harmonic modes
$\ell_{\rm max}={1,5,10}$, and the summation of tensor fluxes up to harmonic modes $\ell_{\rm max}=5, n_{\rm max}=10$.
Other parameters of these EMRI waveform considered here are keeping same with Fig.~\ref{fig:mismatch:observtime}.
}\label{fig:postporb:highermode}
\end{figure*}

\begin{figure*}[htb!]
\centering
\includegraphics[width=\linewidth,trim=1.0 1.0 1.0 1.0,clip]{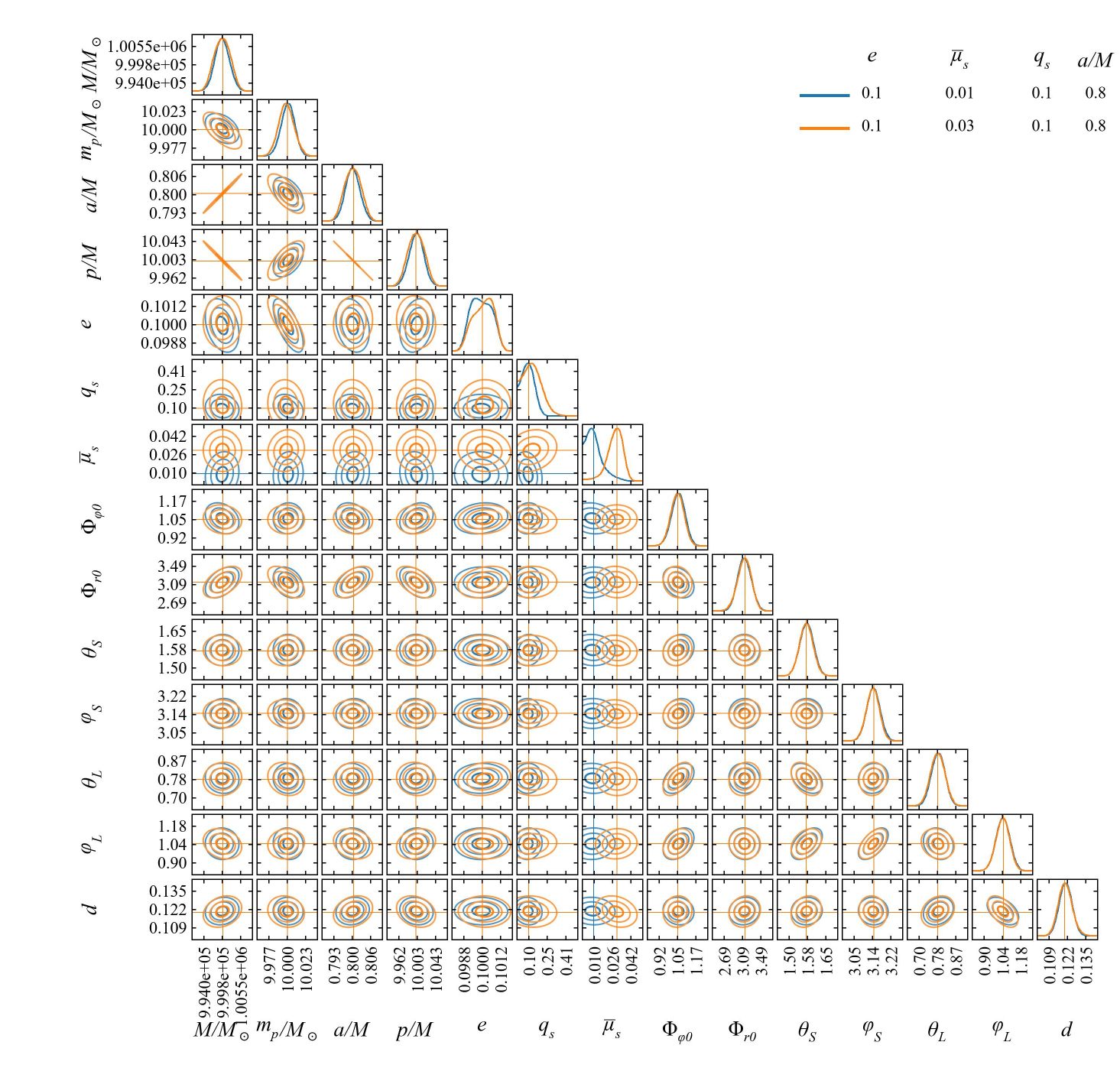}
\caption{
Marginalized posterior distributions from the 14-dimensional MCMC analysis for an eccentric EMRI with \(e=0.1\). The diagonal panels show the one-dimensional marginalized posteriors, while the lower-triangular panels show the corresponding two-dimensional marginalized posteriors. The blue and orange results correspond to injections with \(\bar{\mu}_s=0.01\) and \(0.03\), respectively. The remaining injected parameters are \(M=10^{6}M_{\odot}\), \(m_p=10M_{\odot}\), \(a/M=0.8\), \(p_0/M=10\), and \(q_s=0.1\), with
\(\rho=150\). The reference lines mark the injected values.}\label{fig:MCMC:ecc0d1}
\end{figure*}

\begin{figure*}[htb!]
\centering
\includegraphics[width=\linewidth,trim=1.0 1.0 1.0 1.0,clip]{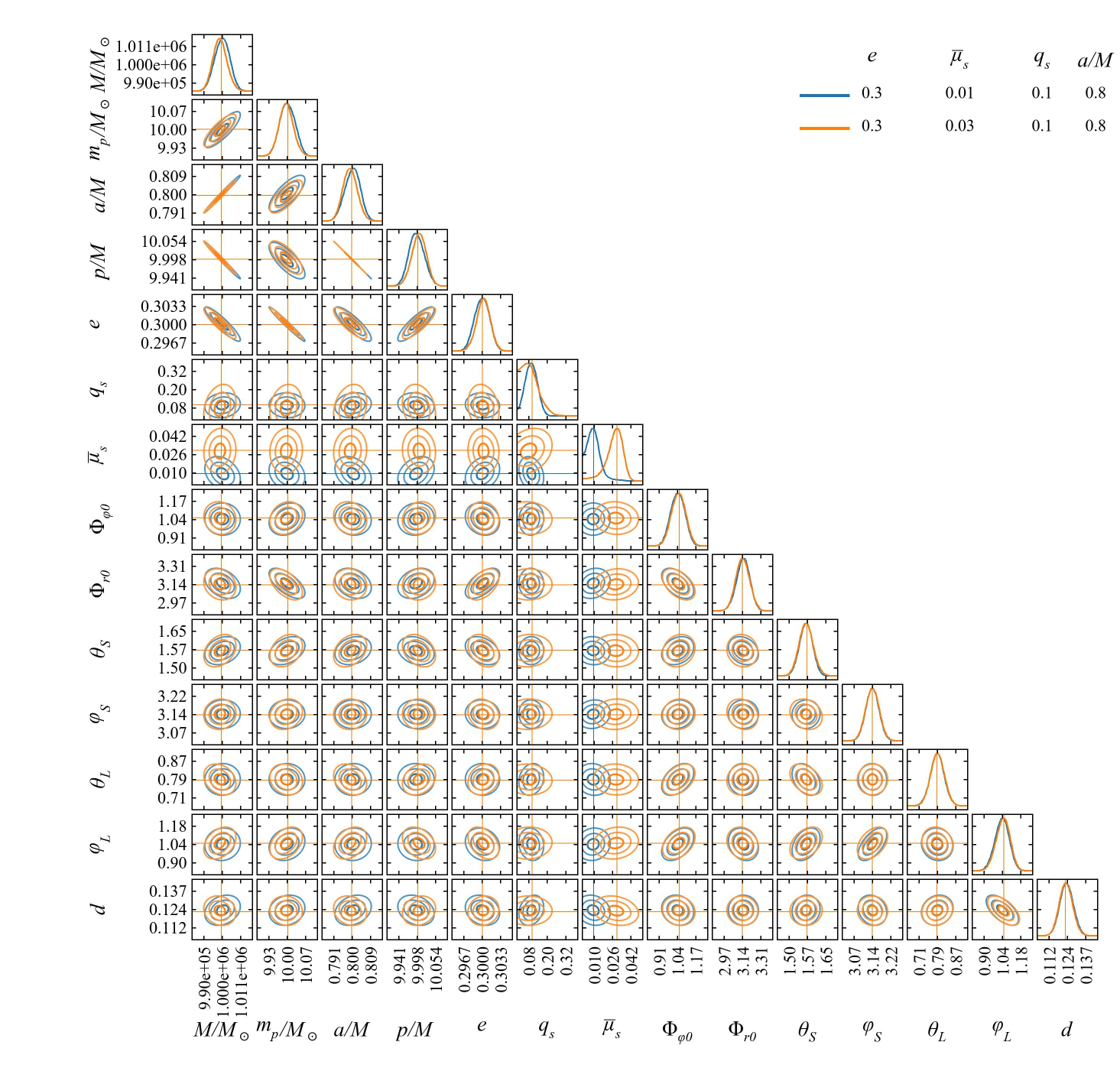}
\caption{Marginalized posterior distributions from the 14-dimensional MCMC analysis for an eccentric EMRI with \(e=0.3\). As in Fig.~\ref{fig:MCMC:ecc0d1}, the diagonal panels show the one-dimensional marginalized posteriors, while the lower-triangular panels display the corresponding two-dimensional
marginalized posterior densities and their nested credible contours. The blue and orange results correspond to injections with
\(\bar{\mu}_s=\{0.01,0.03\}\), respectively. The remaining injected parameters are \(M=10^{6}M_{\odot}\), \(m_p=10M_{\odot}\),
\(a/M=0.8\), \(p_0/M=10\), and \(q_s=0.1\), with the EMRI singal of
SNR \(\rho=150\). The reference lines indicate the
injected parameter values.
}\label{fig:MCMC:ecc0d3}
\end{figure*}

\section{Result}\label{result}
In this section, we first evaluate the energy flux induced by the massive scalar field and compare it with the gravitational flux. We then study the accumulated GW dephasing and waveform mismatch caused by the massive scalar field, and derive the corresponding constraints on the scalar charge and scalar field mass. 
Unless otherwise stated, the sky position of the source and the orientation of the MBH spin are fixed as 
$\theta_S=\pi/3$, $\phi_S=\pi/4$, $\theta_L=\pi/3$, and $\phi_L=\pi/4$. 
For a given set of EMRI parameters, we start the inspiral from an initial semi-latus rectum ($p$) and estimate the constraints on the parameters during last one year of the adiabatic inspiral before plunging into the Kerr MBH. 
Our fiducial binary consists of a $10^6 M_\odot$ MBH and a $10 M_\odot$ secondary compact companion. 
In Fisher-matrix analysis, the luminosity distance $d$ is treated as an adjustable parameter, which is chosen to set the target SNR.

\subsection{Comparison between massive scalar and gravitational flux}\label{wave:discrimination}
To perform adiabatic evolution for eccentric EMRI orbits, we first compute the massive scalar and tensor fluxes by summing over different harmonic modes $(\ell,m,n)$ in Eq.~\eqref{eq:edots}.
Gravitational fluxes are computed with the $\texttt{Teukolsky}$ Mathematica package in the BHPT \cite{BHPT}, which can be generated with high precision for eccentric EMRI orbits.

Among the $(\ell,m,n)$ modes contributing to Eq.~\eqref{eq:edots}, the dominant scalar emission is generated by the $(\ell,m)=(1,1)$ mode. First, we assess the massive-scalar fluxes obtained by summing the radial-harmonic modes ($n$) at the horizon and at infinity for fixed $(\ell,m)$, as shown in Fig.~\ref{fig:bar:energyflux}. We consider four scalar masses, $\bar{\mu}_s=\{10^{-4},10^{-3},10^{-2},10^{-1}\}$, where the summed energy fluxes are defined as $\sum_n \dot{E}_{s,\ell mn}^{H,\infty}$ with $n$ ranging from -10 to 10. 
From Fig.~\ref{fig:bar:energyflux}, we find that the energy fluxes gradually decrease from the dominant $(\ell,m)=(1,1)$ mode as higher spherical-harmonic modes are included. We also observe that the scalar fluxes become smaller as the scalar-field mass increases.

Table~\ref{tab:massivescalar:fluxes} lists the massive scalar fluxes near the horizon and at infinity for a fixed MBH spin $a/M\in[0.1,0.3,0.8]$ and different orbital eccentricities, which contains three cases of scalar mass, $\bar{\mu}_s\in[0.001,0.01,0.1]$.
From this table, one can find that the scalar flux is suppressed by its mass, for example, the scalar flux at infinity generally is larger than one near the horizon.
However, the scalar flux at infinity is much lower than that for the horizon for the more massive scalar fluxes with $\bar{\mu}_s=0.1$ and binary separation $p=15M$. This may be because the asymptotic form of the massive scalar wave resembles $e^{i\sqrt{\omega_{mn}^2-\mu^2} r_\star}$, which decays exponentially at large binary separation. 

We show the massive scalar flux as a function of orbital semi-latus rectum for a fixed MBH-spin $a/M=0.8$ in Fig.~\ref{fig:fluxes}, in which the upper panels consider various values of scalar masses and the lower panels describe the effect of orbital eccentricities. The right panels are the comparison between the massive scalar and  gravitational fluxes.  From the figure, one can observe that the massive scalar fluxes at infinity is suppressed significantly when its mass is larger; these phenomena become more obvious for the larger separation of binary.
Moreover, the massive scalar flux is bigger when the inspiralling orbits are more eccentric, nevertheless the ratio of it and gravitational flux is smaller for keeping same orbital configuration.
From the behaviour of the massive scalar fluxes, we conclude that the deviation of EMRI signals between massive and massless scalar-tensor gravity theories becomes smaller as the scalar mass increases, while the difference relative to the GR case remains more pronounced. This conclusion is in accordance with the computation of dephasing and mismatch in Sec.~\ref{result:inspiral}.
\subsection{Effect of massive scalar field on eccentric inspirals}\label{result:inspiral}
To assess the influence of a massive scalar field on EMRI orbital dynamics and its detectability by LISA, we compute the dephasing and waveform mismatch between EMRIs with massive and massless scalar-charged secondaries (or, equivalently, between scalar-tensor and GR EMRI systems) in this subsection.

Fig.~\ref{fig:dephasing} shows the azimuthal dephasing as a function of LISA observation time, for a fixed scalar charge $q_s = 0.1$. The top panels show results for several scalar masses at fixed initial eccentricity
$e=0.45$, while the bottom panels show results for five orbital eccentricities at fixed scalar mass $\bar{\mu}_s = 0.015$. To assess distinguishability by LISA, we adopt a threshold of $\delta\Psi_\phi = 0.1$ rad as the minimum dephasing detectable by LISA for an EMRI signal with SNR $\rho_{\rm SNR} = 30$. Three distinct orbital azimuthal frequencies are considered: $\Omega_\phi^{\rm mScal}$, corresponding to the inspiral modified by the massive scalar field; $\Omega_\phi^{\rm Scal}$, representing the frequency predicted by massless scalar contribution; and $\Omega_\phi^{\rm GR}$, describing the inspiral in GR. With 6-8 months of observation, these inspirals, modified by scalar mass, generate dephasing larger than the threshold. From the top-right panel, which shows the dephasing between the massive and massless scalar cases, we find that the dephasing first increases and then decreases as the scalar mass increases from $\bar{\mu}_s=0.001$ to $\bar{\mu}_s=0.035$. In contrast, the top-left panel compares the massive scalar scenario with standard GR, where the dephasing increases monotonically with increasing scalar mass. The local scalar-flux interpolation errors are typically $10^{-7}-10^{-3}$, while the corresponding accumulated phase error remains below the adopted LISA threshold of $0.1\,\mathrm{rad}$. The interpolation procedure and its accuracy are examined in Appendix~\ref{sec:interpolation_error}.

To quantify the effect of the massive scalar field on the EMRI waveform, we calculate the mismatches
$\mathcal{M}(h^{\rm mScal},h^{\rm Scal})$ and
$\mathcal{M}(h^{\rm mScal},h^{\rm GR})$, where
$h^{\rm mScal}$, $h^{\rm Scal}$, and $h^{\rm GR}$ denote the
massive-scalar, massless-scalar, and GR waveforms, respectively.
For an EMRI with component masses $(10^6+10)\,M_\odot$,
$q_s=0.01$, $p/M=12$, and $a/M=0.8$, the massive-scalar
waveform becomes distinguishable from the massless-scalar waveform
after approximately $6$--$8$ months, with a threshold
$\bar{\mu}_s\simeq10^{-3}$
$(\mu\sim7.5\times10^{-19}\,\mathrm{eV})$ for $e=0.1$.
The mismatch increases slightly for the larger eccentricity $e=0.4$.
Relative to GR, the mismatch initially increases and then decreases
as the scalar mass increases. This nonmonotonic behavior arises
because modes satisfying $\omega_{mn}<\mu$ are bound and cannot
propagate to infinity, resulting in a suppression of the
massive-scalar flux at larger scalar masses. For a one-year adiabatic inspiral up to the Kerr LSO, the joint
dependence on $(q_s,\bar{\mu}_s)$ gives approximate distinguishability
thresholds of $\bar{\mu}_s\simeq8\times10^{-3}$
$(\mu\sim6\times10^{-18}\,\mathrm{eV})$ relative to the massless-scalar
case and $\bar{\mu}_s\simeq4\times10^{-4}$
$(\mu\sim3\times10^{-19}\,\mathrm{eV})$ relative to GR.
For a fixed scalar mass, the mismatch increases with $q_s$, reflecting
the quadratic dependence of the scalar flux on the scalar charge.
The detailed time-dependent mismatch curves and the corresponding
$(q_s,\bar{\mu}_s)$ contours are presented in
Appendix~\ref{app:D mismatch}.

\begin{table*}[htbp!]
\centering
\begin{tabular}{cccc|cccccccc}
\hline
\hline  
$\rm orbital ~eccentricity$ & $q_s$ &$\ell_{\rm max}$   &$n_{\rm max}$   
& $\sigma_{M}/M$ & $\sigma_{m_p}/m_p$    
&$\sigma_{a}/a$   &$\sigma_{p}/p$ 
&$\sigma_{e}$  & $\sigma_{q_s}/q_s$   
& $\sigma_{\bar{\mu}_s}/\bar{\mu}_s$
\\
\hline
$\rm 0.1 $  & $0.1$ & $1$ & $1$
&$2.36\text{e-3}$    &$2.71\text{e-4}$   
&$3.45\text{e-4}$    &$5.04\text{e-3}$ 
&$4.82\text{e-4}$    &$4.56\text{e-1}$  &$5.28\text{e-1}$  
\\
& & $3$ & $1$
&$2.28\text{e-3}$    &$2.64\text{e-4}$   
&$3.17\text{e-4}$    &$4.76\text{e-3}$ 
&$4.31\text{e-4}$    &$4.23\text{e-1}$  &$4.76\text{e-1}$  
\\
$ $  & $ $ & $1$ & $5$
&$2.27\text{e-3}$    &$2.67\text{e-4}$   
&$3.35\text{e-4}$    &$4.92\text{e-3}$ 
&$4.64\text{e-4}$    &$4.38\text{e-1}$  &$5.17\text{e-1}$  
\\
& & $3$ & $5$
&$2.19\text{e-3}$    &$2.53\text{e-4}$   
&$3.06\text{e-4}$    &$4.54\text{e-3}$ 
&$4.25\text{e-4}$    &$4.17\text{e-1}$  &$4.52\text{e-1}$ 
\\
\hline
$\rm 0.3 $  & $0.1$ & $1$ & $1$
&$2.15\text{e-3}$    &$2.37\text{e-4}$   
&$3.12\text{e-4}$    &$4.46\text{e-3}$ 
&$3.63\text{e-4}$    &$3.55\text{e-1}$  &$3.27\text{e-1}$  
\\
$ $  & $ $ & $3$ & $1$
&$1.62\text{e-3}$    &$1.85\text{e-4}$   
&$2.81\text{e-4}$    &$4.12\text{e-3}$ 
&$3.03\text{e-4}$    &$2.72\text{e-1}$  &$2.43\text{e-1}$  
\\
$ $  & $ $ & $1$ & $5$
&$1.92\text{e-3}$    &$2.14\text{e-4}$   
&$2.97\text{e-4}$    &$4.38\text{e-3}$ 
&$3.45\text{e-4}$    &$2.32\text{e-1}$  &$2.86\text{e-1}$  
\\
$ $  & $ $ & $3$ & $5$
&$1.54\text{e-3}$    &$1.71\text{e-4}$   
&$2.65\text{e-4}$    &$3.53\text{e-3}$ 
&$2.84\text{e-4}$    &$2.63\text{e-1}$  &$1.97\text{e-1}$  
\\
\hline
\hline
\end{tabular}
\caption{Measurement uncertainties for intrinsic EMRI parameters with massive scalar charged secondary are outlined. We consider an EMRI binary with component masses $(M=10^{6}M_\odot,\; m_p=10M_\odot)$, scalar charge $q_s=0.1$ and mass $\bar{\mu}_s=0.15$. 
The inspiral is evolved for one year and the luminosity distance $d$ is chosen such that the network SNR is $150$. The initial orbital phase is $\Phi_{\phi,0}=\Phi_{r,0}=1.0$. Other parameters are keeping same with Fig.~\ref{fig:mismatch:observtime}.
}\label{tab:fim:error}
\end{table*}

\subsection{Constraint on massive scalar fields}
In this subsection, we constrain the scalar charge and scalar field mass using EMRI signals observed by a LISA-like detector within the low-frequency consideration. We first employ the FIM formalism to obtain a computationally efficient estimate of the expected parameter uncertainties over the parameter space defined in \autoref{eq:parameters:vector}. Since the FIM provides only a local Gaussian approximation to the posterior distribution, we subsequently perform a full Bayesian parameter-estimation analysis using MCMC sampling to validate these forecasts and to characterise the posterior distributions more accurately.

For a specified scalar charge $q_s=0.1$, we compute the inspiral by the massive-scalar-induced modifications to the fluxes, construct the quadrupolar waveform via \autoref{amplitude}, and project it onto the detector response using \autoref{antenna}. Although the FIM is evaluated over all parameters in \autoref{eq:parameters:vector}, we report only the uncertainties for intrinsic parameters in \autoref{tab:fim:error}, since the EMRI phase is largely insensitive to extrinsic parameters whose effects are predominantly absorbed by overall amplitude and orientation.
With this assumption, by summing the massive scalar fluxes over different harmonic modes in Eq. \eqref{eq:edots} and computing gravitational fluxes for the same modes, the fractional error for the binary mass and orbital parameters $(M,m_p,a,p,e)$ are at the level of \(\sim10^{-3}-10^{-4}\), and the relative measurement error of the massive scalar parameters ($q_s, \bar{\mu}_s$) are determined within the range of  \(\sim10^{-2}-10^{-1}\), depending on the modes of harmonics. 
One can find that waveforms including EMRI fluxes of the higher order harmonic modes can place a more rigorous constraint on massive scalar field parameters and other intrinsic EMRI parameters.
Comparing with the circular EMRI case \cite{Barsanti:2022vvl}, the relative error on the scalar mass and charge can be improved slightly for orbital eccentricity $e=0.3$, which can reach to the level of $\sigma_{q_s}/q_s\sim26.3\%$ and $\sigma_{\bar{\mu}_s}/\bar{\mu}_s\sim29.7\%$ respectively.

Next, in \autoref{fig:postporb:charge:mass}, we plot the marginal distributions of the scalar charge and mass $(q_s, \bar{\mu}_s)$ describing the two-dimensional matrix, extracting from FIM, considering two cases of scalar mass $\bar{\mu}_s=\{0.01,0.03\}$ and different orbital configurations. Here, we conduct the FIM analysis by computing eccentric inspirals, comprising the maximum harmonic mode $\ell_{\rm max}=3$, $n_{\rm max}=5$ and ignoring higher multipole modes. The probability density becomes increasing when orbital eccentricity is bigger, which is expected as per our prediction that more eccentric EMRI orbits encode more relativistic information about the strong-field regime, and also require flux computations at higher harmonic modes.
Finally, we also discuss the confidence intervals for the posterior distribution of scalar charge $q_s$ and mass $\bar{\mu}_s$ by considering the massive scalar fluxes at different multipole modes $\ell_{\rm max}={1,5,10}$.  As shown in Fig. (\ref{fig:postporb:highermode}), the black dashed lines denote the case of the zero scalar charge and mass described by GR. After including the contribution of multipole harmonic modes, constraints on $q_s$ and $\bar{\mu}_s$ can further be improved with LISA observations. 
Our results indicate that EMRI waveforms from more eccentric orbits provide more rigorous bounds on massive scalar fields.

Finally, we conduct a more rigorous bound/constraint on massive scalar fields using Bayesian parameter estimation with MCMC sampling. Fig.~\ref{fig:MCMC:ecc0d1} shows the posterior distribution of full parameters  for a fixed spin $a/M=0.8$ and orbital eccentricity $e=0.1$, setting the SNR $\rho=150$ by freely adjusting the luminosity distance $d$. The blue and orange lines reflect the waveform posterior for the scalar mass $\bar{\mu}_s=0.01$ and $\bar{\mu}_s=0.03$, respectively.
From the MCMC statistic results in Fig.~\ref{fig:MCMC:ecc0d1}, the constraint of massive scalar field is as follows: the scalar charge is $q_s=0.1_{-9.698\times10^{-2}}^{+8.619\times10^{-2}}$ and scalar mass $\bar{\mu}_s=0.01_{-8.964\times10^{-3}}^{+2.36\times10^{-2}}$ for the injected scalar-mass $\bar{\mu}_s=0.01$, and for another case of $\bar{\mu}_s=0.03$ the constraints are $q_s=0.1_{-4.878\times10^{-2}}^{+9.879\times10^{-2}}$ and $\bar{\mu}_s=0.03_{-2.842\times10^{-2}}^{+4.971\times10^{-3}}$. Their relative measurement error is about $\Delta q_s/q_s=5.652\times10^{-1}$ and $\Delta \bar{\mu}_s/\bar{\mu}_s=6.247\times10^{-1}$ obtained with the $3~\sigma$ inference of  the MCMC computation, which are consistent with the bound results obtained with FIM in~\autoref{tab:fim:error}.
Compared to the case of orbital eccentricity $e=0.3$ in Fig.~\ref{fig:MCMC:ecc0d3}, the level  of constraint in a massive scalar field is slightly improved, such as $q_s=0.1_{-3.386\times10^{-2}}^{+2.369\times10^{-2}}$ and $\bar{\mu}_s=0.03_{-8.984\times10^{-3}}^{+2.338\times10^{-2}}$ for the true value $\bar{\mu}_s=0.01$ of scalar mass. Therefore, the higher harmonics from eccentric EMRI orbits are helpful in giving a more rigorous constraint on the  massive scalar field. Additionally, since the bound of the source parameters deduced from MCMC is a range, to observe the measurement error of each parameter, we list the posterior distribution of all the source parameters in Table~\ref{tab:posterior_sigma_intervals} of the Appendix~\ref{app:error:mcmc}.

\section{Conclusion}\label{result}

EMRIs are one of the most powerful and interesting astrophysical sources for probing fundamental physics in strong-gravity fields. Their high phase sensitivity, accumulated over millions of orbital cycles before plunge, makes them suited for detecting or constraining the presence of additional degrees of freedom beyond GR. Amongst, as studied in the present work, massive scalar fields carry a theoretically motivated and observationally significant role. A nonzero scalar mass reshapes how scalar radiation propagates and how energy escapes the system, which further produces qualitatively distinct signatures in waveforms compared to the vacuum GR and massless scalar scenarios. With the growing interest in eccentric EMRI dynamics, we investigated a systematic relativistic treatment that incorporates massive scalar radiation with finite nonzero orbital eccentricity. Addressing this gap is the principal motivation of the present work.

We studied the effect of a massive scalar field on eccentric equatorial EMRIs around Kerr black holes, and evaluated the prospects for detecting or constraining the scalar charge $q_s$ and dimensionless scalar field mass $\bar{\mu}_s = \mu M$ through future space-based GW detectors such as LISA. Using the BHPT framework and adiabatic approximation, we computed tensor gravitational fluxes and massive scalar energy and angular momentum fluxes by solving the scalar perturbation equation in the Kerr background using frequency-domain decomposition and summing over eccentric-orbit harmonics $(\ell, m, n)$. The flux data generated were represented using CGL interpolants over a three-dimensional grid in $(a/M, u, e)$, enabling adiabatic inspiral evolution from initial orbital configurations to the LSO. Further, waveforms were constructed, and parameter estimation was carried out using FIM analysis and Bayesian inference based on MCMC sampling over the full 14-dimensional parameter space of the EMRI system, and assessing the detection prospects through LISA on measuring the massive scalar field, associated charge and distinguishability from GR with eccentric EMRI.

Below, we list the key findings of the study:

\begin{itemize}
\item As shown in Fig. (\ref{fig:bar:energyflux}), the massive scalar energy flux is dominated by the $(\ell, m) = (1,1)$ mode and decreases monotonically with increasing $(\ell, m)$, while the flux at both the horizon and infinity is progressively suppressed as the scalar mass $\bar{\mu}_s$
increases from $10^{-4}$ to $10^{-1}$. We can also see in Fig. (\ref{fig:fluxes}) that the scalar energy flux at infinity is suppressed by the scalar field mass ($\bar{\mu}_{s}$), with the suppression becoming more pronounced at larger orbital separations. Further, the ratio of scalar to gravitational flux decreases for heavier scalar fields, even if the absolute flux from the more eccentric configurations remains comparatively larger. We notice that the orbital eccentricity plays a crucial role in enhancing the massive scalar signal. We find that more eccentric orbits generate richer harmonic spectra and probe a broader range of the strong-field geometry during each radial cycle, thereby increasing the total scalar flux, which will consequently amplify its effect on the orbital evolution. 

\item The accumulated azimuthal dephasing, compared to GR, crosses the LISA detection threshold of $\delta\Psi_\phi = 0.1$ rad in 6–8 months for source parameters considered here, as presented in Fig. (\ref{fig:dephasing}), with a larger orbital eccentricity consistently giving the highest dephasing. In addition, the dephasing between massive and massless scalar cases is non-monotonic, i.e., it first grows and then shrinks as $\bar{\mu}_{s}$ increases from 0.001 to 0.035, which reflects the tension between scalar mass enhancing low-frequency orbital modifications at intermediate values and suppressing flux propagation at larger values.

\item The waveform mismatch between massive and massless scalar computation crosses the detection threshold ($\sim 10^{-3}$) for scalar masses as small as $\bar{\mu}_s\sim (10^{-3}-10^{-4})$, with the effect modestly strengthened at higher eccentricity ($e$), given the MBH spin $a/M=0.8$. For comparison against GR, the mismatch grows with the increase in scalar mass, as well as the same trend is seen with the increase in initial eccentricities, shown in Fig. (\ref{fig:mismatch:observtime}). Furthermore, the joint mismatch analysis in Fig. (\ref{fig:mismatch:contour}) can distinguish the massive scalar signal from the massless case down to $\bar{\mu}_s\sim 8\times 10^{-3} (\mu\sim6\times10^{-18}\rm eV)$; whereas in the case of GR bound, it turns out to be relatively tighter $\bar{\mu}_{s}\sim4\times 10^{-4} (\mu\sim3\times10^{-19}\rm eV)$. In both scenarios, a higher scalar charge $q_s$ makes the effect in a detectable range, as the flux goes as $q^{2}_s$, indicating the resultant radiation immediately as a deviation.

\item From the marginal distribution of FIM, in Fig. (\ref{fig:postporb:charge:mass}), higher orbital eccentricity ($e$) sharpens the posterior peaks for both ($q_s, \bar{\mu}_{s}$), implying richer information through eccentric EMRI dynamics for massive scalar sector. This improvement is more pronounced for the larger scalar mass case ($\bar{\mu}_s = 0.03$), where the posteriors narrow substantially as eccentricity increases toward $e = 0.5$. Table (\ref{tab:fim:error}) further quantifies this trend: increasing eccentricity from $e = 0.1$ to $e = 0.3$ noticeably reduces the fractional measurement errors on ($q_s, \bar{\mu}_s$), indicating that eccentric orbits provide a measurable improvement in constraining the massive scalar sector with LISA. Thus FIM analysis indicates that LISA can measure intrinsic EMRI parameters such as binary masses, MBH spin, and orbital elements with fractional errors at the level of $10^{-3}-10^{-4}$, and the scalar charge and scalar field mass can be constrained with relative uncertainties of order $10^{-1}-10^{-2}$.

\item The confidence ellipses in Fig. (\ref{fig:postporb:highermode}) show that including higher harmonic contributions $\ell_{\rm max} = (1, 5, 10$) tightens the joint constraints on ($q_s, \bar{\mu}_s$), with the improvement being more notable at $e = 0.5$ than at $e = 0.1$. The GR limit ($q_s = 0, \bar{\mu}_s = 0$, i.e., vertical gray dashed lines) remains well-separated from the posterior bulk in all cases

\item The posterior distributions for four orbital configurations,
\[
(e,\bar{\mu}_s)=(0.1,0.01),\ (0.1,0.03),\ (0.3,0.01),\ \text{and}\ (0.3,0.03),
\]
with a fixed scalar charge \(q_s=0.1\), are shown in
Figs.~\ref{fig:MCMC:ecc0d1} and~\ref{fig:MCMC:ecc0d3}. Depending on the
injected configuration, the MCMC analysis yields fractional uncertainties
in the approximate ranges
\(\Delta\bar{\mu}_s/\bar{\mu}_s\sim10^{-2}\)--\(\mathcal{O}(1)\) and
\(\Delta q_s/q_s\sim10^{-2}\)--\(\mathcal{O}(1)\). These Bayesian constraints
are broadly consistent with the corresponding Fisher-matrix forecasts,
thereby providing an independent cross-check of the parameter-estimation
results. The agreement indicates that LISA could distinguish an eccentric
EMRI with massive-scalar radiation from the corresponding GR signal for the
orbital configurations considered here.

\item {\color{black}We also investigate massive-scalar resonances and their possible floating- or sinking-orbit signatures. For the representative configuration considered, their effects on evolution of ($p, e$) are subtle, making the accumulated azimuthal evolution a more sensitive diagnostic. The related details are presented in Appendix (\ref{sec:resonant_scalar_fluxes}).}

\end{itemize}

In summary, we have shown that massive scalar radiation leaves measurable imprints on eccentric EMRI dynamics and generated gravitational waveforms. Such imprints become more prominent with larger eccentricities. The space-based detector like LISA can potentially detect such a scalar field, if the secondary object carries a nonzero scalar charge, given the constraint on both the scalar charge and the scalar field mass ($q_s, \bar{\mu}_s$). 

Future directions can explore several natural extensions of the present study. As the dephasing, mismatch, and FIM analysis presented here provides an important preliminary and informative picture of how a massive scalar field imprints on eccentric EMRI dynamics beyond GR, a more complete assessment would require accounting for correlations and degeneracies among all source parameters, i.e., fully Bayesian  MCMC analysis with massive scalar-corrected EMRI waveforms for generic orbits (eccentric inclined), incorporating additional orbital frequencies and an enlarged harmonic structure \cite{Gliorio:2026yvh}, potentially stronger constraints on the massive scalar field. Further, considering the secondary to be an extended object, carrying a finite nonzero spin, will provide a rich structure on massive scalar field. In this line, the post-adiabatic corrections to the scalar and gravitational fluxes will play a crucial role in enhancing the accuracy of generated EMRI waveforms \cite{Spiers:2023cva}. Finally, another future direction is to explore the beyond-vacuum scenario by including environmental effects and connecting the constrained scalar charge and mass parameters to ultralight dark matter candidates or modified gravity theories \cite{Kejriwal:2023djc, Yuan:2024duo, OuldElHadj:2024psw, HegadeKR:2025rpr, HegadeKR:2025dur,Duque:2025yfm, Kejriwal:2025upp, Kejriwal:2025jao, Zi:2026zpw, Long:2026dcb, Xia:2026aty}. This would provide a more direct astrophysical interpretation of the resulting bounds and give us meaningful comparisons with constraints from other observational channels. We plan to report on some of these in future work.

\section{Acknowledgments}
We thank Susanna Barsanti and Andrea Maselli for providing very helpful discussion on solving the coefficients of boundary conditions, which improved the numerical stability of the homogeneous solutions from the massive radial equation. T. Z. also thanks Lorenzo Speri for providing very detailed explanation on interpolation techniques in Mathematica. The work is funded by the National Natural Science Foundation of China with Grants No.~12405059, No.~12165013, No.~12505087 and No.~12375049, the Startup Fund for Advanced Talents of Nanchang University with Grant No. 28170256, 
Key Program of the Natural Science Foundation of Jiangxi Province under Grant No. 20232ACB201008 and the Ganpo High-Level Innovative Talent Program. S.K. acknowledges the support and research facilities provided by the Department of Physics, IIT Kharagpur. 

\appendix
\section{Boundary conditions for the massive-scalar perturbation equation}
\label{subsec:scalar_boundary_conditions}

The homogeneous massive scalar solutions are specified by imposing physical
boundary conditions at the black hole horizon and at radial infinity. At the
event horizon \(r=r_+\), with
\begin{equation}
r_+=M+\sqrt{M^2-a^2},
\qquad
\Omega_H=\frac{a}{2Mr_+},
\end{equation}
the physical solution is purely ingoing. With the time dependence
\(e^{-i\omega_{mn}t}\), this boundary condition is imposed as
\begin{equation}
\tilde R^{\rm H}_{\ell mn}(r)
=
e^{-i k_H r_*(r)}
\sum_{j=0}^{N_H}
a^{\rm H}_{j}
\left(r-r_+\right)^j ,
\qquad r\rightarrow r_+ ,
\label{eq:scalar_bc_horizon}
\end{equation}
where
\begin{equation}
k_H=\omega_{mn}-m\Omega_H .
\label{eq:kH_bc}
\end{equation}
The factor \(e^{-i k_H r_*}\) describes a wave falling into the future event
horizon. The expansion coefficients \(a_j^{\rm H}\) are determined recursively
by substituting Eq.~\eqref{eq:scalar_bc_horizon} into the homogeneous radial
equation. The overall normalization is fixed by choosing the zeroth-order
coefficient
\begin{equation}
a_0^{\rm H}=1 .
\end{equation}

At the spatial infinity, propagating massive scalar modes satisfy
\begin{equation}
\omega_{mn}^2>\mu_s^2 ,
\end{equation}
and the asymptotic wave number is
\begin{equation}
k_\infty=\sqrt{\omega_{mn}^2-\mu_s^2}.
\end{equation}
The physical solution is then purely outgoing and is written as
\begin{equation}
\tilde R^{\infty}_{\ell mn}(r)
=
e^{+i k_\infty r_*(r)}
r^{iM\mu_s^2/k_\infty}
\sum_{j=0}^{N_\infty}
\frac{b^{\infty}_{j}}{r^j},
\qquad r\rightarrow \infty .
\label{eq:scalar_bc_infinity}
\end{equation}
The factor \(r^{iM\mu_s^2/k_\infty}\) is the Coulomb-type phase correction
associated with the long-range part of the massive-scalar potential. The
coefficients \(b_j^{\infty}\) are obtained order by order from the homogeneous
equation, with the normalization
\begin{equation}
b_0^{\infty}=1 .
\end{equation}

For subthreshold modes with
\begin{equation}
\omega_{mn}^2<\mu_s^2 ,
\end{equation}
the radial wave number at infinity becomes imaginary. These modes are
evanescent and do not carry scalar radiation to infinity. Defining
\begin{equation}
\kappa_\infty=\sqrt{\mu_s^2-\omega_{mn}^2},
\end{equation}
the physical asymptotic solution is chosen to be exponentially decaying,
\begin{equation}
\tilde R^{\infty}_{\ell mn}(r)
\propto
e^{-\kappa_\infty r_*(r)}
r^{M\mu_s^2/\kappa_\infty}
\left[1+\mathcal{O}\left(r^{-1}\right)\right],
\qquad r\rightarrow\infty .
\label{eq:scalar_bc_evanescent}
\end{equation}
Therefore, only modes satisfying \(\omega_{mn}^2>\mu_s^2\) contribute to the
scalar flux at infinity, whereas the horizon flux is determined by the ingoing
solution in Eq.~\eqref{eq:scalar_bc_horizon}.


Following Ref.~\cite{Barsanti:2022ana}, the truncation orders of the
near-horizon and asymptotic expansions are fixed to
\begin{equation}
N_H=N_\infty=N_{\rm max}=4 .
\label{eq:NH_Ninf_choice}
\end{equation}
This choice provides sufficiently accurate boundary conditions for numerical
integration. In Ref.~\cite{Barsanti:2022ana}, the accuracy was checked by
comparing the dominant \(\ell=m=1\) scalar-flux mode computed with
\(N_{\rm max}=3\) and \(N_{\rm max}=4\). The relative difference was found to be
smaller than \(10^{-10}\%\) over the integration domain
\(2.4M\le r\le15M\) and \(10^{-5}\le \bar\mu_s\le0.4\). We adopt the same
truncation order for the low-eccentricity cases. For larger eccentricities, we
increase \(N_{\rm max}\) when necessary and require the relative change between
two consecutive truncation orders, \(N_{\rm max}\) and \(N_{\rm max}+1\), to be
smaller than \(10^{-10}\). Once this criterion is satisfied, the corresponding
boundary expansion is used to initialize the homogeneous radial solutions.

\begin{table*}[htbp!]
\centering
\begin{tabular}{ccc|ccccc}
\hline
$ \bar{\mu}_s $ & $p/M$  & $e$ &  $\dot{E}^{\ell_{\text{max}}=10}_s$
&$\dot{E}^\text{{int}}_s$ &Relative difference
& $\dot{E}^{\ell_{\text{max}}=9}_s$
& Relative difference\\
  \hline
0.001 &2.5 &0.0  &$7.562\times10^{-4}$  &$7.562\times10^{-4}$ 
& $<10^{-8}\%$ &  $7.562\times10^{-4}$    & $<1\%$ \\
 &8.0 &0.0  &$1.542\times10^{-5}$ &$1.542\times10^{-5}$  & $<10^{-3} \ \%$ 
 & $1.542\times10^{-5}$  & $<10^{-3} \%$  \\
0.01 &2.5  &0.0  &$7.550\times10^{-4}$   & $7.550\times10^{-4}$
& $<10^{-8}\%$ & $7.515\times10^{-4}$ & $<1 \%$  \\
&8.0  &0.0   &$1.440\times10^{-5}$ &$1.440\times10^{-5}$ &  $<10^{-3} \ \%$&$1.440\times10^{-5}$ &$<10^{-3} \%$ \\ 
\hline\hline
$ \bar{\mu}_s $ & $p/M$  & $e$ &  $\dot{E}^{n_{\text{max}}=10}_s$
&$\dot{E}^\text{{int}}_s$ &Relative difference
& $\dot{E}^{n_{\text{max}}=9}_s$
& Relative difference\\
\hline
0.001 &2.5 &0.1  &$5.3781\times10^{-4}$   &$5.3781\times10^{-4}$ 
& $<10^{-8}\%$   &$5.3781\times10^{-4}$     & $<10^{-4}\%$ \\
 &8.0 &0.1  &$4.9115\times10^{-6}$  &$4.9115\times10^{-6}$ & $<10^{-5} \ \%$ 
 & $4.9115\times10^{-5}$  & $<10^{-5} \%$  \\
0.01 &2.5 &0.1  &$5.3874\times10^{-4}$   &$5.3874\times10^{-4}$ 
& $<10^{-4}\%$   &$5.3874\times10^{-4}$     & $<10^{-4}\%$ \\
&8.0 &0.1   &$4.5127\times10^{-6}$  &$4.5127\times10^{-6}$ & $<10^{-5} \ \%$ 
 & $4.5127\times10^{-5}$  & $<10^{-5} \%$  \\
\hline
0.001 &2.5 &0.3  &$5.4732\times10^{-4}$   &$5.4732\times10^{-4}$ 
& $<10^{-8}\%$   &$5.4732\times10^{-4}$     & $<10^{-4}\%$ \\
&8.0 &0.3  &$5.2146\times10^{-6}$  &$5.2146\times10^{-6}$ & $<10^{-5} \ \%$ 
 & $5.2146\times10^{-5}$  & $<10^{-5} \%$  \\
0.01 &2.5 &0.3  &$5.4013\times10^{-4}$   &$5.4013\times10^{-4}$ 
& $<10^{-4}\%$  &$5.4013\times10^{-4}$     & $<10^{-4}\%$ \\
&8.0 &0.3   &$4.5348\times10^{-6}$  &$4.5348\times10^{-6}$ & $<10^{-5} \ \%$ 
 & $4.5348\times10^{-5}$  & $<10^{-5} \%$  \\
\hline
\hline
\end{tabular}
\caption{Relative difference between scalar energy flux (in units of mass-ratio $q^2$) by summing over $\ell_{\text{max}}=\{9,10\}$ and interpolated values based on Chebyshev method are listed for three eccentricities $e=\{0.0,0.1,0.3\}$ and scalar mass $\bar{\mu}_s=\{0.001,0.01\}$. The MBH-spin is set to $a/M=0.9$ and the massive scalar energy fluxes are evaluated over the index $\ell_{\rm max}=10$. The energy fluxes with superscript \text{``int''} denotes the interpolated fluxes.
The column labelled of \text{``Relative difference''} represents the quantity defined as $(\dot{E}^{\ell_{\text{max}}=10}_\text{s}-\dot{E}^{\text{int}}_\text{s})/\dot{E}^{\ell_{\text{max}}=10}_\text{s} \times 100\%$.
}\label{tab:interpolation}
\end{table*}

\begin{figure*}[t]
\centering
\includegraphics[width=7.0in, height=3.5in]{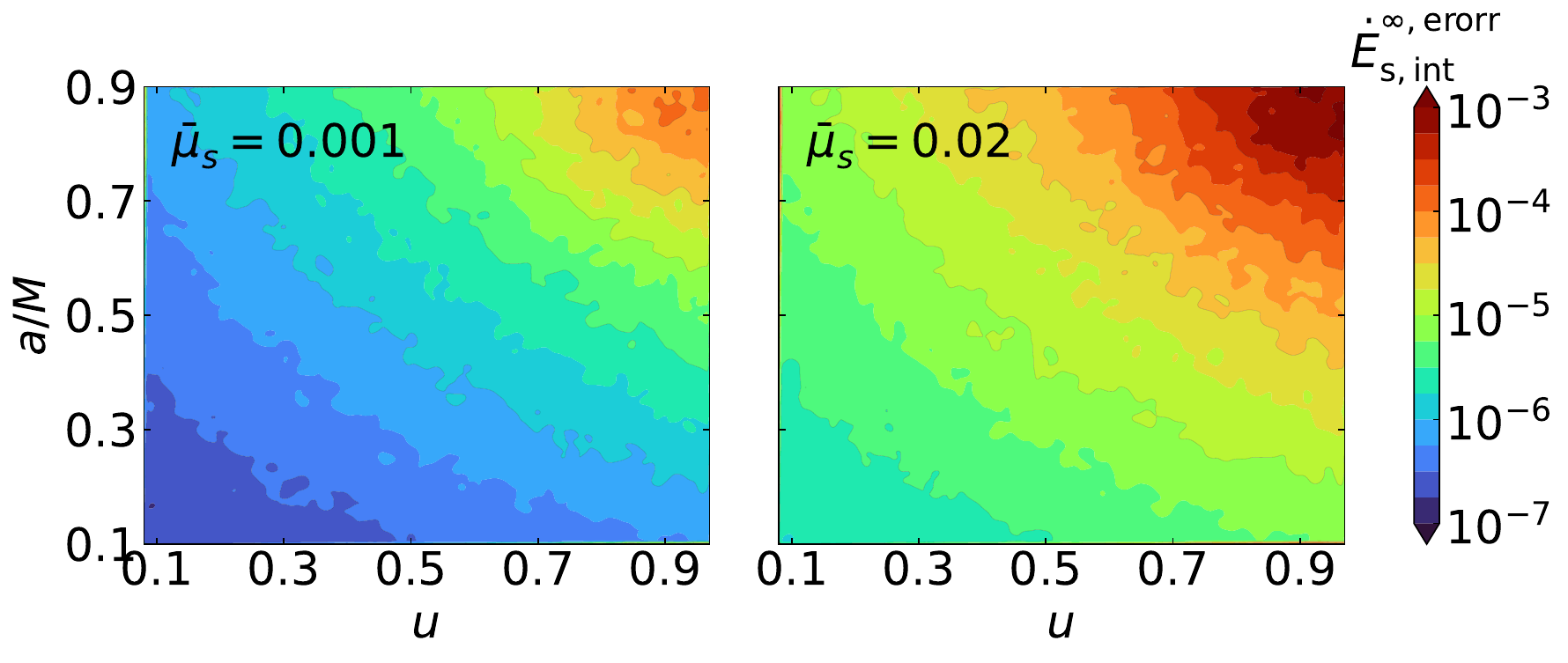}
\caption{Error of EMRI massive scalar energy flux from Chebyshev-interpolated method as a contour of MBH-spin $a/M$ and parameter $u$ is plotted for scalar mass $\bar{\mu}_s=0.001$ in the left panel and $\bar{\mu}_s=0.02$ in right panel. Other intrinsic parameters are set as follows: $q_s=0.1$ and mass-ratio $\epsilon=10^{-5}$, the extrinsic parameters are keeping same with Fig.~\ref{fig:mismatch:observtime}.}\label{fig:scalarflux:error:contour}
\end{figure*}
\begin{figure*}[t]
\centering
\includegraphics[width=4.15in, height=3.15in]{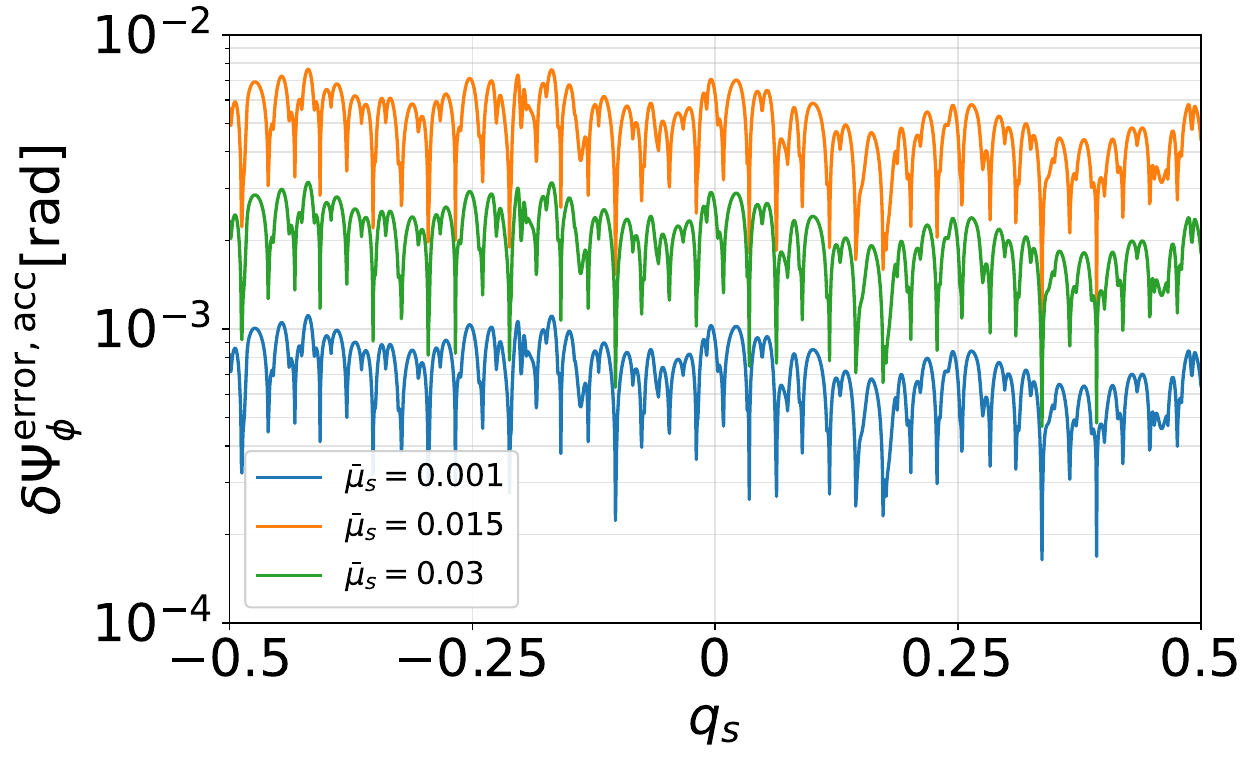}
\caption{Maximum value of accumulated dephasing induced by the error of interpolated scalar energy flux as function of scalar charge is plotted, where the time of orbital inspiral is one-year for three scalar masses $\bar{\mu}_s\in\{0.001,0.015,0.03\}$.}\label{fig:dephasing:acc:error}
\end{figure*}

\section{Accuracy assessment of the interpolated EMRI fluxes}
\label{sec:interpolation_error}
The adiabatic evolution of eccentric EMRIs in this work relies on
interpolated gravitational and scalar energy fluxes constructed with a
Chebyshev-based method. Therefore, it is necessary to quantify the accuracy of
the interpolation and to assess how local flux errors propagate into the
long-duration inspiral phase.

The accuracy of the interpolated fluxes is affected by both the precision of
the discrete flux data on the sampling grid and the interpolation scheme used
to reconstruct the fluxes away from the grid points. We first validate the
massive scalar energy fluxes by comparing the Chebyshev-interpolated results
with the perturbative calculations. The comparison is
summarized in Table~\ref{tab:interpolation}. The first four rows correspond to
circular EMRI orbits with different radii and two representative scalar field
masses, \(\bar{\mu}_s=\{0.001,0.01\}\), using the same orbital configurations
as those considered in Ref.~\cite{Barsanti:2022vvl}. The remaining rows show
eccentric equatorial EMRIs with \(e=\{0.1,0.3\}\), different semi-latus recta
\(p/M\), and the same scalar field masses. In all cases, the scalar flux is
obtained by summing over the relevant multipolar and radial harmonic modes
\((\ell,m,n)\). 
The interpolated massive scalar fluxes agree well with the corresponding
energy fluxes from the perturbative method. The relative differences are typically smaller than \(10^{-4}\%\), indicating that the Chebyshev interpolated method accurately reconstructs the scalar fluxes in the parameter region considered. We also observe massive scalar flux changing trends: the scalar flux is
suppressed as the scalar field mass increases, while eccentric orbits generally
produce larger scalar radiation because additional radial harmonics are
excited. These comparisons provide a direct validation of the interpolated
scalar flux model adopted in the adiabatic evolution.

Then we assess the interpolation accuracy of energy fluxes across the sampling grids of $(a,u,e)$, in which the error of massive scalar fluxes at infinity given by $\dot{E}_{s,\rm int}^{\infty,\rm error}=\dot{E}_s^{\infty,\rm int}-\dot{E}_s^{\infty,\rm per}$, obtained with the interpolation method and perturbation formula, as a function of MBH-spin $a/M$ and $u$ are plotted in Fig.~\ref{fig:scalarflux:error:contour} for the scalar-mass $\bar{\mu}_s=\{0.001,0.02\}$. This precision of flux error has a better stability to perform the adiabatic inspirals on the sampling grids. 
Briefly, as shown in Fig.~\ref{fig:scalarflux:error:contour}, the local interpolation error of the massive scalar energy flux typically lies in the range $10^{-7}-10^{-3}$ across the parameter space $(a/M, u)$, with the largest deviations appearing near the LSO of the rapidly rotating scenario; consequently, we will see below that the resulting maximum accumulated azimuthal phase error $\delta\Psi^{\rm error,acc}_\phi$, shown in Fig.~\ref{fig:dephasing:acc:error}, remains well below the LISA detectability threshold of $\delta\Psi_{\rm min} \sim 0.1$ rad for all scalar charges and masses considered. This confirms that the interpolation uncertainty does not affect the conclusions drawn from the dephasing and mismatch analysis.

Finally, we estimate the interpolation flux error at off-grids, resulting in an accumulative effect of local error on the long-term GW phase. Following the strategy of Ref.~\cite{Khalvati:2025znb}, we propagate the local scalar flux error into the accumulated phase. To the leading order, the interpolation-induced phase error can be estimated as
\begin{equation}
\begin{aligned}
\delta \Psi^{\rm acc}
\simeq
\Bigg|
\int_{p_s}^{p_0}
\omega_{mn}(\tilde a,\tilde p,e, \bar{\mu}_s)
\,
&
\frac{
E'(\tilde a,\tilde p,e, \bar{\mu}_s)
}{
\dot E^2(\tilde a,\tilde p,e, \bar{\mu}_s)
}
\,
\\ &\times
\delta \dot E_s(\tilde a,\tilde p,e, \bar{\mu}_s)dp
\,
\Bigg| ,
\label{eq:phase_error_interpolation}
\end{aligned}
\end{equation}
where
\begin{equation}
\delta \dot E_s(\tilde a,\tilde p,e, \bar{\mu}_s)
=
\dot E_s^{\rm int}(\tilde a,\tilde p,e, \bar{\mu}_s)
-
\dot E_s^{\rm per}(\tilde a,\tilde p,e, \bar{\mu}_s)\;,
\label{eq:scalar_flux_absolute_error}
\end{equation}
with  two dimensionless parameters \(\tilde a=a/M\) and \(\tilde p=p/M\).
Here \(\dot E(\tilde a,\tilde p,e, \bar{\mu}_s)\) is the total energy flux, including both the gravitational and scalar contributions, and
\(E'(\tilde a,\tilde p,e, \bar{\mu}_s)=dE/dp\) is the derivative of the orbital energy with respect to the semi-latus rectum. Equation~\eqref{eq:phase_error_interpolation} follows from linearizing the adiabatic phase evolution with respect to the local flux error.
\begin{figure*}[htb!]
\centering
\includegraphics[width=7.15in, height=3.35in]{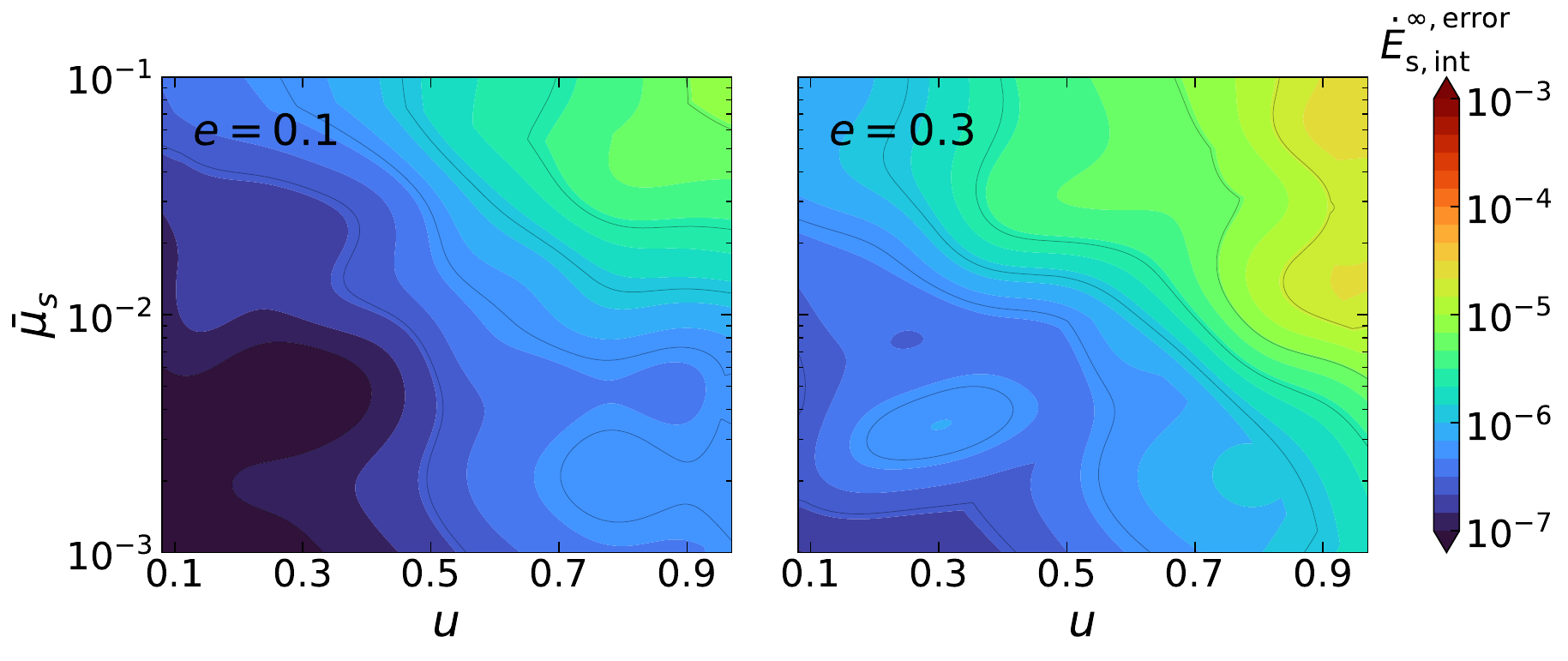}
\caption{Error of EMRI massive scalar energy flux from Chebyshev-interpolated method as a contour of MBH-spin $a/M$ and parameter $u$ is plotted for scalar mass $\bar{\mu}_s=0.001$ in the left panel and $\bar{\mu}_s=0.02$ in right panel. Other intrinsic parameters are set as follows: $q_s=0.1$ and mass-ratio $\epsilon=10^{-5}$.}\label{fig:scalarflux:error:MCMC}
\end{figure*}
Figure~\ref{fig:dephasing:acc:error} shows the resulting maximum accumulated
azimuthal phase error, \(\delta\Psi_\phi^{\rm error,acc}\), for three
representative scalar field masses,
\(\bar{\mu}_s\in\{0.001,0.015,0.03\}\), as a function of the scalar charge
\(q_s\). We find that the accumulated phase error remains small in the
parameter region considered. Its dependence on \(\bar{\mu}_s\) is not
monotonic: the error can initially increase with the scalar field becoming more massive, which is eventually reduced when more massive scalar modes become subthreshold and are suppressed at infinity, as determined with Eq.~\eqref{eq:edots}. In the present interpolation framework, the scalar flux includes modes up to
\(\ell_{\rm max}=5\) and \(n_{\rm max}=5\). The accumulated dephasing-error estimates indicate that the resulting interpolation uncertainty is sufficiently small, which are usually less than the threshold $\delta \Psi_{\rm min}\sim0.1$ radian.
Therefore, our interpolation method provides a better precision for performing the adiabatic evolutions in this work, a systematic extension including
higher multipolar and radial-harmonic modes will be useful for further
improving the accuracy of long-duration eccentric EMRI waveforms. \vspace{-0.2cm}

\section{Interpolation error of EMRI flux in the parameter estimation}\label{app:error:mcmc} \vspace{-0.2cm}
To constrain the massive scalar field using both the FIM and MCMC methods, we construct four
interpolation functions and incorporate them into the \texttt{FEW} framework.
Here, we assess the interpolation accuracy over the sampling grid
$(p/M,e,\bar{\mu}_s)$, fixing the dimensionless black hole spin to
$a/M=0.8$, as adopted in the MCMC analysis. To illustrate the dependence of
the interpolation accuracy on the orbital semi-latus rectum and the scalar-field
mass, Fig.~\ref{fig:scalarflux:error:MCMC} shows the interpolation error
$\dot{E}_{s,\mathrm{int}}^{\infty,\mathrm{error}}
=
\left|
\dot{E}_{s,\mathrm{int}}^{\infty}
-
\dot{E}_{s,\mathrm{per}}^{\infty}
\right|,
$
where \(\dot{E}_{s,\mathrm{int}}^{\infty}\) is obtained from the interpolation
scheme and \(\dot{E}_{s,\mathrm{per}}^{\infty}\) is calculated directly using
black hole perturbation theory. The error is presented as a function of
\((u,\bar{\mu}_s)\) for two representative eccentricities,
\(e=0.1\) and \(e=0.3\). 
Note that when EMRI flux for three sequence of 
$\{\ell_{\rm max},\ell_{\rm max}+1,\ell_{\rm max}+2\}$ is less than mass-ratio $q$, we can generate the EMRI flux data by summing over the maximum value of harmonic, we find that $\ell_{\rm max}= n_{\rm max}=5$ meets our requirements.

From Fig.~\ref{fig:scalarflux:error:MCMC}, one can observe that when the secondary approaches the LSO, corresponding to increasing $u$ in our parameterization, the interpolation error increases slightly. A similar trend is observed as the scalar mass increases. Nevertheless, the maximum interpolation error in the massive scalar flux remains below $10^{-3}$, which is sufficiently small to perform the adiabatic inspiral calculations here. In addition, the interpolation uncertainty remains subdominant to the statistical uncertainties obtained from
the FIM analysis in Table~\ref{tab:fim:error}, demonstrating that the adopted interpolation scheme is sufficiently accurate for the subsequent MCMC parameter estimation.

Our computation details of the MCMC analysis are as follows: the Bayesian inference of the full-parameter posterior distributions was performed on a GPU server equipped with two NVIDIA A100 80~GB PCIe GPUs. The server used CUDA~12.4, with Multi-Instance GPU (MIG) mode disabled. Each MCMC run was assigned to a single A100 GPU and executed with two CPU threads.
For the massive scalar eccentric EMRI model, we considered four parameter
configurations,
\begin{equation}
(e,\bar{\mu}_s)   =
    (0.1,0.01),\,
    (0.1,0.03),\,
    (0.3,0.01),\,
    (0.3,0.03).
\end{equation}
The remaining fiducial parameters were fixed to $M=10^{6}M_{\odot},
    m_p=10M_{\odot},
    a/M=0.8,    
    p/M=10,
    q_s=0.1$ with a \(\mathrm{SNR}\) of
\(150\) and an observation of
\(T_{\mathrm{obs}}=4~\mathrm{years}\).  With this configuration, we can obtain the posterior  of source parameters, specifically their distributions are listed for the $1\sigma$,  $2\sigma$ and  $3\sigma$ inference in Table~\ref{tab:posterior_sigma_intervals}.

\begin{figure*}[t]
\centering
\includegraphics[width=3.5in, height=2.5in]{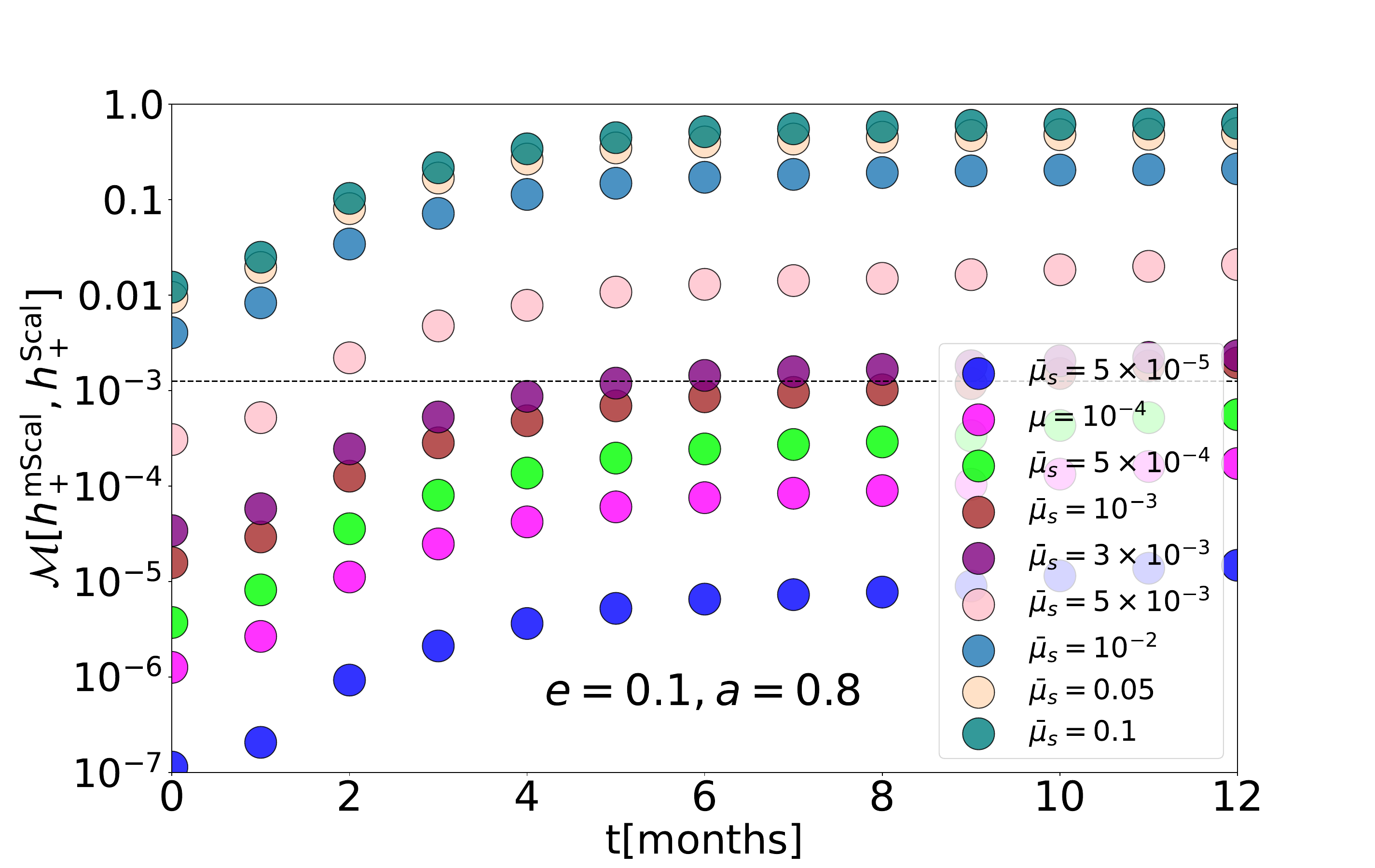}
\includegraphics[width=3.5in, height=2.5in]{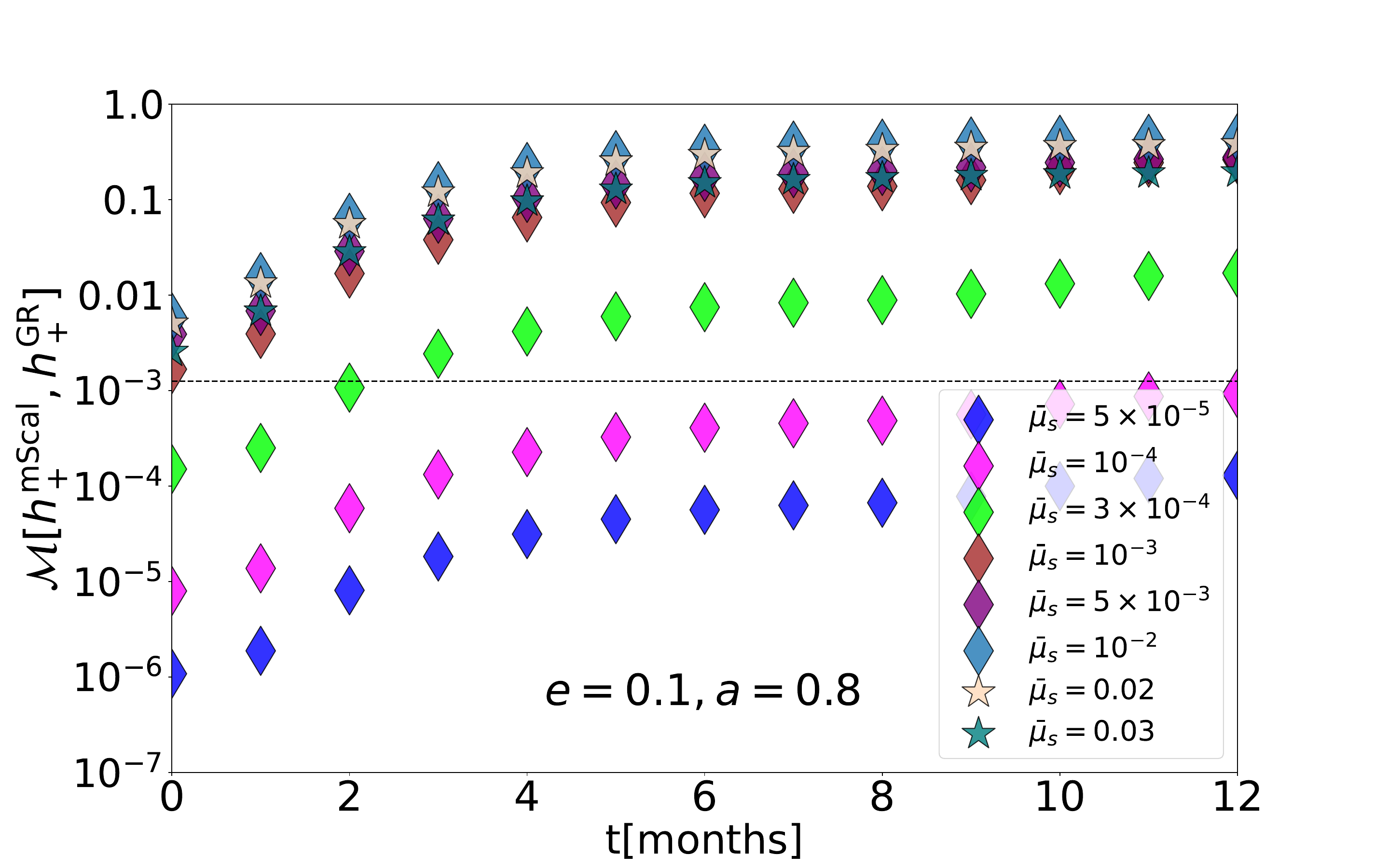}
\includegraphics[width=3.5in, height=2.5in]{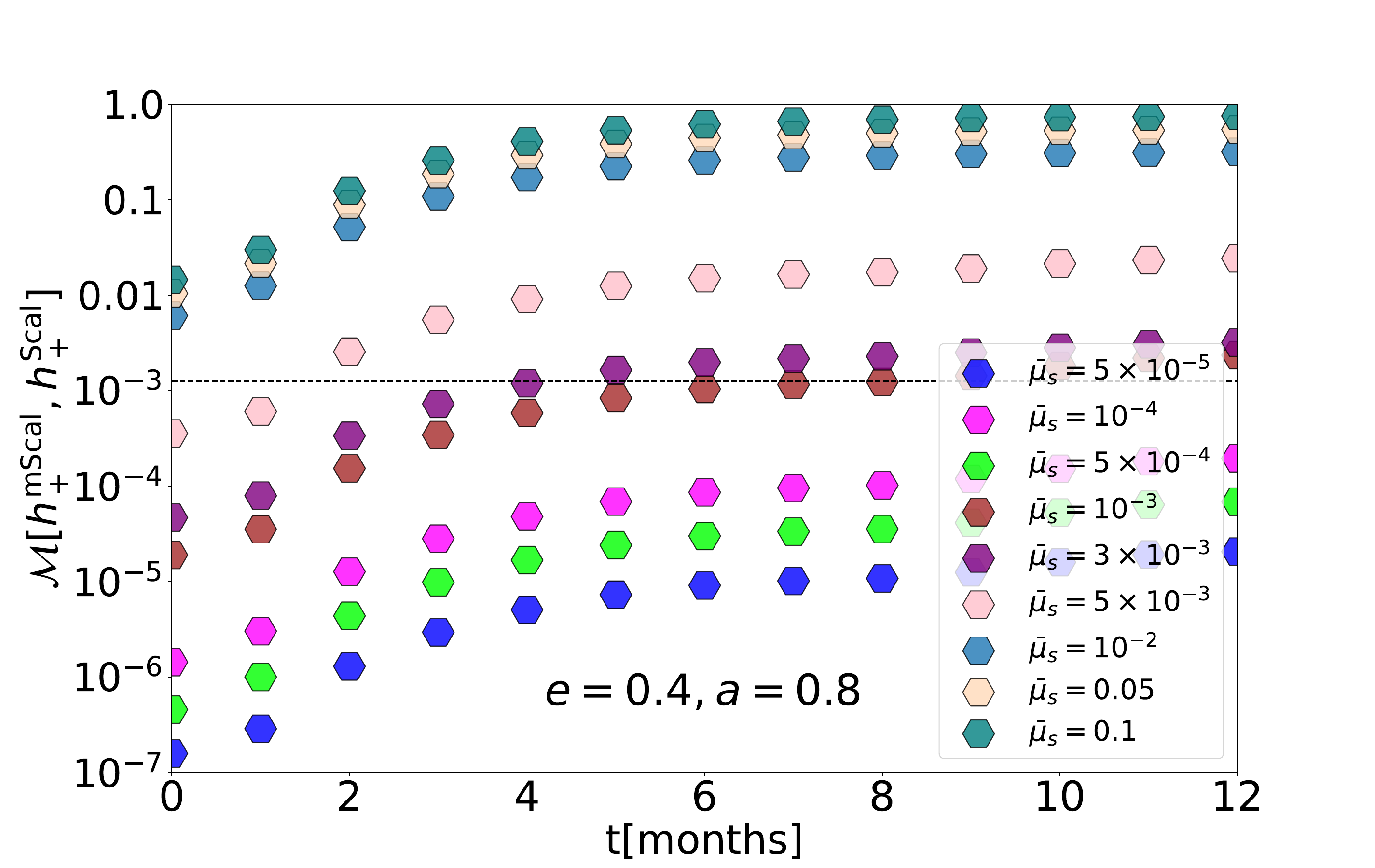}
\includegraphics[width=3.5in, height=2.5in]{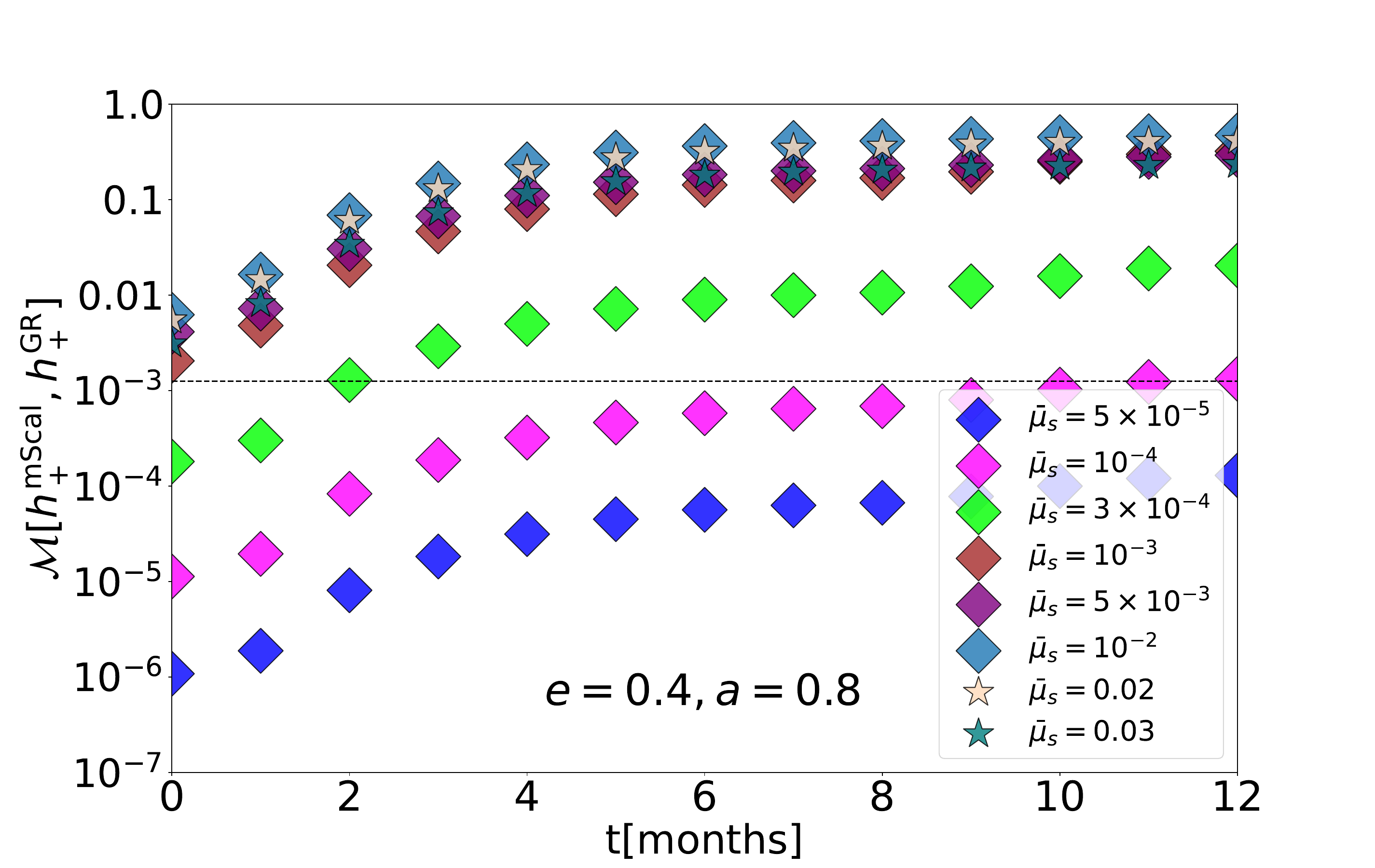}
\caption{Mismatches of EMRI waveforms between massive-scalar and massless-scalar gravitational theories in left panels (or GR in right panels) as the functions of observation time are considered for orbital eccentricity $e\in\{0.1,0.4\}$ and a fixed MBH-spin $a/M=0.8$, including different scalar masses $\bar{\mu}_s=\{5\times10^{-5},10^{-4},5\times10^{-4},10^{-3},3\times10^{-3},5\times10^{-3},10^{-2},0.05,0.1\}$ in the left panel and $\bar{\mu}_s=\{5\times10^{-5},10^{-4},3\times10^{-4},10^{-3},5\times10^{-3},10^{-2},0.02,0.03\}$ in the right panel. Other location parameters are set as the configuration in the opening paragraph of Sec.~\ref{result}.
}\label{fig:mismatch:observtime}
\end{figure*}

{\small
\setlength{\tabcolsep}{4pt}
\renewcommand{\arraystretch}{1.15}
\begin{longtable*}{@{}lcccc@{}}
\caption{Central posterior intervals for the full 14-parameter inference. Other parameters are keeping with Fig.~\ref{fig:MCMC:ecc0d1}.}\label{tab:posterior_sigma_intervals}
\\
\toprule
Parameter & True value & $1\sigma$ confidence interval & $2\sigma$ confidence interval & $3\sigma$ confidence interval \\
\midrule
\endfirsthead
\toprule
Parameter & True value & $1\sigma$ confidence interval & $2\sigma$ confidence interval & $3\sigma$ confidence interval \\
\midrule
\endhead
\midrule
\multicolumn{5}{r}{Continued on next page}\\
\endfoot
\bottomrule
\endlastfoot
\multicolumn{5}{@{}l}{\textbf{Case 1:} $e=0.1$, $\bar{\mu}_s=0.01$, $q_s=0.1$, $a/M=0.8$}\\\hline
$M/M_\odot$ & $1\times 10^{6}$ & $[9.9815\times 10^{5},\,1.0016\times 10^{6}]$ & $[9.962\times 10^{5},\,1.0034\times 10^{6}]$ & $[9.9487\times 10^{5},\,1.0046\times 10^{6}]$ \\
$m_p/M_\odot$ & $10$ & $[9.99446,\,10.0071]$ & $[9.98689,\,10.0129]$ & $[9.9828,\,10.0167]$ \\
$a/M$ & $0.8$ & $[0.797881,\,0.801744]$ & $[0.795867,\,0.803668]$ & $[0.794004,\,0.804926]$ \\
$p/M$ & $10$ & $[9.98919,\,10.0131]$ & $[9.97723,\,10.0256]$ & $[9.96931,\,10.0372]$ \\
$e$ & $0.1$ & $[0.0992871,\,0.100673]$ & $[0.0990513,\,0.100947]$ & $[0.099001,\,0.100995]$ \\
$q_s$ & $0.1$ & $[0.0479769,\,0.132956]$ & $[0.0250526,\,0.165572]$ & $[0.00301256,\,0.188619]$ \\
$\bar{\mu}_s$ & $0.01$ & $[0.00466675,\,0.0161353]$ & $[0.00157624,\,0.0288419]$ & $[0.00103553,\,0.0336453]$ \\
$\Phi_{\phi0}$ & $1.0472$ & $[1.01819,\,1.08528]$ & $[0.976868,\,1.11562]$ & $[0.938531,\,1.1547]$ \\
$\Phi_{r0}$ & $3.14159$ & $[3.0249,\,3.25212]$ & $[2.9317,\,3.36747]$ & $[2.78342,\,3.43121]$ \\
$\theta_S$ & $1.5708$ & $[1.55229,\,1.59339]$ & $[1.53225,\,1.61467]$ & $[1.51542,\,1.63499]$ \\
$\phi_S$ & $3.14159$ & $[3.11931,\,3.16232]$ & $[3.09639,\,3.18514]$ & $[3.06605,\,3.20646]$ \\
$\theta_L$ & $0.785398$ & $[0.764534,\,0.810579]$ & $[0.745439,\,0.83255]$ & $[0.719846,\,0.847196]$ \\
$\phi_L$ & $1.0472$ & $[1.00503,\,1.08416]$ & $[0.962373,\,1.12157]$ & $[0.926599,\,1.15944]$ \\
$d$ & $0.120058$ & $[0.117708,\,0.124152]$ & $[0.114255,\,0.127447]$ & $[0.112048,\,0.130558]$ \\\hline
\addlinespace
\multicolumn{5}{@{}l}{\textbf{Case 2:} $e=0.1$, $\bar{\mu}_s=0.03$, $q_s=0.1$, $a/M=0.8$}\\ \hline
$M/M_\odot$ & $1\times 10^{6}$ & $[9.9761\times 10^{5},\,1.0021\times 10^{6}]$ & $[9.9626\times 10^{5},\,1.0035\times 10^{6}]$ & $[9.9521\times 10^{5},\,1.0042\times 10^{6}]$ \\
$m_p/M_\odot$ & $10$ & $[9.99171,\,10.007]$ & $[9.98405,\,10.0156]$ & $[9.97926,\,10.0197]$ \\
$a/M$ & $0.8$ & $[0.797298,\,0.802398]$ & $[0.79564,\,0.804149]$ & $[0.79469,\,0.804907]$ \\
$p/M$ & $10$ & $[9.98513,\,10.0167]$ & $[9.9742,\,10.0272]$ & $[9.96958,\,10.0329]$ \\
$e$ & $0.1$ & $[0.0993637,\,0.100714]$ & $[0.0990456,\,0.100948]$ & $[0.099003,\,0.100993]$ \\
$q_s$ & $0.1$ & $[0.0512148,\,0.198794]$ & $[0.0123256,\,0.279612]$ & $[0.0011036,\,0.363749]$ \\
$\bar{\mu}_s$ & $0.03$ & $[0.0233331,\,0.033473]$ & $[0.0112918,\,0.0348074]$ & $[0.00157735,\,0.0349774]$ \\
$\Phi_{\phi0}$ & $1.0472$ & $[1.01327,\,1.07816]$ & $[0.978975,\,1.10971]$ & $[0.954013,\,1.13683]$ \\
$\Phi_{r0}$ & $3.14159$ & $[3.02008,\,3.24479]$ & $[2.88811,\,3.33649]$ & $[2.74382,\,3.39763]$ \\
$\theta_S$ & $1.5708$ & $[1.55168,\,1.59125]$ & $[1.53139,\,1.61275]$ & $[1.51476,\,1.63726]$ \\
$\phi_S$ & $3.14159$ & $[3.11914,\,3.15868]$ & $[3.09413,\,3.17848]$ & $[3.07301,\,3.19797]$ \\
$\theta_L$ & $0.785398$ & $[0.761666,\,0.811991]$ & $[0.738238,\,0.836397]$ & $[0.715045,\,0.855683]$ \\
$\phi_L$ & $1.0472$ & $[1.00554,\,1.08074]$ & $[0.962945,\,1.11775]$ & $[0.920954,\,1.14966]$ \\
$d$ & $0.120045$ & $[0.117084,\,0.124251]$ & $[0.114074,\,0.129363]$ & $[0.111091,\,0.134156]$ \\\hline
\addlinespace
\multicolumn{5}{@{}l}{\textbf{Case 3:} $e=0.3$, $\bar{\mu}_s=0.01$, $q_s=0.1$, $a/M=0.8$}\\\hline
$M/M_\odot$ & $1\times 10^{6}$ & $[9.9687\times 10^{5},\,1.004\times 10^{6}]$ & $[9.9347\times 10^{5},\,1.0068\times 10^{6}]$ & $[9.9183\times 10^{5},\,1.0092\times 10^{6}]$ \\
$m_p/M_\odot$ & $10$ & $[9.97588,\,10.0257]$ & $[9.95525,\,10.0428]$ & $[9.94173,\,10.0563]$ \\
$a/M$ & $0.8$ & $[0.797295,\,0.803576]$ & $[0.794025,\,0.806094]$ & $[0.79283,\,0.807856]$ \\
$p/M$ & $10$ & $[9.97786,\,10.0167]$ & $[9.96226,\,10.0369]$ & $[9.9513,\,10.0442]$ \\
$e$ & $0.3$ & $[0.298844,\,0.301009]$ & $[0.298023,\,0.301883]$ & $[0.297203,\,0.302584]$ \\
$q_s$ & $0.1$ & $[0.0661339,\,0.124724]$ & $[0.0467281,\,0.145075]$ & $[0.0135348,\,0.167898]$ \\
$\bar{\mu}_s$ & $0.01$ & $[0.00569767,\,0.0123692]$ & $[0.00183041,\,0.0229061]$ & $[0.00101627,\,0.0333847]$ \\
$\Phi_{\phi0}$ & $1.0472$ & $[0.998355,\,1.07915]$ & $[0.964717,\,1.10667]$ & $[0.931374,\,1.13259]$ \\
$\Phi_{r0}$ & $3.14159$ & $[3.10504,\,3.19463]$ & $[3.04875,\,3.23718]$ & $[2.98978,\,3.26999]$ \\
$\theta_S$ & $1.5708$ & $[1.55026,\,1.58707]$ & $[1.53065,\,1.60415]$ & $[1.51365,\,1.62148]$ \\
$\phi_S$ & $3.14159$ & $[3.12248,\,3.16126]$ & $[3.10384,\,3.18418]$ & $[3.0826,\,3.19768]$ \\
$\theta_L$ & $0.785398$ & $[0.767688,\,0.809761]$ & $[0.741785,\,0.836638]$ & $[0.72303,\,0.855538]$ \\
$\phi_L$ & $1.0472$ & $[0.992854,\,1.07569]$ & $[0.946982,\,1.11333]$ & $[0.923974,\,1.14885]$ \\
$d$ & $0.122823$ & $[0.120596,\,0.12757]$ & $[0.117034,\,0.131094]$ & $[0.113742,\,0.13372]$ \\\hline
\addlinespace
\multicolumn{5}{@{}l}{\textbf{Case 4:} $e=0.3$, $\bar{\mu}_s=0.03$, $q_s=0.1$, $a/M=0.8$}\\\hline
$M/M_\odot$ & $1\times 10^{6}$ & $[9.963\times 10^{5},\,1.0022\times 10^{6}]$ & $[9.9356\times 10^{5},\,1.0061\times 10^{6}]$ & $[9.9162\times 10^{5},\,1.0091\times 10^{6}]$ \\
$m_p/M_\odot$ & $10$ & $[9.97401,\,10.0152]$ & $[9.95338,\,10.04]$ & $[9.9372,\,10.0567]$ \\
$a/M$ & $0.8$ & $[0.796612,\,0.802124]$ & $[0.794422,\,0.805414]$ & $[0.792723,\,0.807913]$ \\
$p/M$ & $10$ & $[9.98687,\,10.021]$ & $[9.96646,\,10.0346]$ & $[9.95093,\,10.045]$ \\
$e$ & $0.3$ & $[0.29938,\,0.301144]$ & $[0.298146,\,0.302082]$ & $[0.29745,\,0.302772]$ \\
$q_s$ & $0.1$ & $[0.0288812,\,0.13227]$ & $[0.00518814,\,0.199537]$ & $[5.3143\times 10^{-4},\,0.290088]$ \\
$\bar{\mu}_s$ & $0.03$ & $[0.022613,\,0.0332435]$ & $[0.00844021,\,0.0347597]$ & $[0.00154158,\,0.0349862]$ \\
$\Phi_{\phi0}$ & $1.0472$ & $[1.00331,\,1.08323]$ & $[0.969134,\,1.11893]$ & $[0.923285,\,1.14695]$ \\
$\Phi_{r0}$ & $3.14159$ & $[3.10107,\,3.19591]$ & $[3.05199,\,3.24419]$ & $[3.01483,\,3.28403]$ \\
$\theta_S$ & $1.5708$ & $[1.54744,\,1.59025]$ & $[1.52885,\,1.61532]$ & $[1.51373,\,1.63562]$ \\
$\phi_S$ & $3.14159$ & $[3.12181,\,3.16339]$ & $[3.1035,\,3.18574]$ & $[3.07925,\,3.2072]$ \\
$\theta_L$ & $0.785398$ & $[0.768734,\,0.811756]$ & $[0.744217,\,0.834528]$ & $[0.729136,\,0.848593]$ \\
$\phi_L$ & $1.0472$ & $[0.998589,\,1.08201]$ & $[0.956776,\,1.11893]$ & $[0.919155,\,1.15377]$ \\
$d$ & $0.122806$ & $[0.120061,\,0.12686]$ & $[0.117444,\,0.130721]$ & $[0.113036,\,0.135626]$ \\
\end{longtable*}
}

\section{Waveform mismatch results}\label{app:D mismatch}
Here, we present the detailed waveform-mismatch results
supporting the discussion in Sec.~III B. We denote the waveforms in the
massive-scalar, massless-scalar, and GR scenarios by
$h^{\rm mScal}$, $h^{\rm Scal}$, and $h^{\rm GR}$, respectively.
We consider the mismatches
$\mathcal{M}(h^{\rm mScal},h^{\rm Scal})$ and
$\mathcal{M}(h^{\rm mScal},h^{\rm GR})$ as functions of the LISA
observation time, orbital eccentricity, scalar mass, and scalar charge.

The left panels of Fig.~\ref{fig:mismatch:observtime} show that the massive-scalar waveform becomes distinguishable from the massless-scalar
waveform after approximately $6$--$8$ months. For $e=0.1$, the corresponding threshold is approximately
$\bar{\mu}_s=10^{-3}$
$(\mu\sim7.5\times10^{-19}\,\mathrm{eV})$, while the mismatch is slightly larger for $e=0.4$. The mismatches generally increase with the scalar mass over the range displayed in these panels. The right panels compare the massive-scalar waveform with the GR waveform. As $\bar{\mu}_s$ increases from $5\times10^{-5}$ to $0.03$,
the mismatch initially increases and then decreases. According to \autoref{eq:edots}, modes for which $\omega_{mn}<\mu$ are bound and cannot propagate to infinity. The resulting suppression of the
massive-scalar flux explains the decrease in the mismatch relative to GR at larger scalar masses. The mismatch is again slightly enhanced for the larger eccentricity.

\begin{figure*}[htb!]
\centering
\includegraphics[width=3.5in, height=2.75in]{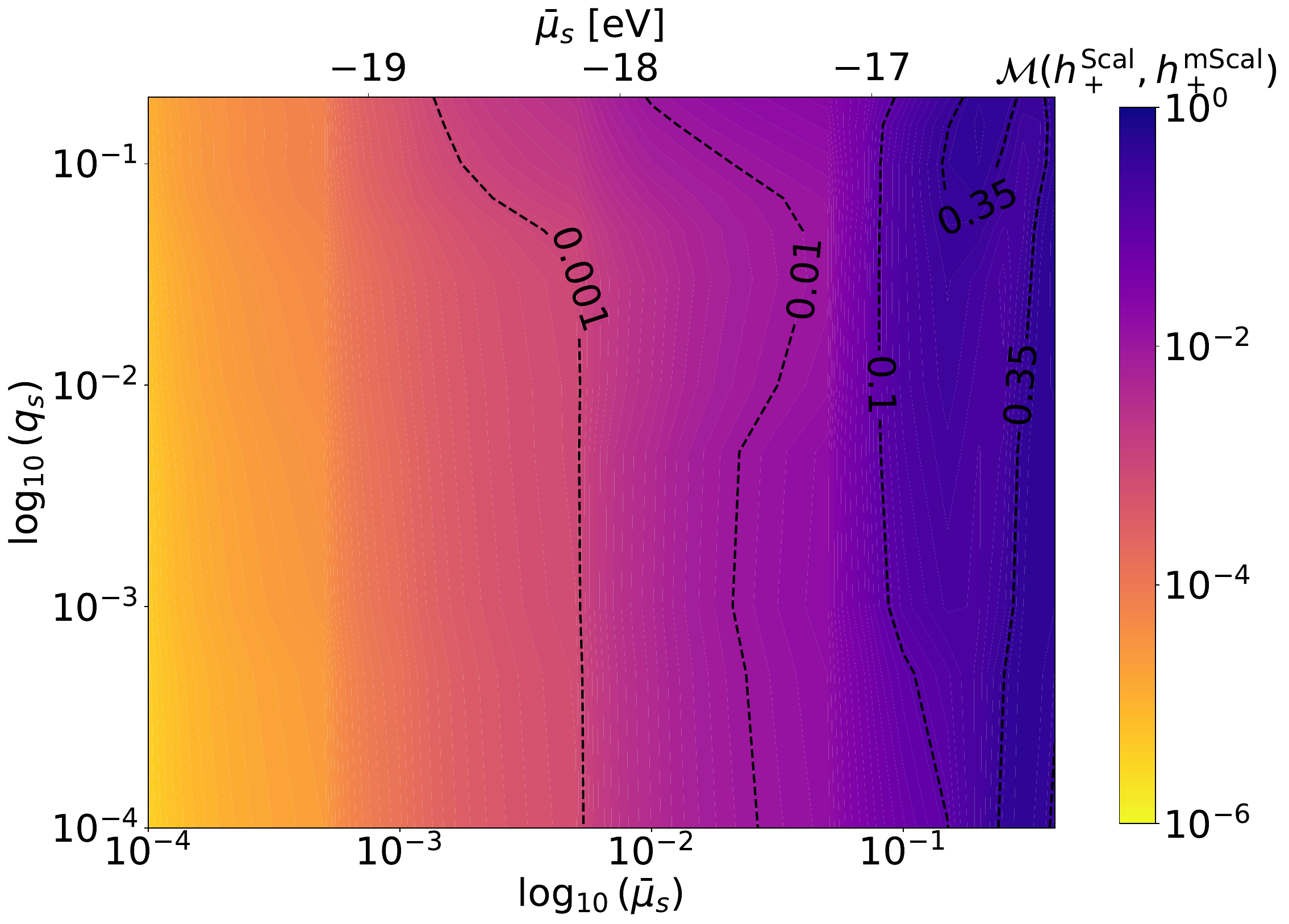}
\includegraphics[width=3.5in, height=2.75in]{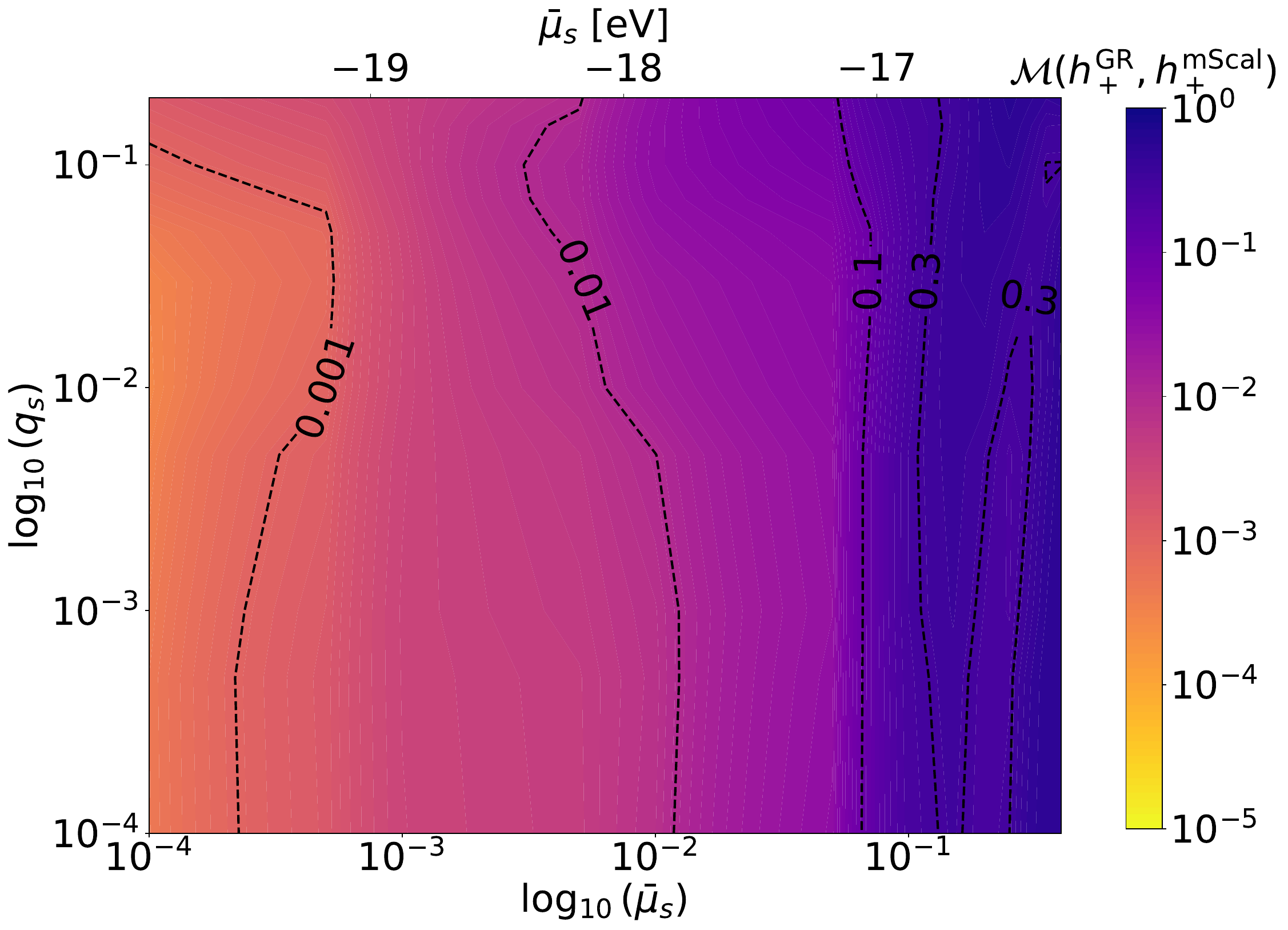}
\caption{Mismatches of EMRI waveforms between massive scalar and massless scalar gravitational theories in the left panel (and GR in the right panel) as functions of parameters $(q_s,\bar{\mu}_s)$ are considered for a fixed MBH-spin $a/M=0.5$ and eccentricity $e=0.3$. The black dashed lines denote the contours of mismatches, including the threshold $\mathcal{M}\sim0.001$ in two subfigures. Three EMRI waveforms are modelled in GR, the massive scalar and massless scalar gravitational theories, the length of waveforms are all set to one year. Other parameters are keeping same with Fig.~\ref{fig:mismatch:observtime}.
}\label{fig:mismatch:contour}
\end{figure*}

We also evaluate the joint dependence of the mismatch on the scalar charge and scalar mass for a one-year observation. The left panel of Fig.~\ref{fig:mismatch:contour} presents
$\mathcal{M}(h^{\rm Scal},h^{\rm mScal})$. The black dashed curves
denote contours of constant mismatch, including the distinguishability
threshold $\mathcal{M}=10^{-3}$. The minimum scalar mass distinguishable
from the massless-scalar case is approximately $\bar{\mu}_s\simeq8\times10^{-3}$
$(\mu\sim6\times10^{-18}\,\mathrm{eV})$. The right panel shows
$\mathcal{M}(h^{\rm GR},h^{\rm mScal})$. For a one-year LISA
observation, the corresponding distinguishable scalar mass is approximately $\bar{\mu}_s\simeq4\times10^{-4}$
$(\mu\sim3\times10^{-19}\,\mathrm{eV})$. For a fixed scalar mass,
the mismatch increases with $q_s$, reflecting the quadratic dependence
of the scalar flux on the scalar charge. The larger mismatch relative to GR than relative to the massless-scalar theory is consistent with the suppression of the massive-scalar flux at larger scalar masses.

\section{Resonant massive-scalar fluxes for eccentric EMRIs}
\label{sec:resonant_scalar_fluxes}

{\color{black}In this section, we investigate the resonant massive-scalar fluxes generated by eccentric EMRIs. Resonances occur when one of the harmonics of the eccentric orbital motion approaches the frequency
of a quasi-bound state of the massive scalar field. For a rotating black hole, a resonantly excited scalar mode may simultaneously satisfy the
superradiant condition, in which the rotational energy is extracted from
the black hole spin. If the resulting negative scalar-energy flux through the
horizon becomes sufficiently large to compensate for the gravitational fluxes of the eccentric orbit, the inspiral can be temporarily stalled,
giving rise to a floating configuration~\cite{Cardoso:2011xi}.
\begin{table*}[t]
\centering
\caption{Mode-by-mode validation of the normalized massive-scalar horizon energy flux against Table~II of Ref.~\cite{Fujita:2016yav}. The columns labeled
``ref'' and ``calc'' denote the numerical results in the reference and present paper, respectively. The quantity $q_s^{\rm crit}$ is the critical scalar charge for a floating configuration. Negative flux at horizon corresponds to
superradiant extraction of black hole rotational energy.
}\label{tab:dEdtH:floating}

\scriptsize
\setlength{\tabcolsep}{2.8pt}
\renewcommand{\arraystretch}{1.08}
\begin{tabular}{c|ccccc|ccccc}
\toprule
& \multicolumn{5}{c}{$\bar{\mu}_s=0.01$}
& \multicolumn{5}{c}{$\bar{\mu}_s=0.10$}
\\
\cmidrule(lr){2-6}\cmidrule(lr){7-11}
$k$
& $p/M$
& $\dot E_{\rm ref}^{s,H}/\mathcal N$
& $\dot E_{\rm calc}^{s,H}/\mathcal N$
& $|\delta_{\dot E}|[\%]$
& $q_s^{\rm crit}$
& $p/M$
& $\dot E_{\rm ref}^{s,H}/\mathcal N$
& $\dot E_{\rm calc}^{s,H}/\mathcal N$
& $|\delta_{\dot E}|[\%]$
& $q_s^{\rm crit}$
\\\hline
\multicolumn{11}{c}{$e=0.1$}\\
\midrule
-1 & 8.242523  & $-1.76\times10^{-3}$ & $-1.757706\times10^{-3}$ & 0.1303 & $2.7\times10^{-1}$
   & 2.839923  & $-2.88\times10^{-3}$ & $-2.878808\times10^{-3}$ & 0.0414 & $2.1$ \\
 0 & 21.204013 & $-4.83\times10^{-1}$ & $-4.825570\times10^{-1}$ & 0.0917 & $1.7\times10^{-3}$
   & 4.305965  & $-1.79\times10^{-1}$ & $-1.791505\times10^{-1}$ & 0.0841 & $1.2\times10^{-1}$ \\
 1 & 32.917260 & $-3.23\times10^{-3}$ & $-3.227677\times10^{-3}$ & 0.0719 & $6.9\times10^{-3}$
   & 6.276878  & $-1.47\times10^{-3}$ & $-1.470055\times10^{-3}$ & 0.0037 & $5.5\times10^{-1}$
\\\hline
\multicolumn{11}{c}{$e=0.3$}\\
\midrule
-1 & 7.905366  & $-1.74\times10^{-2}$ & $-1.737160\times10^{-2}$ & 0.1632 & $1.0\times10^{-1}$
   & 2.803173  & $-2.73\times10^{-2}$ & $-2.732675\times10^{-2}$ & 0.0980 & $7.7\times10^{-1}$ \\
 0 & 19.618724 & $-4.37\times10^{-1}$ & $-4.371177\times10^{-1}$ & 0.0269 & $2.2\times10^{-2}$
   & 4.082557  & $-1.48\times10^{-1}$ & $-1.474677\times10^{-1}$ & 0.3597 & $1.5\times10^{-1}$ \\
 1 & 30.328484 & $-2.62\times10^{-2}$ & $-2.616247\times10^{-2}$ & 0.1432 & $3.1\times10^{-3}$
   & 5.840894  & $-1.29\times10^{-2}$ & $-1.290571\times10^{-2}$ & 0.0442 & $1.29\times10^{-1}$ \\
\\\hline
\multicolumn{11}{c}{$e=0.5$}\\
\midrule
-2 & 3.275859  & $-2.35\times10^{-3}$ & $-2.350455\times10^{-3}$ & 0.0194 & $2.1$
   & 2.279085  & $-2.91\times10^{-2}$ & $-2.910619\times10^{-2}$ & 0.0213 & $1.2$ \\
-1 & 7.186228  & $-5.92\times10^{-2}$ & $-5.923547\times10^{-2}$ & 0.0599 & $7.2\times10^{-2}$
   & 2.718468  & $-7.20\times10^{-2}$ & $-7.195442\times10^{-2}$ & 0.0633 & $5.5\times10^{-1}$ \\
 0 & 16.444447 & $-3.43\times10^{-1}$ & $-3.432192\times10^{-1}$ & 0.0639 & $4.0\times10^{-3}$
   & 3.642199  & $-7.10\times10^{-2}$ & $-7.091754\times10^{-2}$ & 0.1162 & $3.1\times10^{-1}$ \\
 1 & 25.149588 & $-5.76\times10^{-2}$ & $-5.758569\times10^{-2}$ & 0.0248 & $3.4\times10^{-3}$
   & 4.976678  & $-3.15\times10^{-2}$ & $-3.151582\times10^{-2}$ & 0.0502 & $2.3\times10^{-1}$ \\
 2 & 32.823151 & $-1.19\times10^{-2}$ & $-1.189417\times10^{-2}$ & 0.0490 & $3.9\times10^{-3}$
   & 6.406282  & $-5.84\times10^{-3}$ & $-5.843823\times10^{-3}$ & 0.0655 & $3.0\times10^{-1}$ \\
 3 & 39.780705 & $-2.87\times10^{-3}$ & $-2.870050\times10^{-3}$ & 0.0017 & $5.0\times10^{-3}$
   & 7.796703  & $-1.07\times10^{-3}$ & $-1.069621\times10^{-3}$ & 0.0354 & $4.4\times10^{-1}$ \\
 4 & 46.220835 & $-7.66\times10^{-4}$ & $-7.656423\times10^{-4}$ & 0.0467 & $6.6\times10^{-3}$
   & 9.122696  & $-2.13\times10^{-4}$ & $-2.132584\times10^{-4}$ & 0.1213 & $6.8\times10^{-1}$ \\
\\\hline
\multicolumn{11}{c}{$e=0.7$}\\
\midrule
-3 & 2.619283  & $-8.22\times10^{-3}$ & $-8.217904\times10^{-3}$ & 0.0255 & $1.8$
   & 2.222813  & $-3.19\times10^{-2}$ & $-3.189695\times10^{-2}$ & 0.0095 & $1.3$ \\
-2 & 3.304721  & $-2.39\times10^{-2}$ & $-2.389512\times10^{-2}$ & 0.0204 & $6.6\times10^{-1}$
   & 2.334121  & $-3.19\times10^{-2}$ & $-3.185073\times10^{-2}$ & 0.1544 & $1.2$ \\
-1 & 5.959339  & $-1.65\times10^{-1}$ & $-1.648162\times10^{-1}$ & 0.1114 & $6.5\times10^{-2}$
   & 2.562147  & $-1.62\times10^{-2}$ & $-1.624185\times10^{-2}$ & 0.2583 & $1.4$ \\
 0 & 11.670814 & $-1.88\times10^{-1}$ & $-1.879514\times10^{-1}$ & 0.0258 & $1.2\times10^{-2}$
   & 3.011985  & $-3.10\times10^{-3}$ & $-3.104538\times10^{-3}$ & 0.1464 & $2.2$ \\
 1 & 17.377849 & $-7.11\times10^{-2}$ & $-7.115846\times10^{-2}$ & 0.0822 & $7.1\times10^{-3}$
   & 3.722402  & $-2.04\times10^{-2}$ & $-2.035674\times10^{-2}$ & 0.2121 & $5.5\times10^{-1}$ \\
 2 & 22.510881 & $-2.74\times10^{-2}$ & $-2.735752\times10^{-2}$ & 0.1550 & $6.0\times10^{-3}$
   & 4.575213  & $-1.69\times10^{-2}$ & $-1.690234\times10^{-2}$ & 0.0138 & $3.8\times10^{-1}$ \\
 3 & 27.197777 & $-1.18\times10^{-2}$ & $-1.176319\times10^{-2}$ & 0.3120 & $5.7\times10^{-3}$
   & 5.456882  & $-7.46\times10^{-3}$ & $-7.456605\times10^{-3}$ & 0.0455 & $3.7\times10^{-1}$ \\
 4 & 31.550150 & $-5.50\times10^{-3}$ & $-5.504927\times10^{-3}$ & 0.0896 & $5.8\times10^{-3}$
   & 6.322137  & $-3.02\times10^{-3}$ & $-3.020389\times10^{-3}$ & 0.0129 & $4.1\times10^{-1}$ \\
 5 & 35.642671 & $-2.74\times10^{-3}$ & $-2.738820\times10^{-3}$ & 0.0431 & $6.1\times10^{-3}$
   & 7.158618  & $-1.25\times10^{-3}$ & $-1.248266\times10^{-3}$ & 0.1387 & $4.8\times10^{-1}$ \\
 6 & 39.526217 & $-1.43\times10^{-3}$ & $-1.425915\times10^{-3}$ & 0.2857 & $6.5\times10^{-3}$
   & 7.964837  & $-5.37\times10^{-4}$ & $-5.370901\times10^{-4}$ & 0.0168 & $5.6\times10^{-1}$ \\
 7 & 43.236912 & $-7.69\times10^{-4}$ & $-7.686286\times10^{-4}$ & 0.0483 & $7.1\times10^{-3}$
   & 8.742549  & $-2.41\times10^{-4}$ & $-2.406249\times10^{-4}$ & 0.1556 & $6.7\times10^{-1}$ \\
 8 & 46.801404 & $-4.26\times10^{-4}$ & $-4.258161\times10^{-4}$ & 0.0432 & $7.8\times10^{-3}$
   & 9.494295  & $-1.12\times10^{-4}$ & $-1.117465\times10^{-4}$ & 0.2263 & $8.0\times10^{-1}$ \\
 9 & 50.239997 & $-2.41\times10^{-4}$ & $-2.411614\times10^{-4}$ & 0.0670 & $8.7\times10^{-3}$
   & 10.222620 & $-5.35\times10^{-5}$ & $-5.352206\times10^{-5}$ & 0.0412 & $9.6\times10^{-1}$ \\
\bottomrule
\end{tabular}
\end{table*}
More explicitly, the sign of the scalar-flux through the horizon is
controlled by $k_H=\omega_{mn}-m\Omega_H,$
where $\omega_{mn}$ is the frequency of the scalar mode and
$\Omega_H$ is the angular velocity of the black hole horizon. Modes
satisfying $0<\omega_{mn}<m\Omega_H$ have $k_H<0$ and therefore lie in the superradiant regime. The corresponding flux at horizon is negative, indicating a net extraction of rotational energy from the black hole.
\begin{figure*}[htb]
\centering
\includegraphics[width=7.0in, height=3.7in]{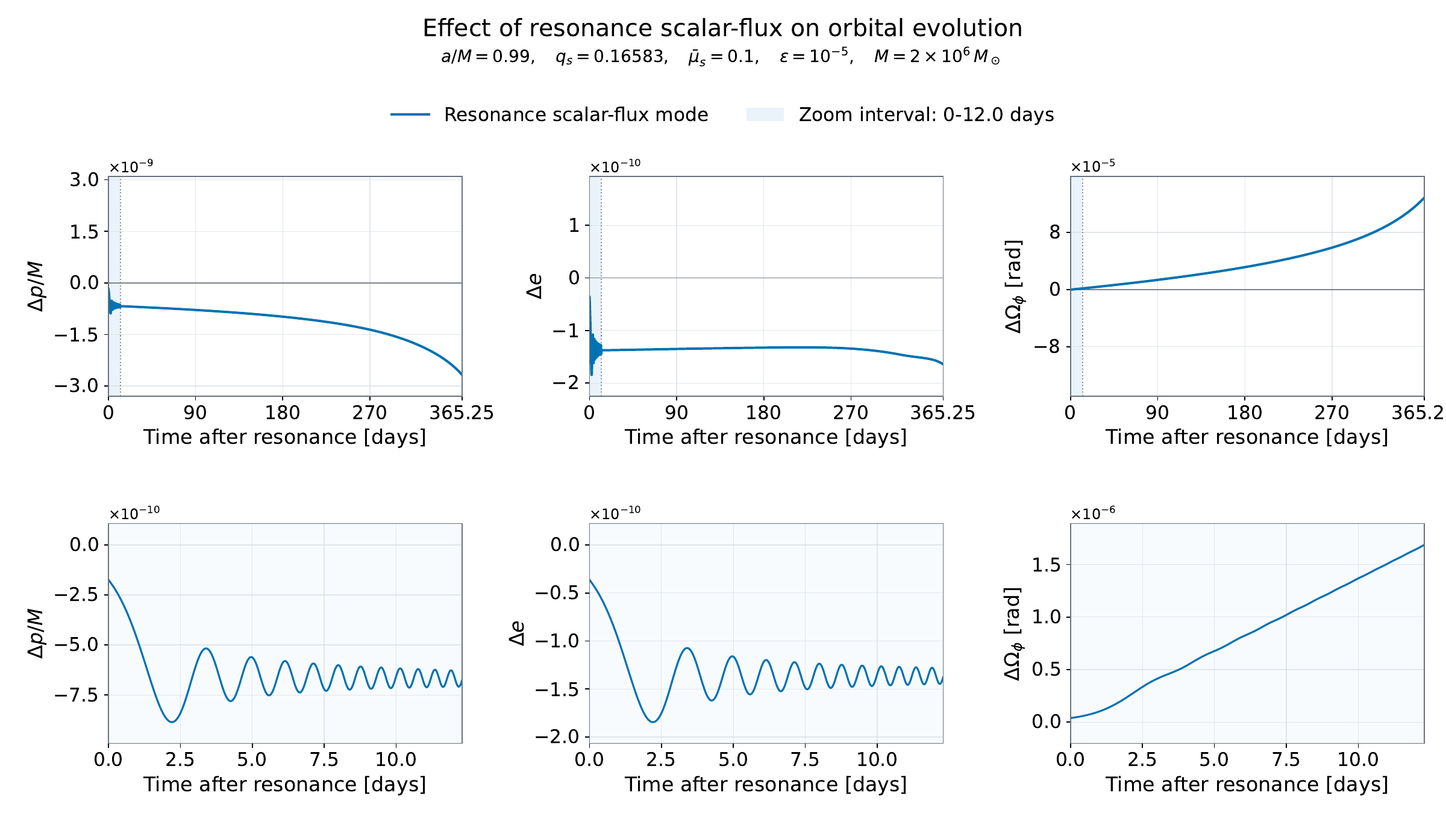}
\caption{Effect of massive-scalar resonance energy flux of BH-horizon on orbital-parameter evolution as function of inspiral time are plotted for scalar mass $\bar{\mu}_s=0.1$ and scalar-charge $q_s=0.16583$. Other intrinsic parameters are set as follows: $q_s=0.1$ and mass-ratio $\epsilon=10^{-5}$.}\label{fig:scalarflux:resonance}
\end{figure*}
To validate our numerical  validity  of the resonant massive-scalar fluxes for eccentric orbits, following the method of Fujita and Cardoso in Ref.~\cite{Fujita:2016yav}, Table~\ref{tab:dEdtH:floating} compares the massive-scalar fluxes at horizon
obtained with our code against the corresponding results reported in
Table~II of ~\cite{Fujita:2016yav}, using the same resonant orbital
configurations. The comparison contains 52 modes spanning
$e=0.1,\,0.3,\,0.5,$ and $0.7$, for two representative values of the
dimensionless scalar mass, $\bar{\mu}_s=0.01$ and $0.1$. We compute
the scalar flux at horizon normalized by $\mathcal N=q_s^2 M^2 m_p^2$
and quantify the mode-by-mode agreement through the absolute relative
deviation
\begin{equation}    
\left|\delta_{\dot E}\right|[\%]
=
100
\left|
\frac{
\dot E_{\rm calc}^{s,H}
-
\dot E_{\rm ref}^{s,H}
}{
\dot E_{\rm ref}^{s,H}
}
\right|.
\end{equation}

After correcting the sign convention consistently between the reference
and calculated fluxes, the numerical agreement is excellent over the
entire set of configurations. Across the 52 modes in
Table~\ref{tab:dEdtH:floating}, the median absolute relative deviation is
$0.0636\%$, while the maximum deviation is $0.3597\%$. In particular,
35 modes agree with the reference values to better than $0.1\%$, 46
modes to better than $0.2\%$, and 50 modes to better than $0.3\%$.
The largest deviation occurs for
$\bar{\mu}_s=0.10$, $e=0.3$, and $k=0$, for which
$\left|\delta_{\dot E}\right|=0.3597\%$. The sub-percent agreement
obtained mode by mode therefore provides a stringent validation of our
numerical calculation of the resonant massive-scalar horizon flux for
eccentric EMRIs.

This table also lists the critical value of scalar charge $q_s^{\rm crit}$ where the negative flux at horizon compensates the positive
gravitational flux losses and hence permit a floating-orbit configuration. The
critical value of scalar-charge exhibits a strong dependence on the scalar mass, eccentricity, and harmonic index. Over the modes considered here,
$q_s^{\rm crit}$ ranges from $1.7\times10^{-3}$ to $2.2$.
The smallest value, $q_s^{\rm crit}=1.7\times10^{-3},$
occurs for $\bar{\mu}_s=0.01$, $e=0.1$, and $k=0$. For the lighter
scalar mass, several resonances require only
$q_s^{\rm crit}\sim10^{-3}$--$10^{-2}$, whereas the
$\bar{\mu}_s=0.10$ configurations generally require substantially larger
charges. In addition, the range of resonant harmonic indices represented
in the table broadens as the eccentricity increases, reflecting the richer
harmonic content of eccentric orbital motion. These results show that the
possibility of a floating configuration depends on higher mode rather
than being controlled by the eccentricity or scalar mass alone.

Finally, we assess the impact of resonant massive-scalar fluxes on the secular evolution of eccentric EMRIs. To isolate the contribution of resonance scalar-flux, we compare two orbital-parameter evolutions. In the first case, denoted by the superscript ``res'', the scalar flux is obtained by summing all relevant multipolar modes, including the contribution of resonant scalar-flux $\dot{E}_{s}^{\rm H,res}$. In the second case, denoted by ``non-res'', the resonant component is removed from the scalar fluxes at horizon, while all remaining nonresonant scalar contributions are retained. We then quantify the accumulated differences in the orbital semilatus rectum, eccentricity and azimuthal orbital frequency according to
\begin{eqnarray}
\Delta p/M &\equiv & \left[p^{\rm non-res}(t)-p^{\rm res}(t)\right]/M\;,\\
\Delta e &\equiv&
e^{\rm non-res}(t)-e^{\rm res}(t)\;,\\
\Delta \Omega_\phi &\equiv&
\Omega_\phi^{\rm non-res}(t)-\Omega_\phi^{\rm res}(t)\;.
\end{eqnarray}
Here, the ``res''  includes the full massive-scalar flux, whereas the ``non-res'' excludes the  contribution of resonant scalar-flux. Figure~\ref{fig:scalarflux:resonance} shows these quantities over a one-year inspiral for a representative configuration with scalar charge $q_s=0.16583$ and dimensionless scalar mass $\bar{\mu}_s=0.1$. For the parameters considered here, both $\Delta p/M$ and $\Delta e$ exhibit small oscillatory features during the early stage of the inspiral, followed by a gradual secular growth as EMRI system evolves. A similar cumulative behavior is observed in $\Delta\Omega_\phi$. These results indicate that, although the resonant scalar flux can produce a discernible modification of the orbital evolution, its effect on the instantaneous orbital elements remains comparatively subtle for the configuration studied here. Consequently, identifying a floating- or sinking-orbit signature solely from the evolution of $p$ and $e$ can be challenging. The cumulative modification of the orbital frequency, and hence of the GW phase, is therefore expected to provide a more sensitive diagnostic of the resonant dynamics.}

\bibliography{0_ref}
\bibliographystyle{utphys1}
\end{document}